\documentclass[10pt,apjl]{emulateapj}

\usepackage{apjfonts}
\usepackage{morefloats}

\newcommand{\teff}{$T_{\rm eff}$} 
\newcommand{\logg}{$\log g$} 
\newcommand{\kms}{km s$^{-1}$}
\newcommand{\fei}{Fe\,{\sc i}}
\newcommand{\feii}{Fe\,{\sc ii}}
\newcommand{\tii}{Ti\,{\sc i}}
\newcommand{\tiii}{Ti\,{\sc ii}}

\usepackage[dvips]{color}

\shorttitle{The Most Metal-Poor Stars. II.} 
\shortauthors{Yong et al.}

\begin{document}

\title{THE MOST METAL-POOR STARS. II. CHEMICAL ABUNDANCES OF 
190 METAL-POOR STARS INCLUDING 10 NEW STARS WITH 
[FE/H] $\le$ $-$3.5\altaffilmark{1,2,3}}

\author{
DAVID YONG\altaffilmark{4},
JOHN E.\ NORRIS\altaffilmark{4}, 
M.\ S.\ BESSELL\altaffilmark{4},
N.\ CHRISTLIEB\altaffilmark{5},
M.\ ASPLUND\altaffilmark{4,6},
TIMOTHY C.\ BEERS\altaffilmark{7,8},
P.\ S.\ BARKLEM\altaffilmark{9}, 
ANNA FREBEL\altaffilmark{10}, AND 
S.\ G.\ RYAN\altaffilmark{11}
}

\altaffiltext{1}{This paper includes data gathered with the 6.5 meter 
Magellan Telescopes located at Las Campanas Observatory, Chile.} 

\altaffiltext{2}{Some of the data presented herein were obtained at the 
W.\ M.\ Keck Observatory, which is operated as a scientific partnership 
among the California Institute of Technology, the University of California 
and the National Aeronautics and Space Administration. The Observatory was 
made possible by the generous financial support of the W.\ M.\ Keck 
Foundation.}

\altaffiltext{3}{Based on observations collected at the 
European Organisation for Astronomical Research in the 
Southern Hemisphere, Chile (proposal 281.D-5015).}

\altaffiltext{4}{Research School of Astronomy and Astrophysics, The
Australian National University, Weston, ACT 2611, Australia;
yong@mso.anu.edu.au, jen@mso.anu.edu.au, bessell@mso.anu.edu.au,
martin@mso.anu.edu.au}

\altaffiltext{5}{Zentrum f\"ur Astronomie der Universit\"at
Heidelberg, Landessternwarte, K{\"o}nigstuhl 12, D-69117 Heidelberg,
Germany; n.christlieb@lsw.uni-heidelberg.de}

\altaffiltext{6}{Max-Planck Institute for Astrophysics, 
Karl-Schwarzschild Str. 1, 85741, Garching, Germany} 

\altaffiltext{7}{National Optical Astronomy Observatory, Tucson, AZ 85719} 

\altaffiltext{8}{Department of Physics \& Astronomy and JINA: Joint
Institute for Nuclear Astrophysics, Michigan State University,
E. Lansing, MI 48824, USA; beers@pa.msu.edu}

\altaffiltext{9}{Department of Physics and Astronomy, Uppsala
 University, Box 515, 75120 Uppsala, Sweden;
 paul.barklem@physics.uu.se}

\altaffiltext{10}{Massachusetts Institute of Technology, 
Kavli Institute for Astrophysics and Space Research, Cambridge, MA 02139, USA;
afrebel@mit.edu} 

\altaffiltext{11}{Centre for Astrophysics Research, School of Physics,
 Astronomy \& Mathematics, University of Hertfordshire, College Lane,
 Hatfield, Hertfordshire, AL10 9AB, UK; s.g.ryan@herts.ac.uk}

\begin{abstract}

We present a homogeneous chemical abundance analysis of 16 elements in
190 metal-poor Galactic halo stars (38 program and 152 
 literature objects). The sample includes 171 stars with [Fe/H] $\le$
 $-$2.5, of which 86 are extremely metal poor, [Fe/H] $\le$ $-$3.0.
 Our program stars include ten new objects with [Fe/H] $\le$
 $-$3.5.  We identify a sample of ``normal'' metal-poor stars and measure the
 trends between [X/Fe] and [Fe/H], as well as the dispersion about
 the mean trend for this sample. Using this mean trend, we identify
 objects that are chemically peculiar relative to ``normal'' stars
at the same metallicity. These chemically unusual stars include
CEMP-no objects, one star with high [Si/Fe], another with high
[Ba/Sr], and one with unusually low [X/Fe] for all elements heavier
than Na.  The Sr and Ba abundances indicate that there may be two
 nucleosynthetic processes at lowest metallicity that are distinct
 from the main $r$-process.  Finally, for many elements, we find a
 significant trend between [X/Fe] versus \teff\ which likely reflects
 non-LTE and/or 3D effects. Such trends demonstrate that care 
 must be exercised when using abundance measurements in metal-poor stars
 to constrain chemical evolution and/or nucleosynthesis predictions.

\end{abstract}


\section{INTRODUCTION}
\label{sec:intro}

 The atmospheres of low-mass stars contain detailed information on the
 chemical composition of the interstellar medium at the time and place
 of their birth. Thus, studies of the most metal-poor stars of the
 Galactic halo arguably offer the best means with which to understand
 the properties of the first stars (e.g.,
 \citealt{chamberlain51,wallerstein63,carney81,bessell84,mcwilliam95,ryan96,norris01,johnson02,cayrel04,beers05,cohen08,lai08,frebel11}). 
	 In recent times, 
         four stars with an iron content less than $\sim1/30,000$ that 
         of the Sun have been discovered -- 
         HE~0107$-$5240 \citep{christlieb02,christlieb04}, 
         HE~1327$-$2326 \citep{frebel05,aoki06}, and 
         HE~0557$-$4840 \citep{norris07},  
         within the Hamburg/ESO Survey (HES; \citealt{hes}), and SDSS 
         J102915+172927 \citep{caffau11}, in the Sloan Digital Sky 
         Survey Data Release 7 \citep{sdss7}.  The abundance patterns 
         of these stars can constrain 
         yields from zero metallicity supernovae (e.g., 
         \citealt{limongi03,umeda05,meynet06,tominaga07,heger10,limongi12}). 

Chemical abundance 
studies of metal-poor stars with increasing accuracy 
and precision have, in some cases, revealed extremely small scatter in 
abundance ratios at low metallicity \citep{cayrel04,arnone05}. 
Such results place strong constraints on the 
yields of the progenitor stars, as well as on the relative contributions of 
intrinsic scatter and measurement errors. 
Meanwhile, parallel studies have started to identify a variety of 
chemically diverse objects 
(e.g., \citealt{aoki05,cohen07,lai09}), 
which may indicate that ``we are beginning to see the 
anticipated and long sought stochastic effects of 
individual supernova events contributing to the 
Fe-peak material within a single star'' \citep{cohen08}. 

As the numbers of metal-poor stars with detailed chemical abundance 
measurements have grown, databases have been 
compiled \citep{saga,saga2,frebel10} which facilitate studies 
of the 
global characteristics of metal-poor stars 
(see also \citealt{roederer09}, who assembled and 
studied a compilation of nearly 700 halo stars). 
However, it is difficult, perhaps impossible, to understand 
whether the trends and dispersions in 
[X/Fe]\footnote{We adopt the usual spectroscopic notations that
[A/B]~$\equiv$ log$_{\rm 10}$(N$_{\rm A}$/N$_{\rm B}$)$_{\star}$~--
log$_{\rm 10}$(N$_{\rm A}$/N$_{\rm B}$)$_{\sun}$, and
that log~$\varepsilon$(B)~=~A(B)~$\equiv$
log$_{\rm 10}$(N$_{\rm B}$/N$_{\rm H}$)~$+$~12.00,
for elements A and B.}
versus [Fe/H] 
are real, or an artefact of an inhomogeneous comparison. 
Moreover, in the absence of a careful, homogeneous analysis, 
subtle effects may be overlooked.

In order to make further progress in this field, there is 
a clear need for a detailed homogeneous chemical abundance 
analysis of a large sample of metal-poor 
stars. To this end, \citet{barklem05} studied 253 stars, presenting 
chemical abundance measurements for some 22 elements. Their study 
included 49 new stars with [Fe/H] $<$ $-$3, but only one of 
which had [Fe/H] $\le$ $-$3.5, thereby highlighting the difficulty 
of finding stars in this metallicity regime. 

This is the second paper in our series on the discovery and analysis
of the most metal-poor stars. Here, we present a
homogeneous chemical abundance analysis for 38 program stars and a
further 152 literature stars. Our program stars include ten new
objects with [Fe/H] $\le$ $-$3.5, and the combined sample includes 86 
extremely metal-poor stars, [Fe/H] $\le$ $-$3.0. To our knowledge,
this represents one of 
the largest homogeneous chemical abundance
analyses of the most metal-poor stars to date based on model
atmosphere analysis of equivalent widths measured in high-resolution,
high signal-to-noise ratio (S/N) spectra. 
The outline of the paper is as follows. 
In Section~\ref{sec:obs}, we describe the analysis of the 
38 program stars. In Section~\ref{sec:firststars}, we compare our 
chemical abundances with those of the {\sc First Stars} 
group \citep{cayrel04,spite05,francois07}. 
In Section~\ref{sec:lit}, we describe our homogeneous 
re-analysis of 207 literature stars; in Section~\ref{sec:comp}, 
we compare these results with the literature values. 
In Section~\ref{sec:nlte}, we consider non-LTE effects. 
Finally, our results, interpretation, and conclusions are presented 
in Sections~\ref{sec:results} and \ref{sec:conc}.

\section{OBSERVATIONS AND ANALYSIS OF 38 PROGRAM STARS}
\label{sec:obs}

\subsection{Stellar Parameters: \teff, $\log g$, $\xi_t$, and [Fe/H]}

In Norris et al. (2012a; Paper I) we describe high-resolution spectroscopic 
observations of 38 program stars (obtained using the Keck, Magellan, and 
VLT telescopes), including the discovery and sample selection, 
equivalent-width measurements, radial velocities, 
and line list. 
Our sample comprises 34 stars original to the present work, 
together with four for which published abundances already 
exist, or which are the subject of analyses currently underway. 
In Paper I, we also describe 
the temperature scale, which consists of spectrophotometry and Balmer-line 
analysis. 
We refer the reader to these works for the details of 
the observational data upon which the present analysis is based.

With effective temperatures, \teff, and equivalent widths in hand, our
analysis proceeded in the following manner.  
Surface gravities\footnote{For 29 program stars, 
the spectrophotometric and Balmer-line analyses provided 
agreement on the evolutionary status: dwarf, subgiant, giant, or horizontal branch. 
For the nine remaining program stars, there was disagreement between 
spectrophotometric and Balmer-line analyses on the evolutionary status: 
dwarf vs.\ subgiant in all cases. We therefore conducted two analyses of 
each of these nine stars, one assuming a 
``dwarf'' gravity and the other assuming a ``subgiant'' gravity.}, $\log
g$, were taken from the $Y^2$ isochrones \citep{y2isochrones}, assuming
an age of 10 Gyr and [$\alpha$/Fe] = +0.3.  We note that changing the
age from 10 Gyr to 13 Gyr (or 7 Gyr) would only introduce a systematic
difference in $\log g$ of $\le$ 0.1 dex.  We also note that these
isochrones only extend down to [Fe/H] = $-$3.5, therefore the
surface gravity we obtain for more metal-poor stars involves a linear
extrapolation, from [Fe/H] = $-$3.5 down to [Fe/H] = $-$4.1, for the
most metal-poor stars in our program sample.  For our four most
metal-poor stars, we note that the average
difference between the surface gravity inferred using [Fe/H] = $-$3.5
(the boundary value of the $Y^2$ isochrones) and the extrapolated
surface gravity using the actual [Fe/H] is 0.06 dex (for the
giant/subgiant case) and 0.01 dex (for the dwarf case).  Initial
estimates of the metallicity came from the medium-resolution follow-up
spectroscopy. 

Model atmospheres were taken from the $\alpha$-enhanced, [$\alpha$/Fe] = +0.4, 
NEWODF grid of ATLAS9 models by \citet{castelli03}. These 
one-dimensional, plane-parallel, local thermodynamic equilibrium (LTE) 
models were computed using a microturbulent velocity of 2 \kms\ 
and no convective overshooting. Interpolation 
within the grid was necessary to produce models with the required 
combination of \teff, $\log g$, and [M/H]. The interpolation software, 
kindly provided by Dr Carlos Allende Prieto, has been used extensively 
(e.g., \citealt{bdp03,s4n}), and involves linear interpolation in 
three dimensions (\teff, 
$\log g$, and [M/H]) to produce the required model. 

The final tool in our analysis kit was 
the 
LTE stellar line-analysis program MOOG 
\citep{moog}. The particular version of MOOG that we used 
includes a proper treatment of continuum scattering (see 
\citealt{sobeck11} for further details). We refer the reader to 
\citet{cayrel04} and \citet{sobeck11} for 
a discussion regarding the importance of Raleigh scattering 
\citep{griffin82} 
at blue wavelengths in metal-poor stars. 

Having computed the abundance for each line, 
the microturbulent velocity was determined, in the usual way, by forcing 
the abundances from \fei\ lines to have no trend with the 
reduced equivalent width, $\log (W_\lambda/\lambda)$. 
The metallicity, [Fe/H], was inferred exclusively from 
\fei\ lines. While we are mindful that such lines are more 
susceptible to non-LTE effects than \feii\ lines (e.g., \citealt{asplund05}), 
we were unable to measure any \feii\ lines for 
a number of program stars. 
Higher-quality spectra are necessary 
to measure additional \feii\ lines in our sample. 
In Section 6 we shall compare iron abundances derived from 
neutral and ionized species for those stars for which data are available.

With an updated estimate of the metallicity, [Fe/H]$_{\rm star}$, we
then compared this value with the metallicity assumed when generating
the model atmosphere, [M/H]$_{\rm model}$.  If the difference exceeded
0.1 dex, we computed an updated model atmosphere with [M/H]$_{\rm
 new}$ = [Fe/H]$_{\rm star}$.  Based on the new metallicity, the
surface gravity was revised
and the star was re-analyzed. 
When this was 
required, we note that the abundance from 
\fei\ lines and the metallicity of the model atmosphere 
converged within one or two iterations. 
That is, the inferred abundance, [Fe/H], is only weakly dependent 
on the metallicity of the model, [M/H], provided the initial guess is close to 
the final value. During the analysis 
process, we removed \fei\ lines for which the abundance 
differed from ($i$) the median abundance by more than 0.5 dex 
or ($ii$) the median abundance by more than 3-$\sigma$. 
(Lines yielding abundances higher or lower 
than the median value by more than 0.5 dex or 3-$\sigma$ were rejected.) 
This criterion resulted in the rejection of a handful of lines for a 
given star. The largest number of rejected lines for a given star was six, 
leaving 33 accepted lines, while the largest fraction of rejected lines 
was three, leaving 15 accepted lines. 

\begin{figure}[t!] 
\epsscale{1.2}
\plotone{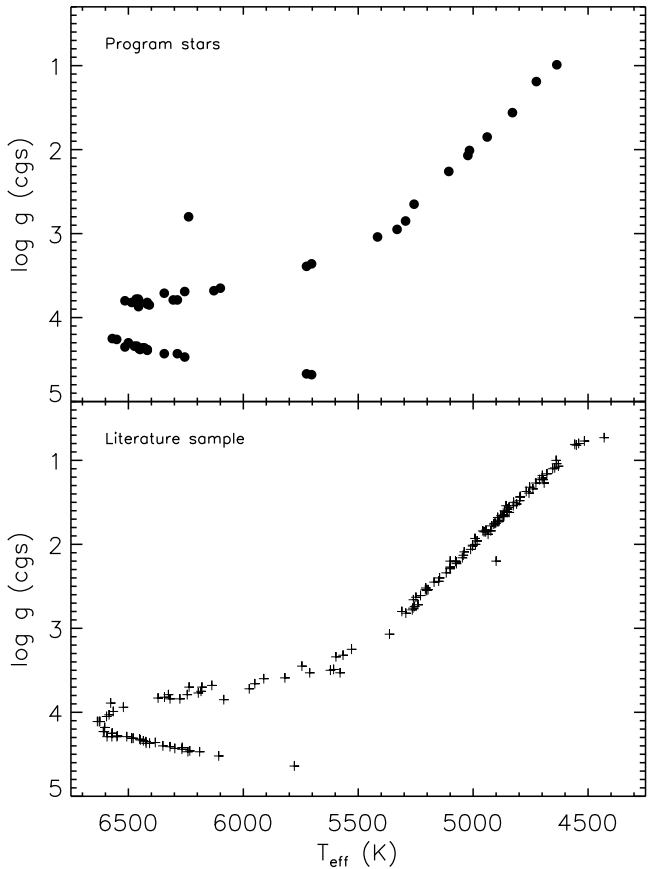} 
\caption{\teff\ vs.\ \logg\ for our sample (upper panel) and for the literature 
sample (lower panel). Note the location of the horizontal branch star (HE~1142$-$1422) in 
the upper panel. 
\label{fig:cmd}}
\end{figure}

In the course of our analysis, another consideration was whether or
not a given Fe line might be blended with CH molecular lines. 
Therefore, we repeated
the entire analysis using a subset of lines which spectrum synthesis
suggests are not blended with CH (see \citealt{norris97,norris10b} for
further details).  For the microturbulent velocity and metallicity,
the results from the two approaches are very similar.  Once we had
measured the [C/Fe] abundance ratio, we adopted the results using the
CH-free line list if the program star was a carbon enhanced metal-poor
(CEMP) object (we applied the \citealt{aoki07} CEMP definition).
Details on the C measurements and CEMP definition are provided in
Sections 2.2 and 7.1.  In Table \ref{tab:param}, we present the
stellar parameters for the program stars. The evolutionary status,
\teff\ vs.\ $\log g$, for the program stars is shown in Figure
\ref{fig:cmd}.

\begin{deluxetable*}{lcccccccc}
\tablecolumns{6} 
\tablewidth{0pc} 
\tabletypesize{\footnotesize}
\tablecaption{Model Atmosphere Parameters and [Fe/H] for the 38 Program Stars \label{tab:param}}
\tablehead{ 
\colhead{Star} & 
\colhead{RA2000\tablenotemark{a}} & 
\colhead{DEC2000\tablenotemark{a}} &
\colhead{\teff} & 
\colhead{$\log g$} & 
\colhead{$\xi_t$} & 
\colhead{[M/H]$_{\rm model}$} &
\colhead{[Fe/H]$_{\rm derived}$} &
\colhead{C-rich\tablenotemark{b}} \\
\colhead{} & 
\colhead{} & 
\colhead{} & 
\colhead{(K)} & 
\colhead{(cgs)} & 
\colhead{(km s$^{-1}$)} & 
\colhead{} &
\colhead{} \\
\colhead{(1)} &
\colhead{(2)} &
\colhead{(3)} &
\colhead{(4)} &
\colhead{(5)} &
\colhead{(6)} &
\colhead{(7)} &
\colhead{(8)} &
\colhead{(9)} 
}
\startdata 
52972-1213-507 & 09 18 49.9 & +37 44 26.8 & 6463 & 4.34 & 1.2 & $-$3.0 & $-$2.98 & 1 \\
53327-2044-515\tablenotemark{c} & 01 40 36.2 & +23 44 58.1 & 5703 & 4.68 & 0.8 & $-$4.0 & $-$4.00 & 1 \\
53327-2044-515\tablenotemark{d} & 01 40 36.2 & +23 44 58.1 & 5703 & 3.36 & 1.5 & $-$4.1 & $-$4.09 & 1 \\
53436-1996-093 & 11 28 13.6 & +38 41 48.9 & 6449 & 4.38 & 1.3 & $-$3.5 & $-$3.53 & 0 \\
54142-2667-094 & 08 51 36.7 & +10 18 03.2 & 6456 & 3.87 & 1.4 & $-$3.0 & $-$2.96 & 0 \\
BS~16545-089 & 11 24 27.5 & +36 50 28.8 & 6486 & 3.82 & 1.4 & $-$3.4 & $-$3.44 & 0 \\
CS~30336-049 & 20 45 23.5 & $-$28 42 35.9 & 4725 & 1.19 & 2.1 & $-$4.1 & $-$4.10 & 0 \\
HE~0049-3948 & 00 52 13.4 & $-$39 32 36.9 & 6466 & 3.78 & 0.8 & $-$3.7 & $-$3.68 & 0 \\
HE~0057-5959 & 00 59 54.0 & $-$59 43 29.9 & 5257 & 2.65 & 1.5 & $-$4.1 & $-$4.08 & 1 \\
HE~0102-1213 & 01 05 28.0 & $-$11 57 31.1 & 6100 & 3.65 & 1.5 & $-$3.3 & $-$3.28 & 0 \\
HE~0146-1548 & 01 48 34.7 & $-$15 33 24.4 & 4636 & 0.99 & 2.1 & $-$3.5 & $-$3.46 & 1 \\
HE~0207-1423 & 02 10 00.7 & $-$14 09 11.1 & 5023 & 2.07 & 1.3 & $-$3.0 & $-$2.95 & 1 \\
HE~0228-4047\tablenotemark{c} & 02 30 33.7 & $-$40 33 54.8 & 6515 & 4.35 & 1.6 & $-$3.8 & $-$3.75 & 0 \\
HE~0228-4047\tablenotemark{d} & 02 30 33.7 & $-$40 33 54.8 & 6515 & 3.80 & 1.7 & $-$3.8 & $-$3.75 & 0 \\
HE~0231-6025 & 02 32 30.6 & $-$60 12 11.2 & 6437 & 4.36 & 1.8 & $-$3.1 & $-$3.10 & 0 \\
HE~0253-1331 & 02 56 06.7 & $-$13 19 27.0 & 6474 & 4.34 & 1.5 & $-$3.0 & $-$3.01 & 0 \\
HE~0314-1739 & 03 17 01.8 & $-$17 28 54.9 & 6570 & 4.25 & 1.1 & $-$2.9 & $-$2.86 & 0 \\
HE~0355-3728\tablenotemark{c} & 03 56 36.5 & $-$44 34 03.4 & 6418 & 4.39 & 1.4 & $-$3.4 & $-$3.41 & 0 \\
HE~0355-3728\tablenotemark{d} & 03 56 36.5 & $-$44 34 03.4 & 6418 & 3.84 & 1.5 & $-$3.4 & $-$3.41 & 0 \\
HE~0945-1435\tablenotemark{c} & 09 47 50.7 & $-$14 49 06.9 & 6344 & 4.43 & 1.2 & $-$3.8 & $-$3.77 & 0 \\
HE~0945-1435\tablenotemark{d} & 09 47 50.7 & $-$14 49 06.9 & 6344 & 3.71 & 1.4 & $-$3.8 & $-$3.78 & 0 \\
HE~1055+0104\tablenotemark{c} & 10 58 04.4 & +00 48 36.0 & 6287 & 4.43 & 1.3 & $-$2.9 & $-$2.87 & 0 \\
HE~1055+0104\tablenotemark{d} & 10 58 04.4 & +00 48 36.0 & 6287 & 3.79 & 1.5 & $-$2.9 & $-$2.89 & 0 \\
HE~1116-0054\tablenotemark{c} & 11 18 47.8 & $-$01 11 19.4 & 6454 & 4.37 & 1.6 & $-$3.5 & $-$3.49 & 0 \\
HE~1116-0054\tablenotemark{d} & 11 18 47.8 & $-$01 11 19.4 & 6454 & 3.80 & 1.6 & $-$3.5 & $-$3.48 & 0 \\
HE~1142-1422 & 11 44 59.2 & $-$14 38 49.6 & 6238 & 2.80 & 3.4 & $-$2.8 & $-$2.84 & 0 \\
HE~1201-1512\tablenotemark{c} & 12 03 37.0 & $-$15 29 33.0 & 5725 & 4.67 & 0.9 & $-$3.9 & $-$3.86 & 1 \\
HE~1201-1512\tablenotemark{d} & 12 03 37.0 & $-$15 29 33.0 & 5725 & 3.39 & 1.5 & $-$3.9 & $-$3.92 & 1 \\
HE~1204-0744 & 12 06 46.2 & $-$08 00 44.1 & 6500 & 4.30 & 1.8 & $-$2.7 & $-$2.71 & 0 \\
HE~1207-3108 & 12 09 54.0 & $-$31 25 10.6 & 5294 & 2.85 & 0.9 & $-$2.7 & $-$2.70 & 0 \\
HE~1320-2952 & 13 22 54.9 & $-$30 08 05.3 & 5106 & 2.26 & 1.5 & $-$3.7 & $-$3.69 & 0 \\
HE~1346-0427\tablenotemark{c} & 13 49 25.1 & $-$04 42 14.8 & 6255 & 4.47 & 1.2 & $-$3.6 & $-$3.57 & 0 \\
HE~1346-0427\tablenotemark{d} & 13 49 25.1 & $-$04 42 14.8 & 6255 & 3.69 & 1.4 & $-$3.6 & $-$3.58 & 0 \\
HE~1402-0523\tablenotemark{c} & 14 04 38.0 & $-$05 38 13.5 & 6418 & 4.38 & 1.0 & $-$3.2 & $-$3.17 & 0 \\
HE~1402-0523\tablenotemark{d} & 14 04 38.0 & $-$05 38 13.5 & 6418 & 3.82 & 1.2 & $-$3.2 & $-$3.19 & 0 \\
HE~1506-0113 & 15 09 14.3 & $-$01 24 56.6 & 5016 & 2.01 & 1.6 & $-$3.5 & $-$3.54 & 1 \\
HE~2020-5228 & 20 24 17.1 & $-$52 19 02.3 & 6305 & 3.79 & 1.4 & $-$2.9 & $-$2.93 & 0 \\
HE~2032-5633 & 20 36 24.9 & $-$56 23 05.8 & 6457 & 3.78 & 1.8 & $-$3.6 & $-$3.63 & 0 \\
HE~2047-5612 & 20 51 22.1 & $-$56 00 52.9 & 6128 & 3.68 & 0.9 & $-$3.1 & $-$3.14 & 0 \\
HE~2135-1924 & 21 38 04.7 & $-$19 11 04.4 & 6449 & 4.37 & 1.2 & $-$3.3 & $-$3.31 & 0 \\
HE~2136-6030 & 21 40 39.5 & $-$60 16 26.4 & 6409 & 3.85 & 2.0 & $-$2.9 & $-$2.88 & 0 \\
HE~2139-5432 & 21 42 42.4 & $-$54 18 42.9 & 5416 & 3.04 & 0.8 & $-$4.0 & $-$4.02 & 1 \\
HE~2141-0726 & 21 44 06.6 & $-$07 12 48.9 & 6551 & 4.26 & 1.5 & $-$2.7 & $-$2.72 & 0 \\
HE~2142-5656 & 21 46 20.4 & $-$56 42 19.1 & 4939 & 1.85 & 2.1 & $-$2.9 & $-$2.87 & 1 \\
HE~2202-4831 & 22 06 05.8 & $-$48 16 53.0 & 5331 & 2.95 & 1.2 & $-$2.8 & $-$2.78 & 1 \\
HE~2246-2410 & 22 48 59.6 & $-$23 54 39.0 & 6431 & 4.36 & 1.5 & $-$3.0 & $-$2.96 & 0 \\
HE~2247-7400 & 22 51 19.4 & $-$73 44 23.6 & 4829 & 1.56 & 2.0 & $-$2.9 & $-$2.87 & 1 \\
\enddata 

\tablenotetext{a}{Coordinates are from the 2MASS database \citep{2mass}.}
\tablenotetext{b}{1 = CEMP object, adopting the \citet{aoki07} definition and 0 = C-normal (see Section 7.1 for details).}
\tablenotetext{c}{For this set of results, a dwarf gravity is assumed (see Section 2.1 for details).}
\tablenotetext{d}{For this set of results, a subgiant gravity is assumed (see Section 2.1 for details).}

\end{deluxetable*}

 We note here that for the nine stars for which we conducted 
 separate analyses using a dwarf gravity and a subgiant gravity, the average 
   difference in iron abundance between the dwarf and subgiant analyses
   is $\langle$[Fe/H]$_{\rm dwarf}$ $-$ 
   [Fe/H]$_{\rm subgiant}\rangle$ = 0.02 $\pm$ 0.01 dex ($\sigma$ = 
   0.03); the relative abundance is  $\langle$[X/Fe]$_{\rm dwarf}$ $-$ 
   [X/Fe]$_{\rm subgiant}\rangle$ = 0.05 $\pm$ 0.02 dex ($\sigma$ = 0.16). 
 Given these modest differences, for a given star we average the
 abundance ratios [Fe/H] and [X/Fe] for the dwarf and subgiant
 cases. Unless noted otherwise, we use these average values for a
 given star throughout the remainder of the paper.  We
 report the values for the individual analyses in the relevant tables.

\subsection{Element Abundances}

The abundances of atomic lines were computed using our measured
equivalent widths and $\log gf$ values from Paper I, final model
atmospheres, and MOOG. Lines of Sc\,{\sc ii}, Mn\,{\sc i}, Co\,{\sc
  i}, and Ba\,{\sc ii} are affected by hyperfine splitting.  In our
analysis, we treated these lines appropriately, using the data 
from \citet{kurucz95}.  Additionally, Ba\,{\sc ii} lines are affected
by isotopic splitting. Our Ba abundances were computed assuming the
\citet{mcwilliam98} $r$-process isotopic composition and 
  hyperfine splitting.  For a restricted
  number of elements in some stars, we determined upper limits to the
  chemical abundance, based on equivalent-width limits presented in
  Paper I. 

Given the low metallicities of our sample and the S/N of 
  our spectra, we are not well-positioned to measure the abundance of 
  oxygen.  That said, we have determined the abundance (or its upper 
  limit) of this element for six C-rich (i.e., CEMP) stars in our 
  sample, which we shall discuss further in Paper IV (Norris 
  et al.\ 2012b, in preparation).  For HE~2139$-$5432, the oxygen 
abundance was derived from analysis of the 7771.94\AA, 7774.17\AA, and 
7775.39\AA\ lines.  The measured equivalent widths for 
  these lines are 20.1 m\AA, 18.7 m\AA, and 12.2 m\AA\ and the adopted $\log gf$ 
  values are 0.324, 0.174, and $-$0.046 respectively. Thus, we obtained 
an LTE abundance of A(O) = 7.82 ($\sigma$ = 0.05) for HE~2139$-$5432, 
which corresponds to [O/Fe] = +3.15 (we adopt the 
\citealt{asplund09} solar abundances). 
For HE~0146--1548 and HE~1506$-$0113, the O limits ([O/Fe] $<$ +1.63
and +2.32, respectively) were determined 
using an equivalent width of $<$10m\AA\ for 
the 6300.30\AA\ line, adopting $\log gf$ = $-$9.820. 
Finally, for 53327-2044-515, HE~0057-5959, and HE~1201$-$1512, the O
limits ([O/Fe] $<$ +2.81, +2.77, and +2.64, respectively) were determined
using an equivalent width of 10 m\AA\ for the 7771.94\AA\ line (where for
53327-2044-515 and HE~1201$-$1512 the abundance value for each star is
the mean of the low- and high-gravity solutions).

\begin{figure}[t!]
\epsscale{1.2}
\plotone{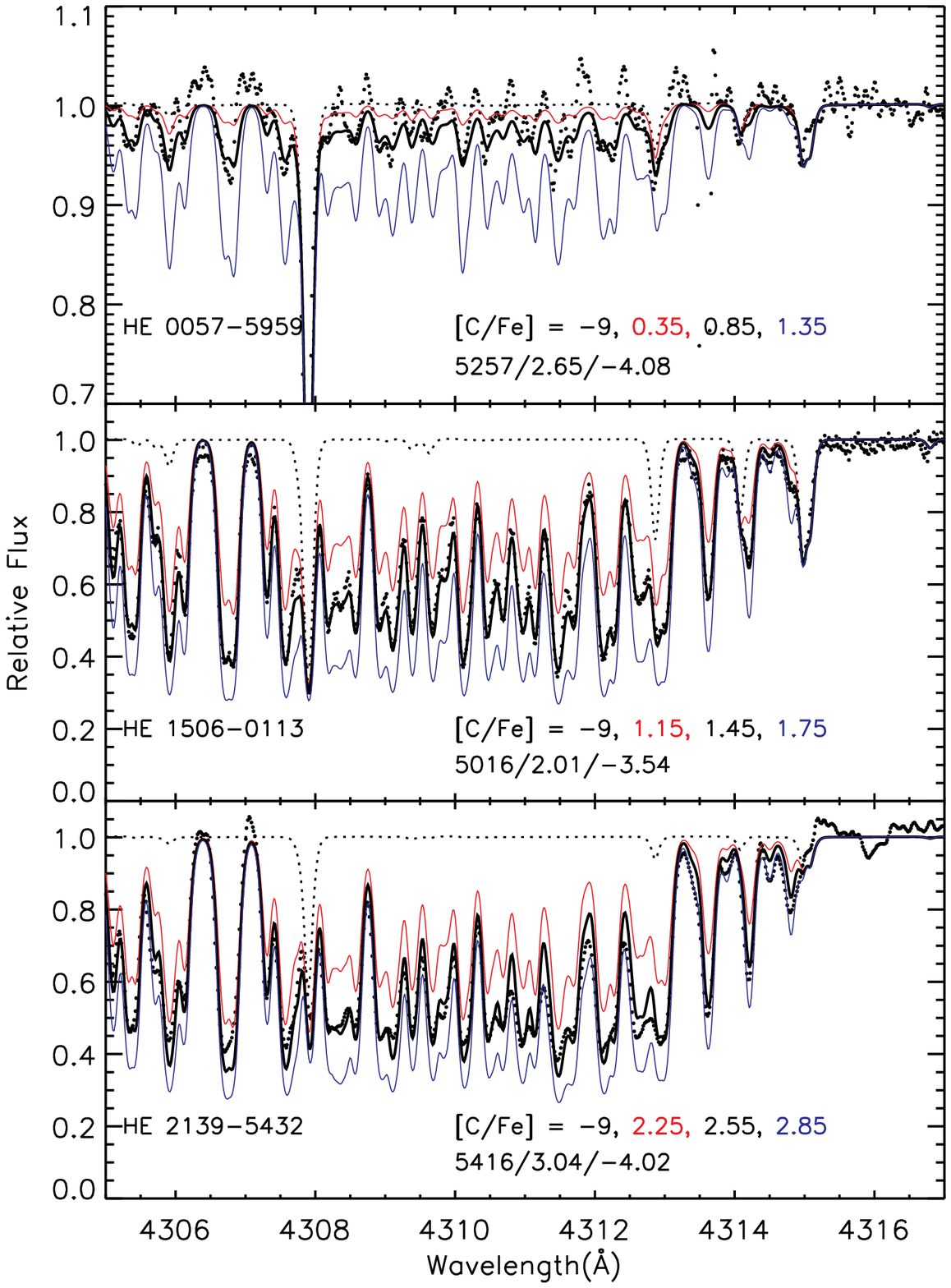} 
\caption{Comparison of observed (filled dots) and synthetic spectra  in the 
region 4305\AA\ to 4317\AA. Synthetic spectra with no C, [C/Fe] = $-$9, are 
shown as thin dotted lines. The best-fitting synthetic spectra are the thick 
black lines. Unsatisfactory fits are shown as red and blue thin lines. 
The stellar parameters, \teff/$\log g$/[Fe/H] are shown. 
\label{fig:ch1}}
\end{figure}

\begin{figure}[t!]
\epsscale{1.2}
\plotone{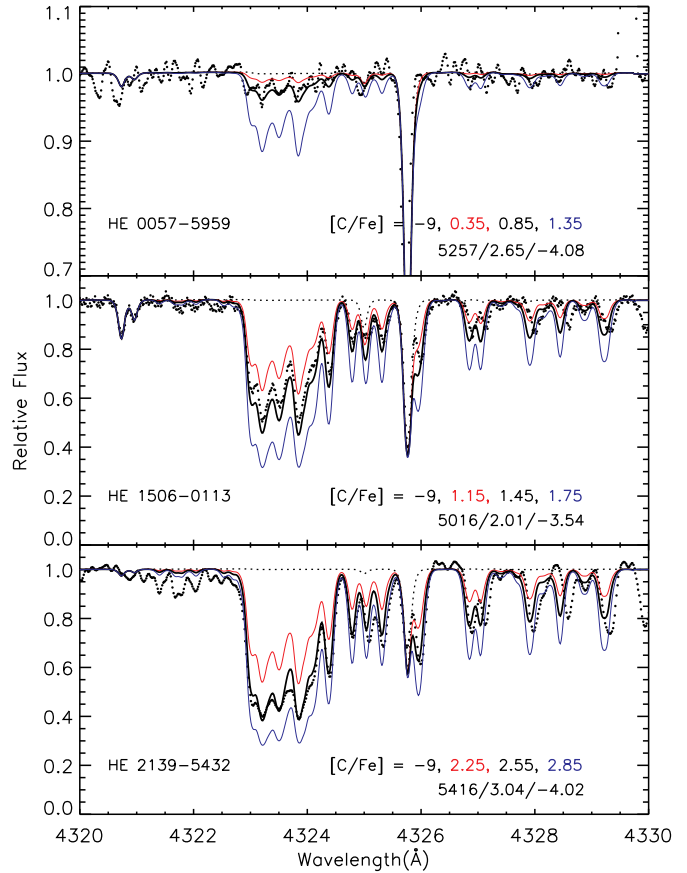} 
\caption{Same as Figure \ref{fig:ch1}, but for the region 4320\AA\ to 4330\AA. 
\label{fig:ch2}}
\end{figure}

\begin{figure}[t!] 
\epsscale{1.2}
\plotone{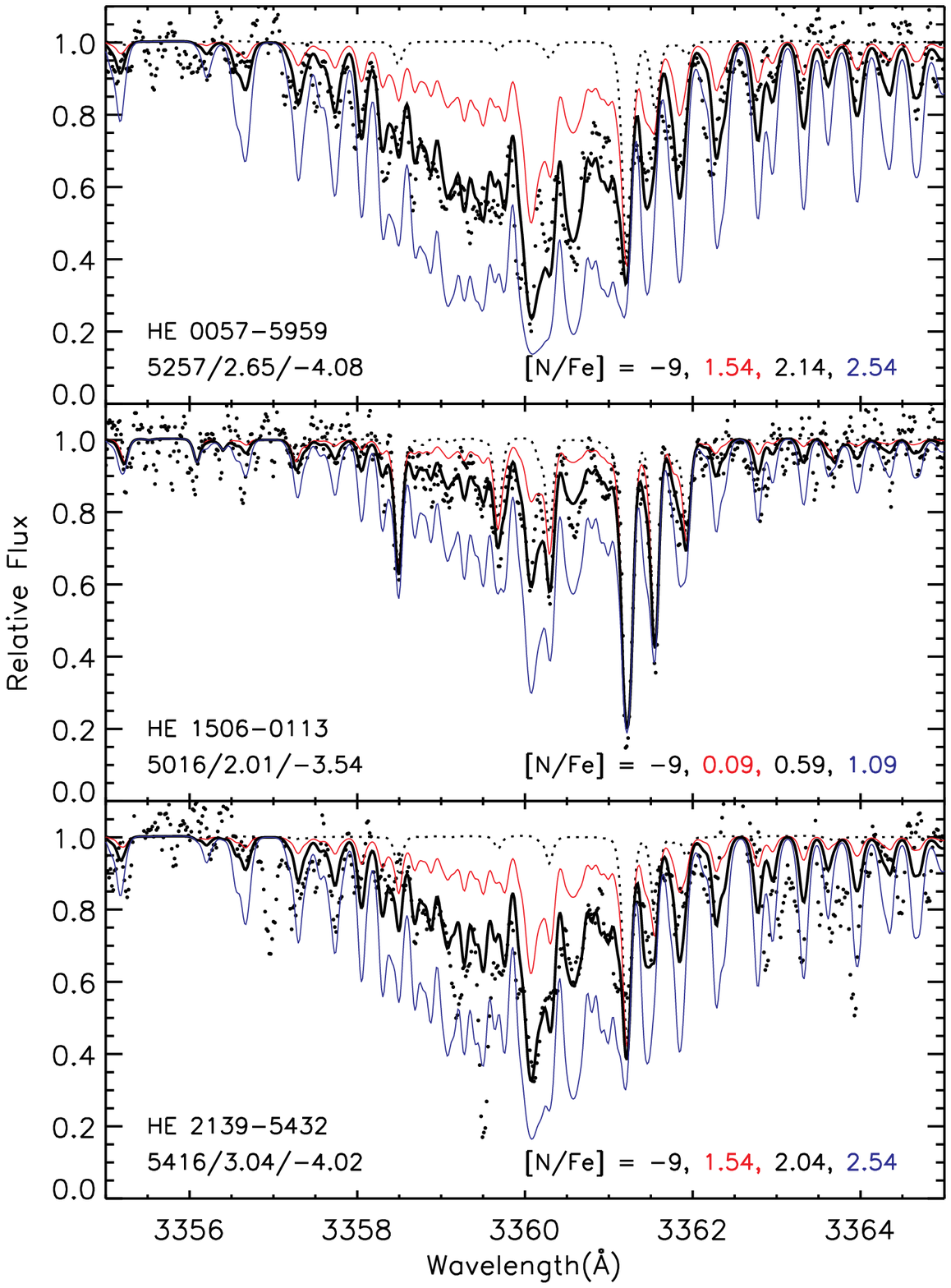} 
\caption{Same as Figure \ref{fig:ch1}, but for [N/Fe] in the region 3355\AA\ 
to 3365\AA. 
\label{fig:nh1}}
\end{figure}

For C and N, abundances (or upper limits) were determined
from analysis of the (0-0) and (1-1) bands of the $A-X$ electronic
transitions of the CH molecule (4290\AA\ to 4330\AA) and the NH
molecule (3350\AA\ to 3370\AA).  We compared synthetic spectra,
generated using MOOG, with the observed spectra and adjusted the input
abundance until the two spectra matched. 
The macroturbulent broadening was determined using a Gaussian 
representing the combined effects of the instrumental profile, 
atmospheric turbulence, and stellar rotation. The width of this 
Gaussian was estimated during the course of the spectrum synthesis 
fitting, and the C and N abundances are thus (slightly) 
sensitive to the adopted broadening. 
Following the analysis
described in \citet{norris10}, the CH line list was that compiled by
B.\ Plez, T.\ Masseron, \& S.\ Van Eck (2009, private
communication). We used a dissociation energy of 3.465 eV.  The
abundances of C and O are coupled through the CO molecule.  In the
absence of an O abundance measurement, we arbitrarily adopted a
halo-like value of [O/Fe] = +0.4, and note that for our program stars,
the derived C abundance is insensitive to the adopted O abundance 
(i.e., 
for a handful of stars, we adopted [O/Fe] = 0.0 and [O/Fe] = +1.5, and 
the derived C abundance was unchanged).  
The NH line list was the same as in \citet{johnson07}, in which the
Kurucz-$gf$ values were reduced by a factor of two, and the dissociation
energy was 3.450 eV. Given the low S/N at these wavelengths, we
smoothed the observed spectra with a 5-pixel boxcar function to
increase the S/N (at the expense of spectral resolution).  The N
abundance was adjusted until the synthetic spectra matched the
observed spectra.  In Figures \ref{fig:ch1}, \ref{fig:ch2}, and
\ref{fig:nh1}, we show examples of the spectrum synthesis.

As with the analysis of Fe lines, we repeated the element abundance 
analysis using the subset of lines believed to be unaffected by 
CH blends, combined with the stellar parameters obtained using the 
set of ``CH clean'' Fe lines. 
Depending on whether the star was found to be a CEMP object, we
adopted the element abundances associated with the relevant element
abundance analysis.  
In particular, we note that our line list has only one Si line, 
3905.52\AA, and that line is excluded in the ``CH clean'' set 
of lines. Therefore, there are CEMP stars with Si equivalent-width 
measurements in Paper I without Si abundance measurements. 
For these CEMP stars, we tried to measure Si abundances (or limits) 
from spectrum synthesis of the 4102.94\AA\ Si line. 
We present abundance ratios in Table \ref{tab:cn}
(C and N) and Table \ref{tab:abund} (Na to Ba).  The 
adopted solar abundances for all elements were from
\citet{asplund09}.

\begin{deluxetable}{lrrrr}
\tabletypesize{\footnotesize}
\tablecolumns{5} 
\tablewidth{0pc} 
\tablecaption{C and N Abundances for the 38 Program Stars \label{tab:cn}}
\tablehead{ 
\colhead{Star} & 
\colhead{A(C)} & 
\colhead{[C/Fe]} & 
\colhead{A(N)} &
\colhead{[N/Fe]} \\
\colhead{(1)} &
\colhead{(2)} &
\colhead{(3)} &
\colhead{(4)} &
\colhead{(5)} 
}
\startdata 
52972-1213-507 & 8.26 & 2.82 & \ldots & \ldots \\
53327-2044-515\tablenotemark{a} & 5.56 & 1.13 & \ldots & \ldots \\
53327-2044-515\tablenotemark{b} & 5.91 & 1.57 & \ldots & \ldots \\
53436-1996-093 & $<$6.56 & $<$1.66 & \ldots & \ldots \\
54142-2667-094 & $<$6.86 & $<$1.39 & \ldots & \ldots \\
BS~16545-089 & $<$6.76 & $<$1.77 & \ldots & \ldots \\
CS~30336-049 & $<$4.56 & $<$0.23 & 4.7 & 0.97 \\
HE~0049-3948 & $<$6.56 & $<$1.81 & $<$6.55 & $<$2.40 \\
HE~0057-5959 & 5.21 & 0.86 & 5.9 & 2.15 \\
HE~0102-1213 & $<$6.46 & $<$1.31 & \ldots & \ldots \\
HE~0146-1548 & 5.81 & 0.84 & \ldots & \ldots \\
HE~0207-1423 & 7.86 & 2.38 & \ldots & \ldots \\
HE~0228-4047\tablenotemark{a} & $<$6.56 & $<$1.88 & \ldots & \ldots \\
HE~0228-4047\tablenotemark{b} & $<$6.66 & $<$1.98 & \ldots & \ldots \\
HE~0231-6025 & $<$6.96 & $<$1.64 & \ldots & \ldots \\
HE~0253-1331 & $<$7.06 & $<$1.64 & \ldots & \ldots \\
HE~0314-1739 & $<$7.06 & $<$1.49 & \ldots & \ldots \\
HE~0355-3728\tablenotemark{a} & $<$7.16 & $<$2.14 & \ldots & \ldots \\
HE~0355-3728\tablenotemark{b} & $<$7.36 & $<$2.34 & \ldots & \ldots \\
HE~0945-1435\tablenotemark{a} & $<$6.36 & $<$1.70 & \ldots & \ldots \\
HE~0945-1435\tablenotemark{b} & $<$6.46 & $<$1.81 & \ldots & \ldots \\
HE~1055+0104\tablenotemark{a} & $<$6.76 & $<$1.20 & \ldots & \ldots \\
HE~1055+0104\tablenotemark{b} & $<$6.96 & $<$1.42 & \ldots & \ldots \\
HE~1116-0054\tablenotemark{a} & $<$6.66 & $<$1.72 & $<$7.25 & $<$2.91 \\
HE~1116-0054\tablenotemark{b} & $<$6.86 & $<$1.92 & $<$7.25 & $<$2.91 \\
HE~1142-1422 & $<$7.16 & $<$1.57 & $<$7.50 & $<$2.51 \\
HE~1201-1512\tablenotemark{d} & 5.71 & 1.14 & $<$5.20 & $<$1.23 \\
HE~1201-1512\tablenotemark{d} & 6.11 & 1.6 & $<$5.20 & $<$1.29 \\
HE~1204-0744 & $<$7.26 & $<$1.55 & \ldots & \ldots \\
HE~1207-3108 & $<$5.46 & $<-$0.27 & 5.55 & 0.42 \\
HE~1320-2952 & $<$5.26 & $<$0.52 & $<$5.00 & $<$0.86 \\
HE~1346-0427\tablenotemark{a} & $<$5.96 & $<$1.10 & \ldots & \ldots \\
HE~1346-0427\tablenotemark{b} & $<$6.16 & $<$1.31 & \ldots & \ldots \\
HE~1402-0523\tablenotemark{a} & $<$6.76 & $<$1.50 & $<$6.50 & $<$1.84 \\
HE~1402-0523\tablenotemark{b} & $<$6.86 & $<$1.62 & $<$6.60 & $<$1.96 \\
HE~1506-0113 & 6.36 & 1.47 & 4.9 & 0.61 \\
HE~2020-5228 & $<$7.16 & $<$1.66 & $<$7.20 & $<$2.30 \\
HE~2032-5633 & $<$7.16 & $<$2.36 & $<$6.80 & $<$2.60 \\
HE~2047-5612 & $<$6.66 & $<$1.37 & $<$6.60 & $<$1.91 \\
HE~2135-1924 & $<$6.86 & $<$1.74 & \ldots & \ldots \\
HE~2136-6030 & $<$7.26 & $<$1.71 & \ldots & \ldots \\
HE~2139-5432 & 7.01 & 2.59 & 5.9 & 2.08 \\
HE~2141-0726 & $<$7.26 & $<$1.55 & \ldots & \ldots \\
HE~2142-5656 & 6.51 & 0.95 & 5.5 & 0.54 \\
HE~2202-4831 & 8.06 & 2.41 & \ldots & \ldots \\
HE~2246-2410 & $<$6.86 & $<$1.39 & $<$7.15 & $<$2.28 \\
HE~2247-7400 & 6.26 & 0.7 & \ldots & \ldots \\
\enddata 

\tablenotetext{a}{For this set of results, a dwarf gravity is assumed (see Section 2.1 for details).}
\tablenotetext{b}{For this set of results, a subgiant gravity is assumed (see Section 2.1 for details).}

\end{deluxetable}

\begin{deluxetable}{lrcccr}
\tabletypesize{\small}
\tablecolumns{6} 
\tablewidth{0pc} 
\tablecaption{Chemical Abundances (Na-Ba) for the Program Stars \label{tab:abund}}
\tablehead{ 
\colhead{Star} & 
\colhead{A(X)} & 
\colhead{N$_{\rm lines}$} & 
\colhead{s.e.$_{\rm log \epsilon}$\tablenotemark{a}} & 
\colhead{Total Error\tablenotemark{b}} & 
\colhead{[X/Fe]} \\
\colhead{(1)} &
\colhead{(2)} &
\colhead{(3)} &
\colhead{(4)} &
\colhead{(5)} &
\colhead{(6)} 
}
\startdata
 \noalign{\vskip +0.5ex}
 \multicolumn{6}{c}{Na} \cr
 \noalign{\vskip  .8ex}%
 \hline
 \noalign{\vskip-2ex}\\
 52972-1213-507                 &    4.86 &      2 &     0.32 &     0.33 &    1.60 \\ 
 53327-2044-515\tablenotemark{c}&    2.35 &      1 &   \ldots &   \ldots &    0.11 \\ 
 53327-2044-515\tablenotemark{d}&    2.32 &      1 &   \ldots &   \ldots &    0.17 \\ 
 53436-1996-093                 &    2.58 &      1 &   \ldots &   \ldots & $-$0.13 \\ 
 54142-2667-094                 &  \ldots & \ldots &   \ldots &   \ldots &  \ldots 
\enddata

\tablenotetext{a}{Standard error of the mean.} 
\tablenotetext{b}{Total error is the quadratic sum of the updated random error and the systematic error (see Section 2.3 for details).} 
\tablenotetext{c}{Dwarf gravity is assumed (see Section 2.1 for details).}
\tablenotetext{d}{Subgiant gravity is assumed (see Section 2.1 for details).}

\tablenotetext{e}{Abundances, or limits, were determined from 
spectrum synthesis of the 4102.94\AA\ Si line.}
\tablenotetext{f}{Abundances limits were determined from 
the 4077.71\AA\ Sr line.}
\tablenotetext{g}{Abundances limits were determined from 
the 4554.03\AA\ Ba line.}

\tablerefs{ 
Note. Table 3 is published in its entirety in the electronic edition of The Astrophysical Journal. A portion is shown here for guidance regarding its form and content.
}

\end{deluxetable}

\subsection{Abundance Uncertainties}

Our abundance measurements are subject to uncertainties in the model 
parameters. We estimated these uncertainties to be \teff\ $\pm$ 100K, 
\logg\ $\pm$ 0.3 dex, $\xi_t$ $\pm$ 0.3 \kms, and 
[M/H] $\pm$ 0.3 dex. To determine the abundance errors, we 
repeated our analysis varying our parameters, one at a time, assuming 
that the errors are symmetric for positive and negative changes. 
We present these abundance uncertainty estimates in Table \ref{tab:parvar},  
in which the final column is the accumulated error when the four 
uncertainties are added quadratically. Strictly speaking, quadratic 
addition is appropriate if the errors are fully independent. Additionally, 
our approach neglects covariances, and we refer the interested reader to 
\citet{mcwilliam95}, \citet{johnson02}, and \citet{barklem05} 
for a more detailed discussion.  
Note that the contribution 
from the uncertainties in [M/H] to the total error budget is small, in general. 
Therefore, our condition to re-compute a model atmosphere only 
when $|$[M/H]$_{\rm model}~-~$[Fe/H]$_{\rm star}| > 0.1$ dex does not 
adversely affect our results. 

\begin{deluxetable}{lrrrrc}
\tabletypesize{\scriptsize}
\tablecolumns{6} 
\tablewidth{0pc} 
\tablecaption{Abundance Errors from Uncertainties in Atmospheric Parameters 
\newline 
for the 38 Program Stars \label{tab:parvar}}
\tablehead{ 
\colhead{Species} & 
\colhead{$\Delta$\teff} &
\colhead{$\Delta\log g$} &
\colhead{$\Delta\xi_t$} &
\colhead{$\Delta${\rm [M/H]}} &
\colhead{$\Delta${\rm [X/Fe]}} \\
\colhead{} &
\colhead{(100 K)} &
\colhead{(0.3 dex)} &
\colhead{(0.3 \kms)} &
\colhead{(0.3 dex)} &
\colhead{(dex)}  \\
\cline{2-6} 
\colhead{(1)} &
\colhead{(2)} &
\colhead{(3)} &
\colhead{(4)} &
\colhead{(5)} &
\colhead{(6)} 
}
\startdata 
 \noalign{\vskip +0.5ex}
 \multicolumn{6}{c}{52972-1213-507} \cr
 \noalign{\vskip  .8ex}%
 \hline
 \noalign{\vskip-2ex}\\
     $\Delta$[Na/Fe] &    0.00 & $-$0.08 &    0.00 &    0.04 &    0.09 \\
     $\Delta$[Mg/Fe] & $-$0.01 & $-$0.05 &    0.00 &    0.04 &    0.06 \\
     $\Delta$[Al/Fe] &  \ldots &  \ldots &  \ldots &  \ldots &  \ldots \\
     $\Delta$[Si/Fe] &  \ldots &  \ldots &  \ldots &  \ldots &  \ldots \\
     $\Delta$[Ca/Fe] &    0.01 & $-$0.07 &    0.00 & $-$0.03 &    0.08 
\enddata

\tablenotetext{a}{Dwarf gravity is assumed (see Section 2.1 for details).}
\tablenotetext{b}{Subgiant gravity is assumed (see Section 2.1 for details).}

\tablerefs{ 
Note. Table 4 is published in its entirety in the electronic edition of The Astrophysical Journal. A portion is shown here for guidance regarding its form and content.
}

\end{deluxetable}

To obtain the total error estimates given in Table \ref{tab:abund}, 
we follow \citet{norris10}. We replace the random error in 
Table \ref{tab:abund} (s.e.$_{\log \epsilon}$) by 
max(s.e.$_{\log \epsilon}$, 0.20/$\sqrt{N_{\rm lines}}$) where the 
second term is what would be expected for a set of $N_{\rm lines}$ 
with a dispersion of 0.20 dex (a conservative estimate for the 
abundance dispersion of \fei\ lines). The total error is obtained 
by quadratically adding this updated 
random error with the systematic error in 
Table \ref{tab:parvar}. 
Finally, we note that this 1D LTE analysis is subject to abundance 
uncertainties from 3D and non-LTE effects \citep{asplund05}.

\subsection{Comparison using Different Model Atmospheres} 

For a subset of the program stars, we computed abundances 
using the MARCS \citep{marcs} grid of model atmospheres. 
Three stars were chosen to sample a reasonable range 
of stellar parameters: two giants and one main sequence turn-off 
star. 
In Table \ref{tab:marcs} 
we show the abundance differences for A(X), in the sense 
\citet{castelli03} $-$ MARCS. 
For these representative objects, 
the maximum abundance difference was $\Delta$A(X) = 0.05 dex, 
and the minimum abundance difference was $\Delta$A(X) = $-$0.02 dex. 
When considering the ratio [X/Fe], we note that the maximum 
abundance difference was $\Delta$[X/Fe] = 0.02 dex, and the minimum 
abundance difference was $\Delta$[X/Fe] = $-$0.01 dex. 
We regard these abundance differences, $\Delta$A(X) and 
$\Delta$[X/Fe], to be small, especially 
when compared to the abundance uncertainties and errors in Tables 
\ref{tab:abund} and \ref{tab:parvar}. 
Therefore, we do not expect the choice of model atmosphere grid, 
\citet{castelli03} or MARCS \citep{marcs}, 
to significantly alter our abundance 
results or subsequent interpretation. 
However, we note that a more complete chemical abundance analysis would 
require, amongst other things, 
tailored models with appropriate CNO abundances. 

\begin{deluxetable}{lccc}
\tablecolumns{5} 
\tablewidth{0pc} 
\tablecaption{Abundance Differences Between the \citet{castelli03} and MARCS 
\citep{marcs} 
Model Atmospheres for Three Representative Stars \label{tab:marcs}}
\tablehead{ 
\colhead{Species} & 
\colhead{HE~0057-5959\tablenotemark{a}} &
\colhead{HE~1320-2952\tablenotemark{b}} &
\colhead{HE~2032-5633\tablenotemark{c}} \\ 
\colhead{(1)} & 
\colhead{(2)} &
\colhead{(3)} &
\colhead{(4)}  
}
\startdata 
$\Delta$A(Na) & 0.03 & 0.03 & $-$0.01 \\
$\Delta$A(Mg) & 0.01 & 0.03 & $-$0.01 \\
$\Delta$A(Al) & \ldots & 0.02 & $-$0.01 \\
$\Delta$A(Si) & \ldots & 0.03 & 0.00 \\
$\Delta$A(Ca) & 0.01 & 0.02 & $-$0.02 \\
$\Delta$A(Sc) & 0.03 & 0.03 & \ldots \\
$\Delta$A(Ti\,{\sc i}) & 0.02 & 0.02 & \ldots \\
$\Delta$A(Ti\,{\sc ii}) & 0.02 & 0.04 & 0.01 \\
$\Delta$A(Cr) & 0.01 & 0.02 & \ldots \\
$\Delta$A(Mn) & \ldots & 0.02 & \ldots \\
$\Delta$A(Fe\,{\sc i}) & 0.02 & 0.03 & $-$0.01 \\
$\Delta$A(Fe\,{\sc ii}) & \ldots & \ldots & \ldots \\
$\Delta$A(Co) & \ldots & 0.01 & \ldots \\
$\Delta$A(Ni) & 0.02 & 0.02 & \ldots \\
$\Delta$A(Sr) & 0.03 & 0.05 & \ldots \\
$\Delta$A(Ba) & 0.03 & 0.04 & \ldots \\
\enddata

\tablenotetext{a}{HE~0057-5959: \teff\ = 5257K, $\log g$ = 2.65, [Fe/H] = $-$4.08}
\tablenotetext{b}{HE~1320-2952: \teff\ = 5106K, $\log g$ = 2.26, [Fe/H] = $-$3.69}
\tablenotetext{c}{HE~2032-5633: \teff\ = 6457K, $\log g$ = 3.78, [Fe/H] = $-$3.63}

\end{deluxetable}

\section{COMPARISON WITH THE {\sc FIRST STARS} ABUNDANCE SCALE}
\label{sec:firststars}

In the context of elemental-abundance determinations in metal-poor 
stars, 
the {\sc First Stars} group 
(e.g., \citealt{cayrel04,spite05,francois07,bonifacio09}) 
obtained very high-quality spectra 
and conducted a careful analysis, such that 
their derived abundances for all
elements exhibit  
very small scatter about the mean trends with metallicity. 
As already mentioned, \citet{cayrel04} highlight the importance 
of correct treatment of continuum scattering to 
``allow proper interpretation of the blue regions of the spectra''. 
Therefore, we regard the following as an important test of our  
analysis of metal-poor stars: namely, using the \citet{cayrel04} 
line list and atmospheric parameters, does our combination of 
model atmospheres (\citeauthor{castelli03}) 
and line-analysis software (MOOG) reproduce their 
abundances from the OSMARCS model atmospheres \citep{gustafsson75} 
and the Turbospectrum \citep{alvarez98} line-analysis 
software. 

\begin{figure*}
\epsscale{1.0}
\plotone{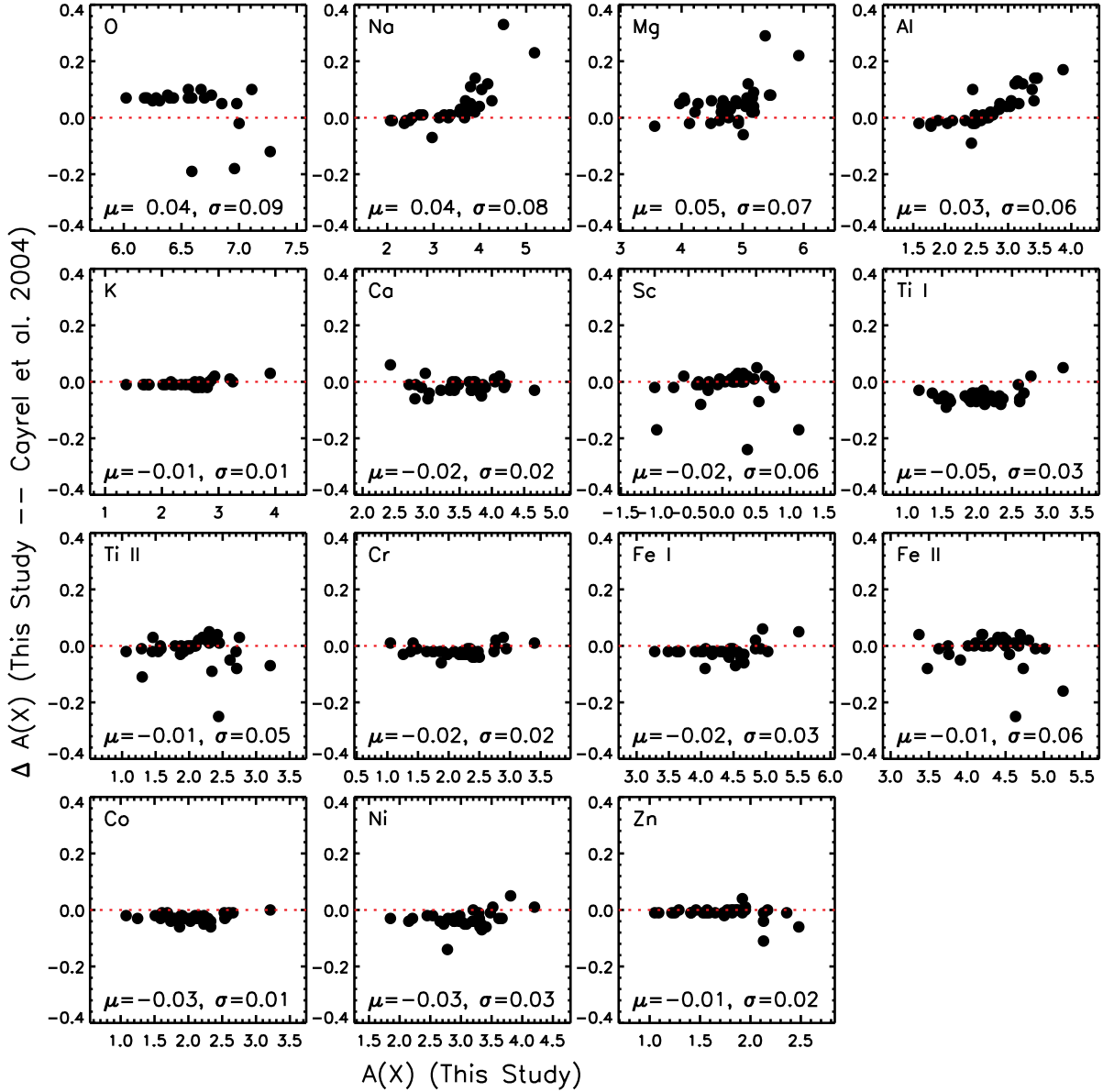} 
\vspace{5mm} 
\caption{A star-by-star 
comparison between the \citet{cayrel04} abundances 
and our abundances when using their line list and atmospheric 
parameters. 
The numbers at the bottom of each panel are the 
mean (Our Analysis $-$ Cayrel) and the dispersion. 
\label{fig:caycomp}}
\end{figure*}

In Figure \ref{fig:caycomp}, we compare our 
abundances with those of \citet{cayrel04}, star by star, and 
find excellent agreement. (For O, Na, and Mg, there are a handful of outliers,  
and on closer examination we find one outlier common to all three panels. This 
object is the most metal-rich and one of the warmest stars in the sample.) 
We note that our abundances were produced 
using the updated version of MOOG with appropriate continuum scattering 
routines. We conducted a similar test 
using the 2009 version of MOOG available on the 
web\footnote{http://verdi.as.utexas.edu/moog.html}.  
Although that version of MOOG, which does not include the newer continuum 
scattering routines, was also able to provide a good match 
to the \citet{cayrel04} abundances, the version we employed produced 
superior agreement. As expected, the abundance 
differences between \citet{cayrel04}  
and the 2009 version of MOOG exhibited a strong wavelength dependence 
towards the blue. 
Therefore, the results of this test show that our ``machinery'' 
(\citeauthor{castelli03} models atmospheres and the MOOG spectrum 
synthesis program) reproduces the \citet{cayrel04} abundances, when adopting the 
same atmospheric parameters. 

For C and N, we conducted the following test. In the course of our
observing campaigns, described in Paper I, we obtained high-S/N spectra
of the metal-poor standards BD $-$18$^\circ$ 5550, CD $-$38$^\circ$ 245, and
CS~22892$-$052.  Using the \citet{cayrel04} stellar parameters, we
derived [C/H] and [N/H] from our spectra.  (Our Keck spectrum of
BD $-$18$^\circ$ 5550 did not include the 3360\AA\ NH lines and for 
CD $-$38$^\circ$ 245, \citealt{spite05} only give an upper limit for C.)  For [C/H], our
abundances were within 0.02 dex and 0.08 dex of the \citet{spite05}
values for BD $-$18$^\circ$ 5550 and CS~22892$-$052, respectively.  For [N/H],
our abundances were within 0.05 dex and 0.10 dex of the \citet{spite05}
values for CD $-$38$^\circ$ 245 and CS~22892$-$052, respectively.  We consider
these abundance differences ({\sc This Work} $-$ {\sc
 \citeauthor{spite05}}) to be small, and therefore regard the results
of this test as a demonstration that our C and N abundances are on the
\citet{spite05} scale, when using their stellar parameters.

Finally, we compared our Sr and Ba 
abundances to those of \citet{francois07}, who measured 
relative abundances for some some 16 neutron-capture elements using 
spectrum synthesis. 
For the metal-poor standards BD $-$18$^\circ$ 5550, 
CD $-$38$^\circ$ 245, and CS~22892$-$052, 
the abundance differences $\Delta$[Sr/H] 
({\sc This Study} $-$ {\sc \citeauthor{francois07}}) 
are $-$0.08 dex, $-$0.09 dex, and $-$0.01 dex, respectively, while 
the abundance differences $\Delta$[Ba/H] are +0.06 dex, $-$0.02 dex, and 0.00 
dex respectively. 
Again, we consider these differences to be small, and hence this comparison 
demonstrates that our Sr and Ba abundances are on the \citet{francois07} 
scale, when using their stellar parameters. 

\section{RE-ANALYSIS OF THE LITERATURE SAMPLE}
\label{sec:lit}

Having completed the analysis of the program stars, we then 
sought to undertake a homogeneous re-analysis of 
all extremely 
metal-poor Galactic stars with [Fe/H] $\le$ $-$3.0. 
We queried the SAGA database \citep{saga} for 
all stars with [Fe/H] $\le$ $-$2.9, with the aim of re-analyzing those 
stars using the published equivalent widths, but with 
our analysis procedures and techniques. 
The rationale for choosing [Fe/H] = $-$2.9 as the cutoff 
was that we were hoping to find as many stars as possible 
with [Fe/H] $\le$ $-$3.0 on our scale, some of which may have higher 
published metallicities. At the same 
time, we needed to ensure a manageable sample. 

At the time of our SAGA query (2 Feb 2010), the database had been last
updated on 2 Sep 2009. Our query returned 196 stars with
[Fe/H] $\le$ $-$2.9; from this list of stars we identified 16
references 
  in each of which reliable equivalent widths had been 
  published for several stars, based on high-quality spectra. These 
16 references included a large number of stars in the SAGA database 
with [Fe/H] $\le$ $-$3.0 (notable exceptions include stars unique to 
\citealt{barklem05}, BD +44$^\circ$ 493 with [Fe/H] = $-$3.7 [\citealt{ito09}], 
        the C-rich dwarf G77$-$61 with [Fe/H] = $-$4.03 
        [\citealt{plez05}], SDSS J102915+172927 with [Fe/H]$_{\rm 
        1D~LTE}$ = $-$4.73 [\citealt{caffau11}], and recent 
        papers by \citealt{hollek11} and \citealt{sbordone12}).  
        It is also worth 
        noting that this is a study of Galactic stars, and so we did 
        not consider any of the growing number of stars in dwarf 
        galaxies with [Fe/H] $\le$ $-$3, even though we have analyzed 
        several such objects using very similar techniques to those in 
        this series of papers (e.g., \citealt{norris10b,norris10}). 
        The 16 references we selected contained some 207 stars, 
many 
more metal-rich than [Fe/H] = $-$3.0. 
Nevertheless, we analyzed all 207 stars in the following manner. 

We defined ``giants'' as those stars with $\log g$ $<$ 3, and
``dwarfs'' as those stars with $\log g$ $>$ 3, where the surface
gravities were taken from the literature sources\footnote{ Clearly our
  definition of ``dwarfs'' will include many subgiants and stars near
  the base on the giant branch. Nevertheless, we needed to define a
  boundary to separate dwarfs from giants and to then apply the
  different color-temperature relations to determine effective
  temperatures.  Throughout the rest of the paper, these definitions
  for dwarfs and giants apply.}.  With the exception of the three most
metal-poor stars, which we shall discuss below, we then used the
published photometry, reddenings, and metallicities, [Fe/H], together
with the infrared flux method (IRFM) metallicity-dependent
color-temperature relations adopting \citet{casagrande10} for the
dwarfs and \citet{ramirez05} for the giants, to determine effective
temperatures.  We note that the \citet{ramirez05} calibration is valid
only for [Fe/H] $> -$4.0. For a small number of stars, our
\teff\ involve a small extrapolation down to [Fe/H] = $-$4.2.  For a
subset of these literature stars, observations and analysis using the
spectrophotometric procedures described in Paper I yielded \teff. We refer to these
\teff\ as the ``Bessell temperatures''.  For these stars with both
``Bessell temperatures'' and IRFM temperatures, we found average
\teff\ offsets of +19K $\pm$ 42K (Bessell $-$ \citealt{casagrande10}) 
and $-$45K $\pm$ 19K (Bessell $-$ \citealt{ramirez05}) 
from the IRFM for dwarfs and giants, respectively.  (We note that
obtaining these offsets involved an iterative process since the
derived \teff\ are weakly dependent on the adopted metallicity and the
adopted metallicity changed as we employed the updated \teff.)
Finally, we took into account the 51K average difference between the
``Bessell temperatures'' and our final temperatures, as determined from
our 38 program stars (see Paper I).  In summary, we applied corrections of
$\Delta$\teff\ = $+$19K $+$51K = $+$70K to the \citet{casagrande10}
IRFM \teff\ for dwarf stars, and $\Delta$\teff\ = $-$45K $+$51K = $+$6K
to the \citet{ramirez05} IRFM \teff\ for giant stars.

Using our line list, presented in Paper I, we adopted
the literature equivalent widths for lines in common with our list, and
ignored lines that were not in common.  This ensured that the $\log
gf$ values were homogeneous.  As demonstrated in Paper I, our
equivalent-width measurements are on the same scale as various
literature studies.  For the surface gravity, we followed our analysis
procedure for the program stars in which $\log g$ was determined from
the $Y^2$ isochrones \citep{y2isochrones}, assuming the revised \teff,
an age of 10 Gyr, and [$\alpha$/Fe] = +0.3.
The published surface gravities were used for dwarf/subgiant discrimination. 
The microturbulent velocity was determined in the usual way, by 
forcing the abundances from \fei\ lines to show no trend with 
reduced equivalent width. During this process, we removed \fei\ lines 
having abundances that differed ($i$) from the median value by more than 0.5 dex 
or ($ii$) from the median abundance by more than 3-$\sigma$, 
as for our program stars, and the rejection applied to lines 
yielding abundances higher or lower than the median value. 
The average number of lines rejected per star was 4, and the 
average number of lines per star was 49. The largest number of rejected 
lines in a given star was 30 (of a total of 84 lines), and this star 
also has the highest fraction of rejected lines, 36\%. 
The result of this first-pass analysis was a revised estimate of the 
metallicity, [Fe/H]. 

We then repeated the analysis using the updated metallicity. 
That is, we determined updated \teff\ using the 
IRFM calibrations, which are (weakly) sensitive to the assumed 
metallicity. In this second iteration, 
we again used the $Y^2$ isochrones and the 
published surface gravity for dwarf/subgiant classification, noting once more 
that this involves extrapolation beyond [Fe/H] = $-$3.5 
(down to [Fe/H] = $-$4.2 for the most metal-poor object in the literature 
sample).  
With these revised \teff\ and updated surface gravities, we computed
new model atmospheres with the appropriate stellar parameters (\teff,
$\log g$, and [M/H] = [Fe/H]). For \teff\ and $\log g$, the revised
values were generally very close to the initial values.  The
microturbulent velocity was determined and \fei\ outliers removed
using the criteria outlined above. The results of this second-pass
analysis were final stellar parameters and metallicities for the
literature sample, which are presented in Table \ref{tab:param2}.  (For
a small number of stars, a third iteration was necessary to ensure
that the derived metallicity was sufficiently close to the value used
to generate the model atmosphere, within 0.3 dex.)  
The evolutionary status,
\teff\ vs.\ $\log g$, for the literature sample is shown in the lower
panel of Figure \ref{fig:cmd}.
We note that for our 38 program
stars, the smallest number of \fei\ lines measured in a given star was
14.  Therefore, for the literature sample, we discarded stars in which
there were fewer than 14 \fei\ lines.  

\begin{deluxetable*}{lccccccccc}
\tabletypesize{\scriptsize}
\tablecolumns{9} 
\tablewidth{0pc} 
\tablecaption{Model Atmosphere Parameters and [Fe/H] for the Literature Sample\label{tab:param2}}
\tablehead{ 
\colhead{Star} & 
\colhead{RA2000\tablenotemark{a}} &
\colhead{DEC2000\tablenotemark{a}} &
\colhead{\teff} & 
\colhead{$\log g$} & 
\colhead{$\xi_t$} & 
\colhead{[M/H]$_{\rm model}$} & 
\colhead{[Fe/H]$_{\rm derived}$} & 
\colhead{C-rich\tablenotemark{b}} & 
\colhead{Source} \\
\colhead{} & 
\colhead{} &
\colhead{} &
\colhead{(K)} & 
\colhead{(cgs)} & 
\colhead{(\kms)} & 
\colhead{} & 
\colhead{} & 
\colhead{} \\
\colhead{(1)} & 
\colhead{(2)} &
\colhead{(3)} &
\colhead{(4)} & 
\colhead{(5)} & 
\colhead{(6)} & 
\colhead{(7)} & 
\colhead{(8)} & 
\colhead{(9)} & 
\colhead{(10)} 
}
\startdata 
 CS~22957-022 & 00 01 45.5 & $-$05 49 46.6 & 5146 & 2.40 & 1.5 & $-$2.9 & $-$2.92 & 0 &        14 \\
 CS~29503-010 & 00 04 55.4 & $-$24 24 19.3 & 6570 & 4.25 & 1.3 & $-$1.0 & $-$1.00 & 1 &         3 \\
 CS~31085-024 & 00 08 27.9 &   +10 54 19.8 & 5778 & 4.64 & 0.3 & $-$2.8 & $-$2.80 & 0 &        14 \\
 BS~17570-063 & 00 20 36.2 &   +23 47 37.7 & 6233 & 4.46 & 0.8 & $-$3.0 & $-$2.95 & 0 &         5 \\
 HE~0024-2523 & 00 27 27.7 & $-$25 06 28.2 & 6635 & 4.11 & 1.2 & $-$2.8 & $-$2.82 & 0 &         6 
\enddata


\tablerefs{
1 = \citet{aoki02}; 
2 = \citet{aoki06};
3 = \citet{aoki07}; 
4 = \citet{aoki08}; 
5 = \citet{bonifacio07,bonifacio09}; 
6 = \citet{carretta02,cohen02}; 
7 = \citet{cayrel04}; 
8 = \citet{christlieb04}; 
9 = \citet{cohen04}; 
10 = \citet{cohen06}; 
11 = \citet{cohen08}; 
12 = \citet{frebel07}; 
13 = \citet{honda04}; 
14 = \citet{lai08}; 
15 = \citet{norris01}; 
16 = \citet{norris07}; 
\\
\\
Note. Table 6 is published in its entirety in the electronic edition of The Astrophysical Journal. A portion is shown here for guidance regarding its form and content.
}

\tablenotetext{a}{Coordinates are from the 2MASS database \citep{2mass}.}
\tablenotetext{b}{1 = CEMP object, adopting the \citet{aoki07} definition and 0 = C-normal (see Section 7.1 for details).}

\end{deluxetable*}

Following the procedures outlined above, 
we then 
determined element abundances 
using the published equivalent widths 
(only lines in common with our line list), final model atmospheres, 
and MOOG. Lines affected by hyperfine and/or isotopic splitting 
were treated appropriately. Chemical abundances for the literature sample 
are presented in Table \ref{tab:abund2}, where [C/Fe] and [N/Fe] are 
the values taken from the literature,   
but [X/Fe] for X = Na to Ba are 
recomputed on our homogeneous scale. 
As described in Yong et al. (2012; Paper III), 
we chose not to update the [C/Fe] 
and [N/Fe] abundances using our revised metallicities 
via [C/Fe]$_{\rm New}$ = [C/Fe]$_{\rm Literature}$ $-$ ([Fe/H]$_{\rm
   This~study}$ $-$ [Fe/H]$_{\rm Literature}$), since this approach only incorporates 
changes to the metallicity and does not include any 
changes to the C and/or N abundances. 
Furthermore, such an update does not affect our results or interpretation. 

Following the procedure we adopted for the analysis of the 
program stars, we repeated the entire analysis of the literature 
sample using only a 
subset of lines believed to be free from CH blends. 
Depending on the published [C/Fe] abundance ratio and the subsequent 
CEMP classification, we adopted the final stellar parameters and 
chemical abundances as determined using the appropriate line list and 
analysis. 

The final sample of literature stars was reduced 
from 207 to 152 stars by the averaging of the 
results of stars having multiple analyses into a single set of 
abundances and removal of stars with fewer than 14 \fei\ lines. 
In all cases, there was excellent agreement for a 
given abundance ratio [X/Fe] for stars having multiple analyses. 

For the three HES stars with [Fe/H] < $-$4.5, 
HE~0107$-$5240 \citep{christlieb02,christlieb04}, 
HE~1327$-$2326 \citep{frebel05,aoki06}, and 
HE~0557$-$4840 \citep{norris07}, 
we did not attempt to re-derive effective temperatures or 
surface gravities. These stars lie in a metallicity regime in 
which the IRFM color-temperature relations are not calibrated.  
Therefore, we adopted the published stellar parameters 
(\teff, $\log g$, $\xi_t$, and [M/H]) and computed metallicities 
and chemical abundances using lines in common with our line list.  
Furthermore, we retain HE~1327$-$2326, despite the fact that only 
four \fei\ lines (from a total of 7) were in common with our line list. 

In summary, we have computed metallicities and chemical abundances for
some 16 elements in 190 metal-poor Galactic stars (38 program stars
and 152 literature stars). This is a homogeneous analysis with stellar
parameters (\teff, $\log g$, $\xi_t$), metallicities, 
atomic data, solar abundances, and therefore abundance ratios, [X/Fe],
all on the same scale.  For convenience, Table \ref{tab:app} 
  includes coordinates, stellar parameters and
  abundance ratios for all of the program stars and literature stars
  presented here. 

\section{COMPARISON WITH PREVIOUS STUDIES} 
\label{sec:comp} 

\begin{figure}[t!]
\epsscale{1.2}
\plotone{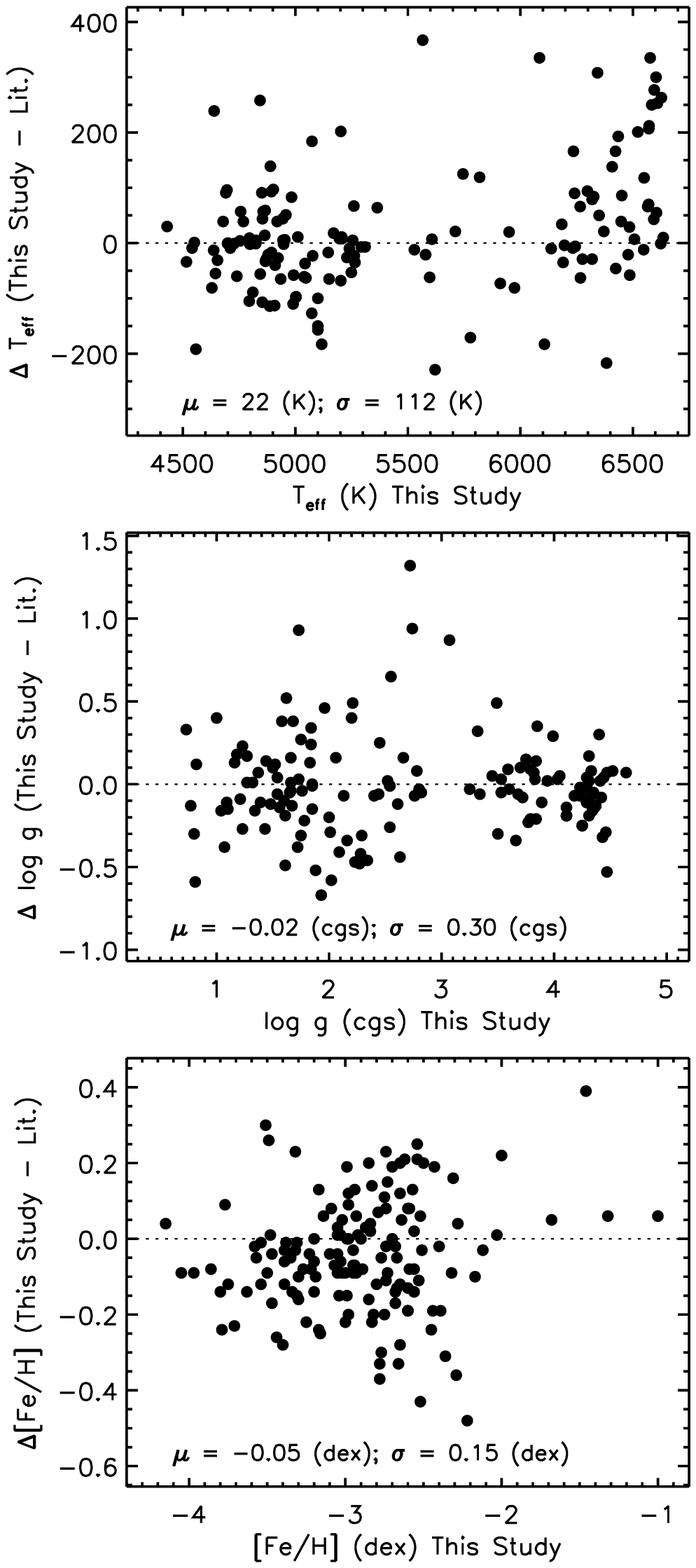} 
\caption{A star-by-star comparison of stellar parameters, \teff\ (upper panel), 
$\log g$ (middle panel), and [Fe/H] (lower panel), for 
our re-analysis of the literature sample 
and the original literature values. 
The numbers at the bottom of each panel are the 
mean (Our Analysis $-$ Literature) and the dispersion. 
\label{fig:lit_param_comp}}
\end{figure}

In Figure \ref{fig:lit_param_comp}, we compare our stellar parameters 
(\teff, $\log g$, and [Fe/H]) with the literature values. (In this 
comparison we exclude the three HES stars with [Fe/H] < $-$4.5, since we 
adopted their published \teff\ and $\log g$.) 
For all parameters, our revised values are, on average, in good agreement with 
the literature values. This is perhaps not surprising, given that many 
of the literature studies adopt similar approaches to determine these 
parameters. When comparing our values with the literature, the dispersion 
is comparable to our estimates of our internal uncertainties in 
stellar parameters. 

\citet{schorck09} and \citet{li10} published metallicities for 
HES stars based on medium-resolution spectra. For our combined 
sample (program stars and literature stars), there are 12 stars 
in common with \citet{schorck09} and \citet{li10}. 
Recall that there 
are program stars for which we conducted analyses assuming a 
dwarf gravity and a subgiant gravity. For the 
purposes of this comparison, we regard each analysis as an independent 
measurement. Thus, for these 12 stars we have 18 [Fe/H] measurements. 
As noted in Paper III, our metallicities 
are lower than theirs by 0.26 $\pm$ 0.06 dex ($\sigma$ = 0.27). 

We then compared the abundance ratios [X/Fe], for X = Na to Ba, between 
the values computed in this study and the literature values. 
Our assumption was that our revised [X/Fe] abundance ratios would be 
reasonably close to the literature values. Any differences in 
abundance ratios would presumably be driven by differences in the 
stellar parameters, model atmospheres, line-analysis software, 
atomic data, and/or solar abundances. As discussed, the stellar parameters 
are in good agreement. Similarly, we have shown that different 
model atmospheres and line-analysis software introduce only very small 
abundance differences. Finally, we do not anticipate large differences 
to arise from the solar abundances or atomic data. 
Therefore, for each element in each star, we plotted 
the abundance differences (This Study $-$ Literature). We fitted these
differences for a given element with a Gaussian, and measured the 
full width at half maximum (FWHM), which ranged from
0.05 dex to 0.14 dex, with a mean of 0.10 dex.  We then eliminated
those stars in which the abundance differences exceeded max(0.50 dex,
3-$\sigma$) from the average difference for a particular element 
(the average difference for a given element ranged from $-$0.10 dex 
to +0.17 dex). That
is, based on the abundance differences, we removed particular elements
from a given star.  For example, a star may have an anomalous [Mg/Fe]
value, which is then removed. If all other elements in that star have
[X/Fe] ratios sufficiently close to the literature values, then those
abundance ratios would be retained.  For a given element, this
resulted in fewer than 7 stars being rejected, and we speculate that
many of these outliers may be due to errors in the 
tables of equivalent widths, of which there are some 18,000
values. The abundance outliers are not included in Table
\ref{tab:abund2}, nor in any other table or figure.  Note that the
three HES stars with [Fe/H] < $-$4.5 were included in this analysis, 
and there were no abundance outliers among these objects.

\begin{deluxetable*}{lrrrrrrrrrrrr} 
\tabletypesize{\tiny}
\tablecolumns{12} 
\tablewidth{0pc} 
\tablecaption{Chemical Abundances (C-Ba) for the Literature Sample \label{tab:abund2}}
\tablehead{ 
\colhead{Star} & 
\colhead{[C/Fe]\tablenotemark{a}} &
\colhead{[N/Fe]\tablenotemark{a}} &
\colhead{[Na/Fe]} &
\colhead{[Mg/Fe]} &
\colhead{[Al/Fe]} &
\colhead{[Si/Fe]} &
\colhead{[Ca/Fe]} &
\colhead{[Sc/Fe]} &
\colhead{[Ti\,{\sc i}/Fe]} &
\colhead{[Ti\,{\sc ii}/Fe]} &
\colhead{[Cr/Fe]} & 
\colhead{[Mn/Fe]} \\ 
\colhead{(1)} &
\colhead{(2)} &
\colhead{(3)} &
\colhead{(4)} &
\colhead{(5)} &
\colhead{(6)} &
\colhead{(7)} &
\colhead{(8)} &
\colhead{(9)} &
\colhead{(10)} &
\colhead{(11)} &
\colhead{(12)} & 
\colhead{(13)} 
}
\startdata 
 CS~22957-022 &     0.16 &     0.21 &  \ldots &    0.21 &  \ldots &  \ldots &    0.27 &    0.07 &    0.30 &    0.32 & $-$0.31 & $-$0.68  \\
 CS~29503-010 &     1.07 &  $<$1.28 & $-$0.04 & $-$0.06 &  \ldots &  \ldots &    0.11 &  \ldots &    0.28 &    0.21 &  \ldots &  \ldots  \\
 CS~31085-024 &     0.36 & $<-$0.24 &  \ldots &    0.08 &  \ldots &    0.39 &    0.28 &    0.33 &    0.33 &    0.50 & $-$0.12 & $-$0.60  \\
 BS~17570-063 &     0.40 &   \ldots &  \ldots &    0.28 &  \ldots &    0.51 &    0.33 &    0.77 &    0.60 &    0.70 & $-$0.14 &  \ldots  \\
 HE~0024-2523 &   \ldots &   \ldots &  \ldots &    0.82 & $-$0.49 &  \ldots &    0.54 &    0.25 &    0.63 &    0.35 & $-$0.38 &  \ldots  
\enddata

\tablenotetext{a}{For [C/Fe] and [N/Fe], the values are taken from the literature reference.} 


\tablerefs{ 
Note. Table 7 is published in its entirety in the electronic edition of The Astrophysical Journal. A portion is shown here for guidance regarding its form and content.
}

\end{deluxetable*}

\begin{figure*}
\epsscale{1.0}
\plotone{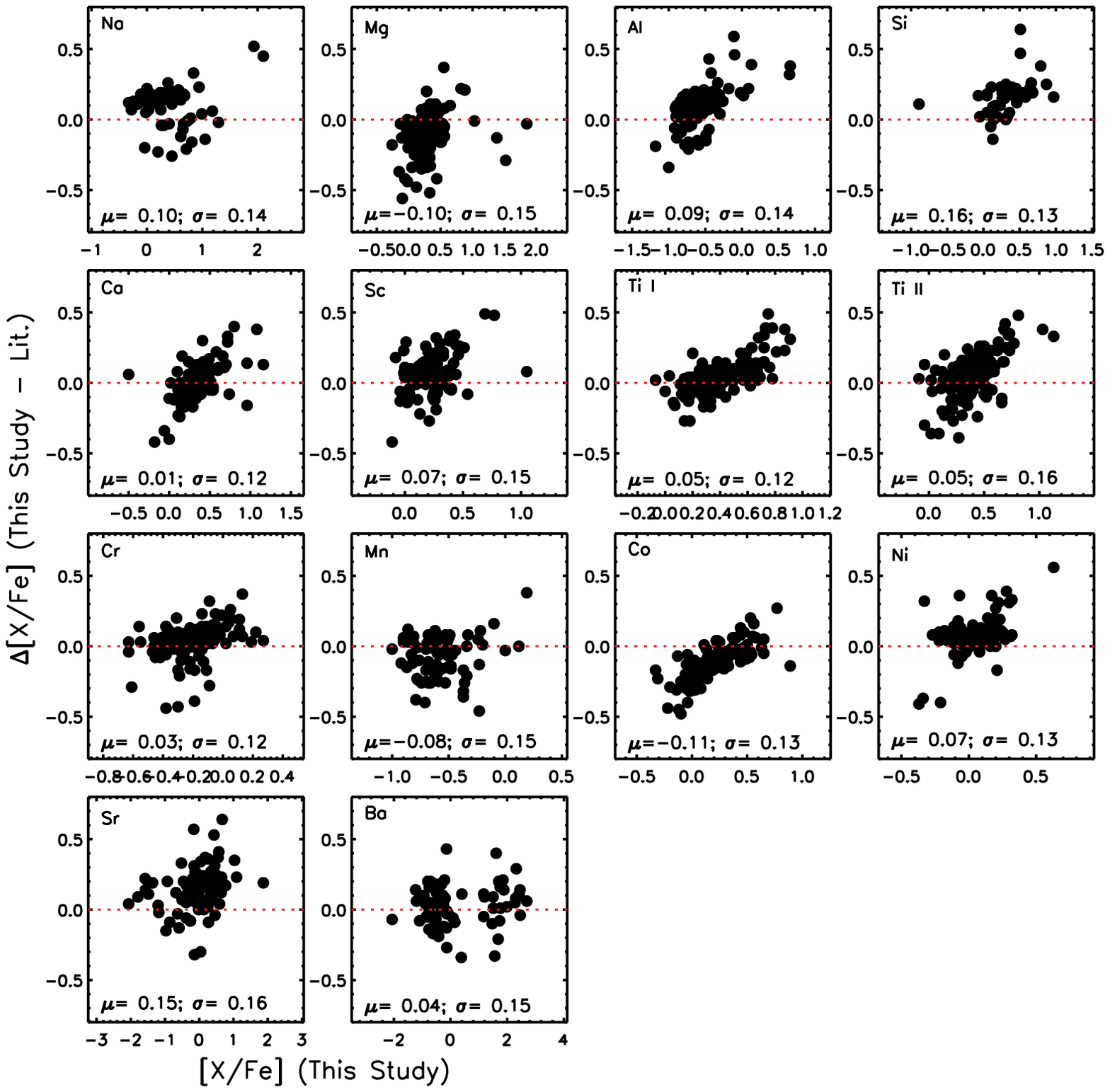} 
\vspace{5mm} 
\caption{A star-by-star comparison of abundance ratios, [X/Fe], for 
our re-analysis of the literature sample 
and the original literature values. 
The numbers at the bottom of each panel are the 
mean (Our Analysis $-$ Literature) and the dispersion. 
\label{fig:lit_xfe_comp}}
\end{figure*}

In Figure \ref{fig:lit_xfe_comp}, we compare abundance ratios [X/Fe], 
for X = Na to Ba, between the values re-computed in this study and 
the literature values. For all elements, the average values are 
in good agreement. 
The dispersions are comparable to 
the uncertainties given in Table \ref{tab:abund} and, as 
shown 
in Section 7.2 below, to the dispersions about the mean trend when plotting 
[X/Fe] vs.\ [Fe/H]. 

\begin{figure}[t!]
\epsscale{1.2}
\plotone{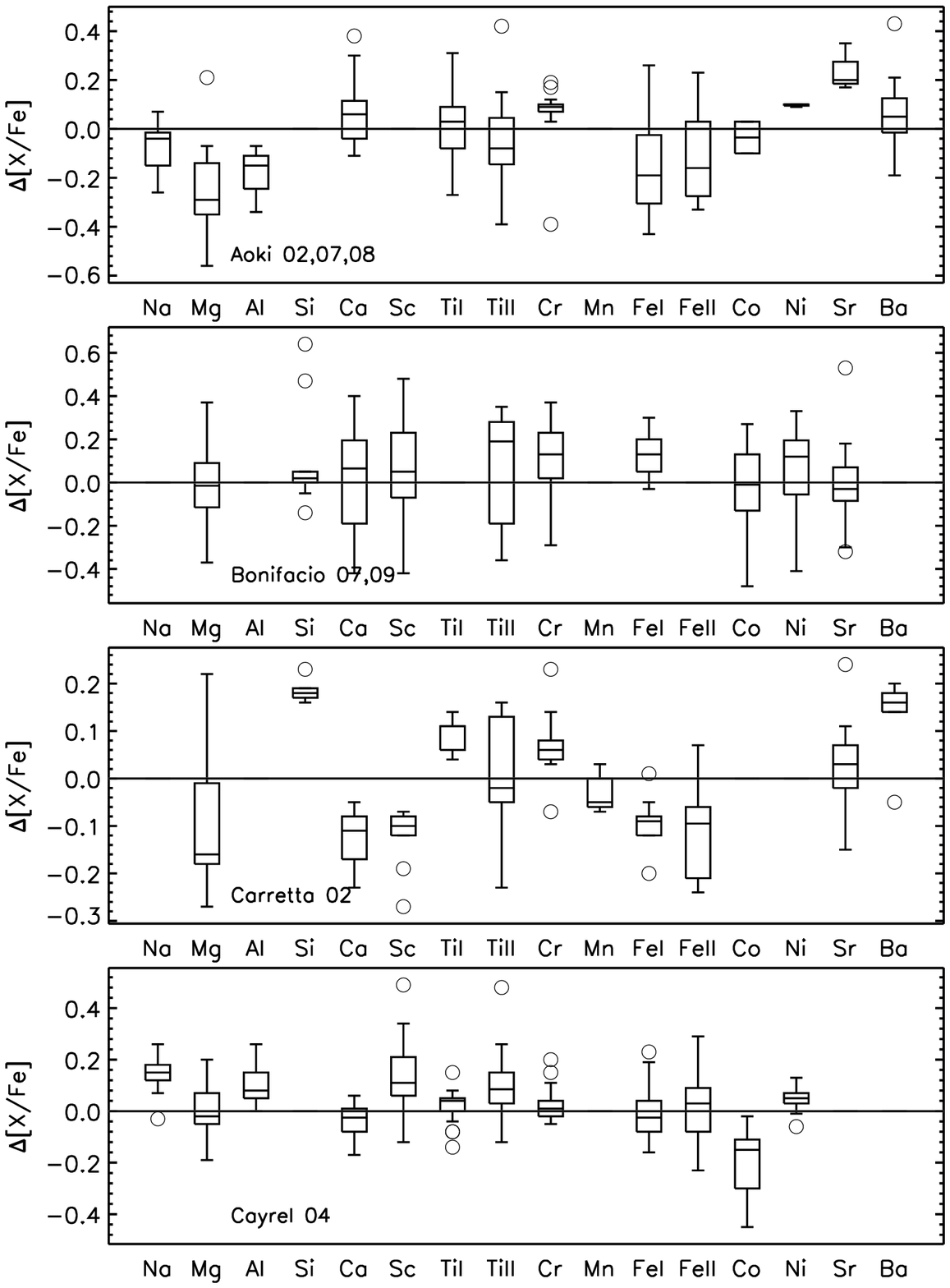} 
\caption{Boxplots illustrating the abundance differences $\Delta$[X/Fe] 
and $\Delta$[Fe/H] 
(This Study $-$ Literature) for the various literature references. 
The box defines the interquartile range, the median is identified, 
the whiskers extend to the maximum (or minimum) or 1.5 times 
the interquartile range, and circles indicate outliers. 
\label{fig:lit_xfe_compa}}
\end{figure}

\begin{figure}[t!]
\epsscale{1.2}
\plotone{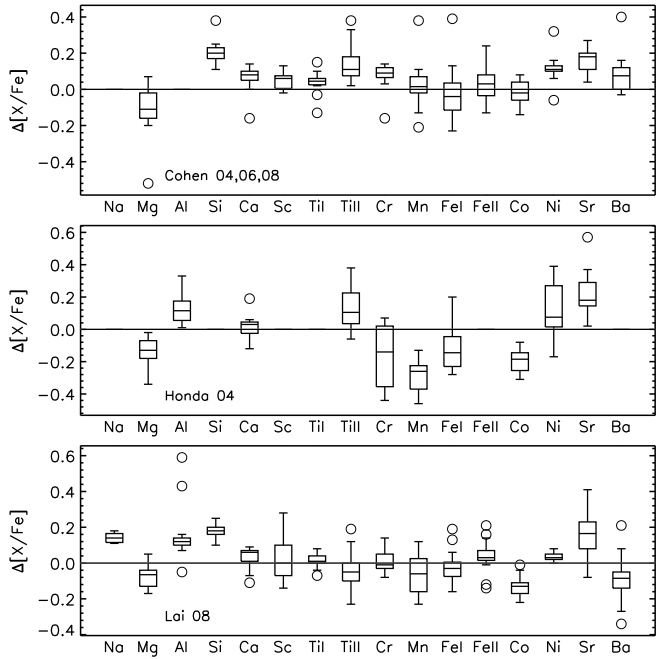} 
\caption{Same as Figure \ref{fig:lit_xfe_compa}, but for the 
next set of literature references. 
\label{fig:lit_xfe_compb}}
\end{figure}

\begin{figure}[t!]
\epsscale{1.2}
\plotone{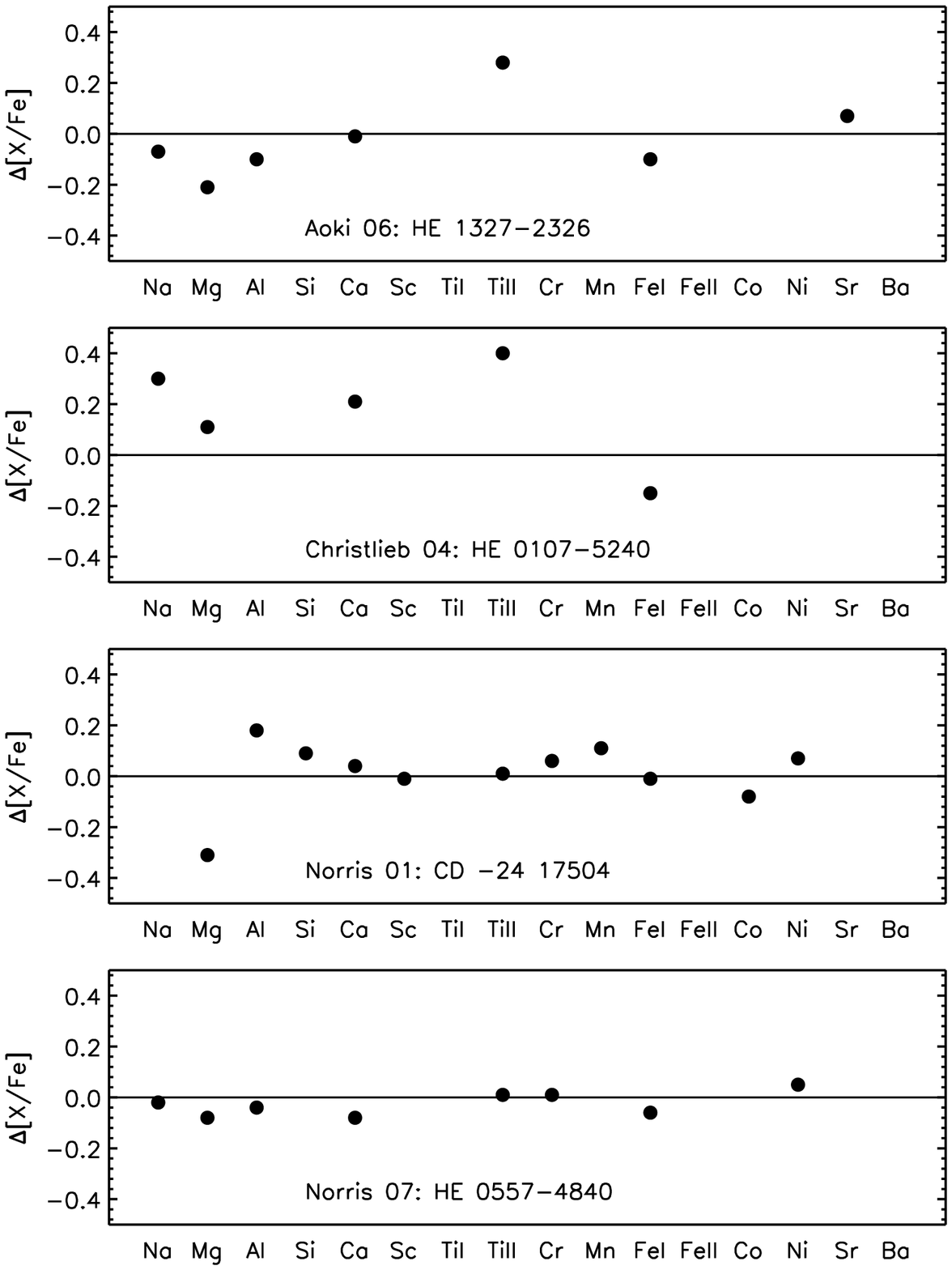} 
\caption{Abundance differences $\Delta$[X/Fe] and 
$\Delta$[Fe/H] (This Study $-$ Literature) for 
individual stars. 
\label{fig:lit_xfe_compc}}
\end{figure}

In Figures \ref{fig:lit_xfe_compa} to \ref{fig:lit_xfe_compc}, we show
the abundance differences $\Delta$[X/Fe] (This study $-$ Literature)
for each literature reference, or set of references.  From these
figures, any systematic abundance offsets between our re-analysis and
the original literature abundances would be readily seen. Since we
have already eliminated the handful of outliers as described above, 
it is not surprising that our revised abundances are generally in 
good agreement with the literature values. 
We shall not seek to
understand the reasons for differences in a given element in a
particular analysis, except to say that the cause is almost certainly
due to the stellar parameters, solar abundance, and/or atomic data. It
is re-assuring that for the three HES stars with [Fe/H] < $-$4.5, our
1D LTE abundances are in good agreement with the published 1D LTE
abundances, despite the fact that the published values were based on
model atmospheres with appropriate CNO abundances, in contrast to our
analysis which assumed scaled solar abundances, 
but with [$\alpha$/Fe] = +0.40. 

\section{NON-LTE EFFECTS}
\label{sec:nlte}

Our analysis tools (1D model atmospheres and spectrum synthesis code) assume 
LTE and, therefore, this analysis is subject 
to systematic uncertainties from non-LTE as well as 
3D (granulation) effects \citep{asplund05}. 
We now offer some comments on the role of 
non-LTE effects, and refer the reader to 
work by Asplund and collaborators regarding 3D effects. 
As discussed extensively in the literature, comparison of the 
abundances of Fe from neutral and ionized species provide a check 
on the presence of departures from LTE and/or the adopted surface gravity. 
In the event of differences in the abundance from neutral and ionized 
species, 
and in the absence of trigonometric distances (e.g., \citealt{nissen97}) 
and model-insensitive \teff\ measurements, 
it is difficult to gauge the relative contributions of 
non-LTE effects or errors in the surface gravity to the 
abundance discrepancy. Our surface gravities, 
at least for the program stars, were informed by spectrophotometry and 
from Balmer-line analysis. That is, both techniques used to derive \teff\ 
required estimates of $\log g$ (and [Fe/H]), and were therefore 
sensitive to the surface gravity. To explore the degree of 
non-LTE effects, we shall assume (in this subsection) 
that any abundance differences between 
neutral and ionized species reflect non-LTE effects rather than surface-gravity 
errors. 

\begin{figure}[t!]
\epsscale{1.2}
\plotone{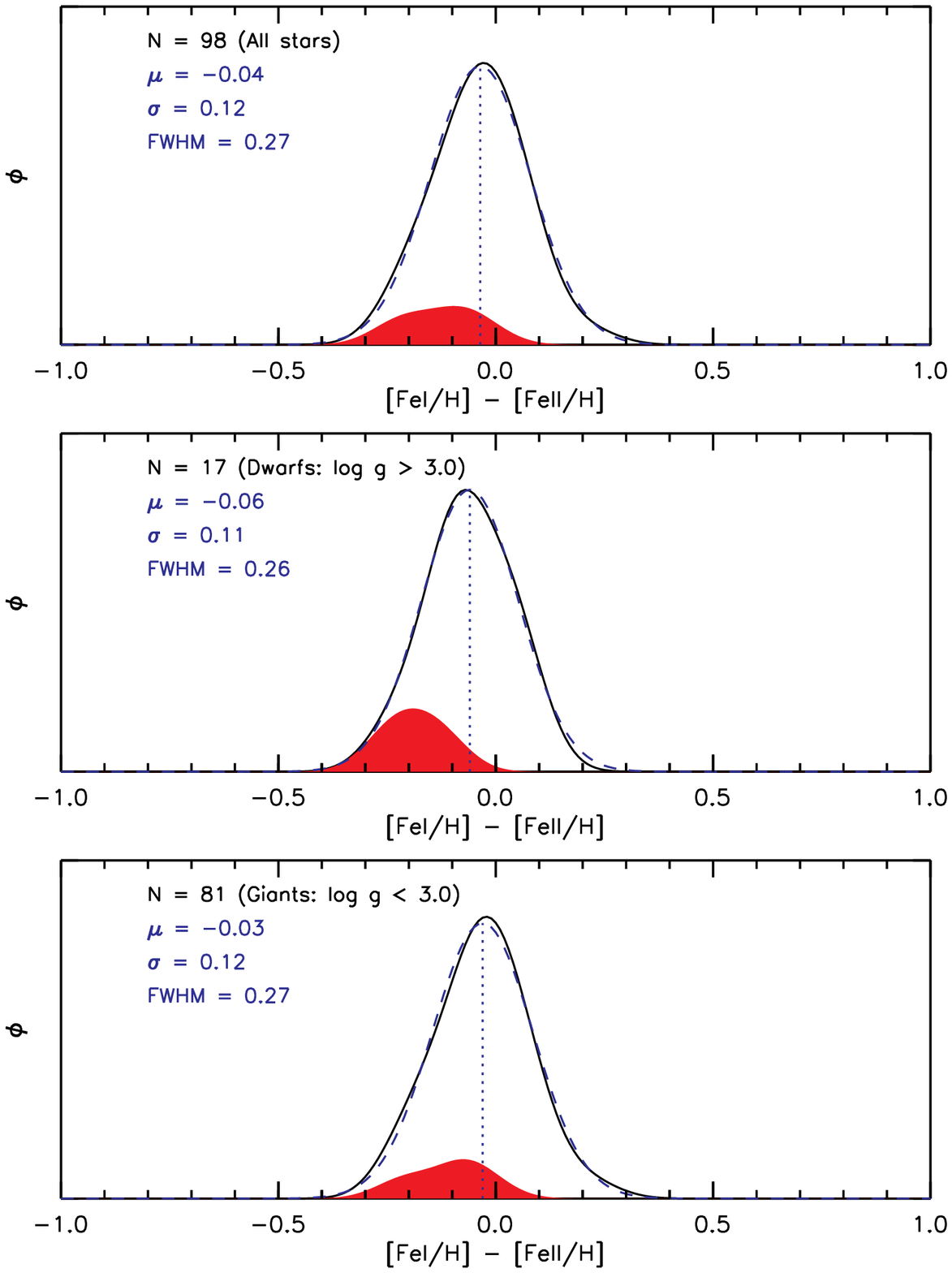} 
\caption{Generalized histograms showing the abundance difference 
[\fei/H] $-$ [\feii/H] for N $\ge$ 2 lines 
and [Fe/H] $\le$ $-$2.5. 
The red histograms represent the relevant subset of our 38 program stars.  
The upper panel shows all 
stars, the middle panel only dwarfs ($\log g$ $>$ 3.0), and 
the lower panel only giants ($\log g$ $<$ 3.0). 
In each panel 
we show the number of stars and the $\mu$ and $\sigma$ for the 
Gaussian ($ae^{-\frac{(x-\mu)^2}{2\sigma^2}}$) fit (blue dashed line) to the data, 
as well as the FWHM. 
\label{fig:nlte}}
\end{figure}

In Figure \ref{fig:nlte}, we plot the difference between the abundance
from \fei\ and \feii\ lines. We only consider stars with two or more
\feii\ lines. In this figure, we use generalized histograms, in which
each data point (i.e., each star) is replaced by a unit Gaussian of
width 0.15 dex. The Gaussians are then summed to produce a
realistically smoothed histogram. By fitting a Gaussian to this
histogram, we can measure the center ($\mu$) and 
width of the distribution (FWHM or dispersion, $\sigma$).  In this figure, we
consider (a) all stars, (b) dwarfs ($\log g$ $>$ 3.0), and (c) giants
($\log g$ $<$ 3.0). In all panels, the Gaussian fit to the generalized
histogram is centered at 
[\fei/H] $-$ [\feii/H] $\sim -0.04$ dex,
and the FWHM and 
dispersion of the Gaussian fit are 0.27 dex and 0.12 dex, 
respectively.  We note that this dispersion of 0.12 dex is
smaller than the average ``total error'' for \fei\ (0.13 dex) and
\feii\ (0.17 dex), added in quadrature (0.21 dex),  
suggesting that the width of the Gaussian is smaller than 
that expected from the \fei\ and \feii\ measurement errors 
(and assuming that the \fei\ and \feii\ errors are fully 
independent). 

The dominant non-LTE mechanism in late-type metal-poor stars 
is overionization, and abundances derived from \fei\ lines in 
LTE are expected to be underestimated 
(e.g., \citealt{thevenin99,mashonkina11,bergemann12,lind12}). 
Therefore, if non-LTE effects were at play in our sample, 
we would expect the mean (LTE) value of [\fei/H] $-$ [\feii/H] to be negative. 
For our sample of program stars (red generalized histograms in 
Figure \ref{fig:nlte}), there is evidence for overionization 
(our sample is, on average, more metal-poor than the full sample, and 
\citealt{lind12} show that the degree of overionization 
is a function of metallicity). 
However, as noted, the full sample suggests that 
$<$[\fei/H]$>$ $\simeq$ $<$[\feii/H]$>$. 
We are not suggesting that non-LTE effects are not at play in 
our sample. Instead, we 
can only say that the abundances from \fei\ and \feii\ are, on average, 
in agreement for both dwarfs and giants for the adopted gravities. The number of stars 
with [\fei/H] $<$ [\feii/H] is similar to the number with 
[\fei/H] $>$ [\feii/H] for both dwarfs and giants. 

\begin{figure}[t!]
\epsscale{1.2}
\plotone{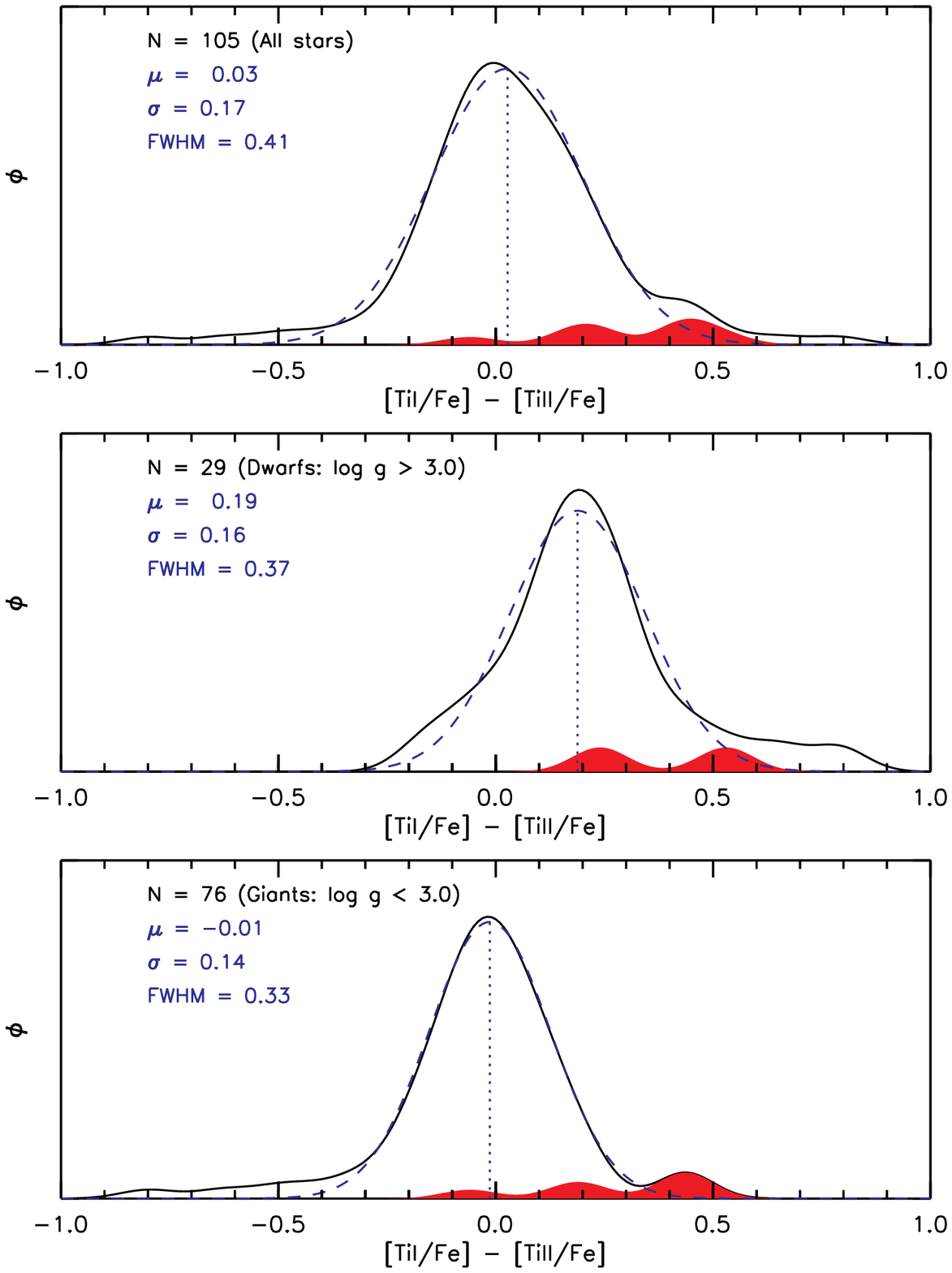} 
\caption{Same as Figure \ref{fig:nlte}, but for 
[\tii/Fe] $-$ [\tiii/Fe]. 
\label{fig:nlte2}}
\end{figure}

Measurements of the Ti abundance from neutral and ionized species 
permit an alternative view of the possible non-LTE overionization 
\citep{bergemann11}. 
Given the difference in ionization potentials (6.8 eV for \tii\ 
and 7.9 eV for \fei), one might naively 
expect overionization to affect Ti to a larger degree than for Fe. 
In Figure \ref{fig:nlte2}, we plot the difference between 
the abundance from \tii\ and \tiii\ lines. We again use 
generalized histograms, in which each data point is replaced by a 
unit Gaussian of width 0.15 dex. Only stars with two or more 
\tii\ and \tiii\ lines are considered; we fit a Gaussian 
to the generalized histogram to quantify the 
center ($\mu$) and width (FWHM or dispersion, $\sigma$). 
In this figure, we consider (a) all stars, (b) dwarfs ($\log g$ $>$ 3.0), 
and (c) giants ($\log g$ $<$ 3.0). For all stars and for the 
giant sample, the Gaussians are centered near 0. However, for the dwarfs, 
the Gaussian is centered at +0.19 dex. This indicates that for the 
dwarfs, the abundance from \tii\ exceeds that from \tiii, a result 
not seen for Fe. Such a 
discrepancy between the abundance from neutral and ionized species  
has the opposite sign compared with that expected from non-LTE overionization. 
The largest dispersion is 0.16 dex (for the dwarfs), and this value 
is smaller than the ``total error'' for \tii\ (0.16 dex) and \tiii\ (0.16 dex),  
added in quadrature (0.23 dex). 
This again suggests that the width of the Gaussian is smaller than 
that expected from the \tii\ and \tiii\ measurement errors. 

Non-LTE effects can also manifest as trends between the abundance from
\fei\ lines and the lower excitation potential ($\chi$). Other studies
of metal-poor stars (e.g., \citealt{cayrel04,cohen08,lai08}) find a
negative trend between the \fei\ abundance and $\chi$, which could be
due to systematic errors in the $\log gf$ values, non-LTE
effects, or temperature errors. 
While revision of the
temperature scale could alleviate this trend, the magnitude of the
required correction (a $\sim$200K reduction in \teff\ for most cases, 
but a considerably larger reduction in \teff\ for several stars)
exceeds our estimate of the uncertainty in \teff. 
We also find such a trend for both the program stars and the literature stars. 
When considering all lines, the program stars show
an average trend of $-$0.04 dex/eV ($\sigma$ = 0.05); 
the literature sample, as reanalyzed here, shows an identical
average trend and dispersion.
(When considering the dwarf and giant samples separately, there is no
difference between the two populations.)   

Our adopted \teff\ do not rely on the excitation balance of \fei\ lines. 
\citet{hosford10} and 
\citet{lind12} have demonstrated that excitation temperatures are 
susceptible to non-LTE effects. 
\citet{asplund01} showed that IRFM \teff\ are little dependent on 3D 
effects, although a more systematic investigation would be welcome. 
Had we employed
excitation temperatures (and their corresponding surface gravities and
microturbulent velocities), we would have obtained 
different metallicities and abundance ratios [X/Fe].  We find that
excluding lines with $\chi$ $<$ 1.2 eV decreases the average trend between
excitation potential and \fei\ abundance, as seen in previous studies
of metal-poor stars.  When considering only lines with $\chi$ $\ge$
1.2 eV, the average trends are 0.01 dex/eV ($\sigma$ = 0.20) and
$-$0.03 dex/eV ($\sigma$ = 0.11) for the program stars and literature
stars, respectively. For the program stars, had we included only those
lines with $\chi >$ 1.2 eV, the average [Fe/H] would be lower by only
0.01 $\pm$ 0.01 dex ($\sigma$ = 0.06 dex).
A more detailed assessment of the role and magnitude of non-LTE effects
is beyond the scope of the present paper, though 
we shall touch upon the matter again at several points 
in the subsections that follow. 

\section{RESULTS}
\label{sec:results}

\subsection{CEMP Objects} 

\begin{figure}[t!]
\epsscale{1.2}
\plotone{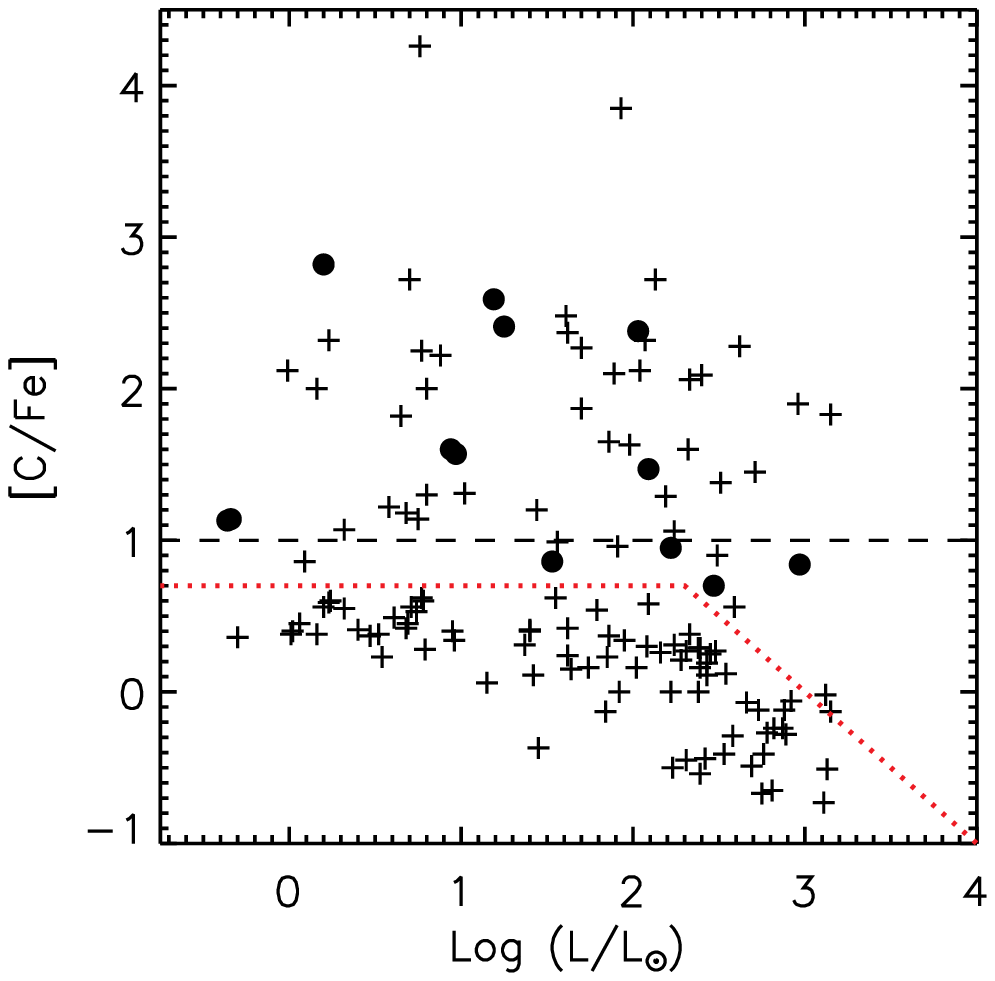} 
\caption{[C/Fe] vs.\ $\log (L/L_\odot)$ for the program stars (circles) 
and the literature sample (plus signs). (Only detections are plotted.) 
The dashed line shows the \citet{beers05} 
CEMP definition, [C/Fe] = +1.0, while the red dotted line shows the 
\citet{aoki07} CEMP definition. 
\label{fig:clsun}}
\end{figure}

We adopted the \citet{aoki07} definition 
for CEMP stars, which accounts for nucleosynthesis and 
mixing in evolved giants, namely 
(i) [C/Fe] $\ge$ +0.70, for $\log (L/L_\odot) \le 2.3$ and (ii) [C/Fe] $\ge$ 
+3.0 $-$ $\log (L/L_\odot)$, for $\log (L/L_\odot) > 2.3$. 
The original definition 
proposed by \citet{beers05} 
is [C/Fe] $\ge$ +1.0. 
In Figure \ref{fig:clsun} 
we plot [C/Fe] vs.\ $\log (L/L_\odot)$, showing 
both CEMP definitions. 
We refer the reader to Paper III for more discussion 
of the CEMP fraction 
(and the metallicity distribution function)
at lowest metallicity. 
(Throughout the present paper, we use CEMP and C-rich
interchangeably.) 
None of our program stars are C-normal objects. This is likely due 
to selection biases and that we could only obtain [C/Fe] limits 
for many program stars. 

\subsection{Abundance Trends [X/Fe] vs.\ [Fe/H]}

\begin{figure*}
\epsscale{1.0}
\plotone{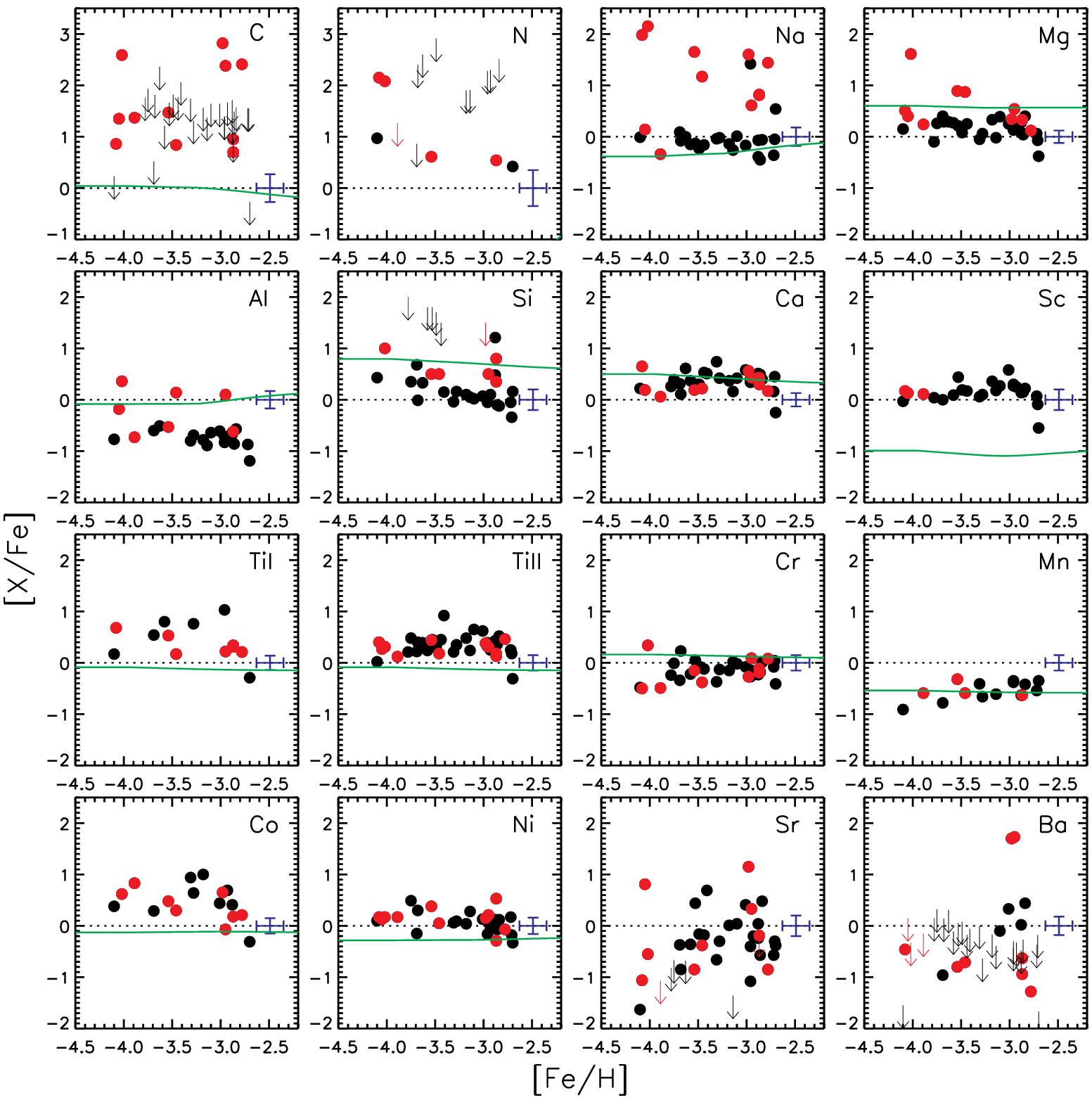} 
\vspace{8mm} 
\caption{[X/Fe] vs.\ [Fe/H] for the program stars. 
Stars with [C/Fe] $\ge$ +1.0 are marked in red. The solid green 
line represents the predictions from \citet{kobayashi06}. A 
representative error bar is shown in each panel. 
\label{fig:ab2}}
\end{figure*}

In Figure \ref{fig:ab2}, we plot abundance ratios 
[X/Fe] vs.\ [Fe/H] for the 38 program stars. 
The CEMP stars are marked in red in each panel. 
The green line in this figure represents the predictions from 
the Galactic chemical enrichment models of 
\citet{kobayashi06}, which will also be discussed in Sections 7.5 and 7.7. 
For all elements, the abundance dispersion exceeds the measurement 
uncertainty. In particular, Na, Sr, and Ba exhibit very large dispersions,  
as do C and N. 
Before we consider the complete sample (program stars + literature stars), 
we note that the abundances of Mg and Si appear to be lower than the 
canonical halo value of [$\alpha$/Fe] = +0.4. 
Furthermore, for a given element, the outliers are often, 
but not always, CEMP objects. 
However, for many of our stars we could only obtain upper limits for the 
C abundance. It would be interesting to re-assess whether the 
abundance outliers are always CEMP objects by either (a) restricting the 
sample to those stars with C measurements and/or (b) obtaining higher-quality 
spectra to convert upper limits for C into detections. 

\begin{figure*}[t!] 
\epsscale{0.90}
\plotone{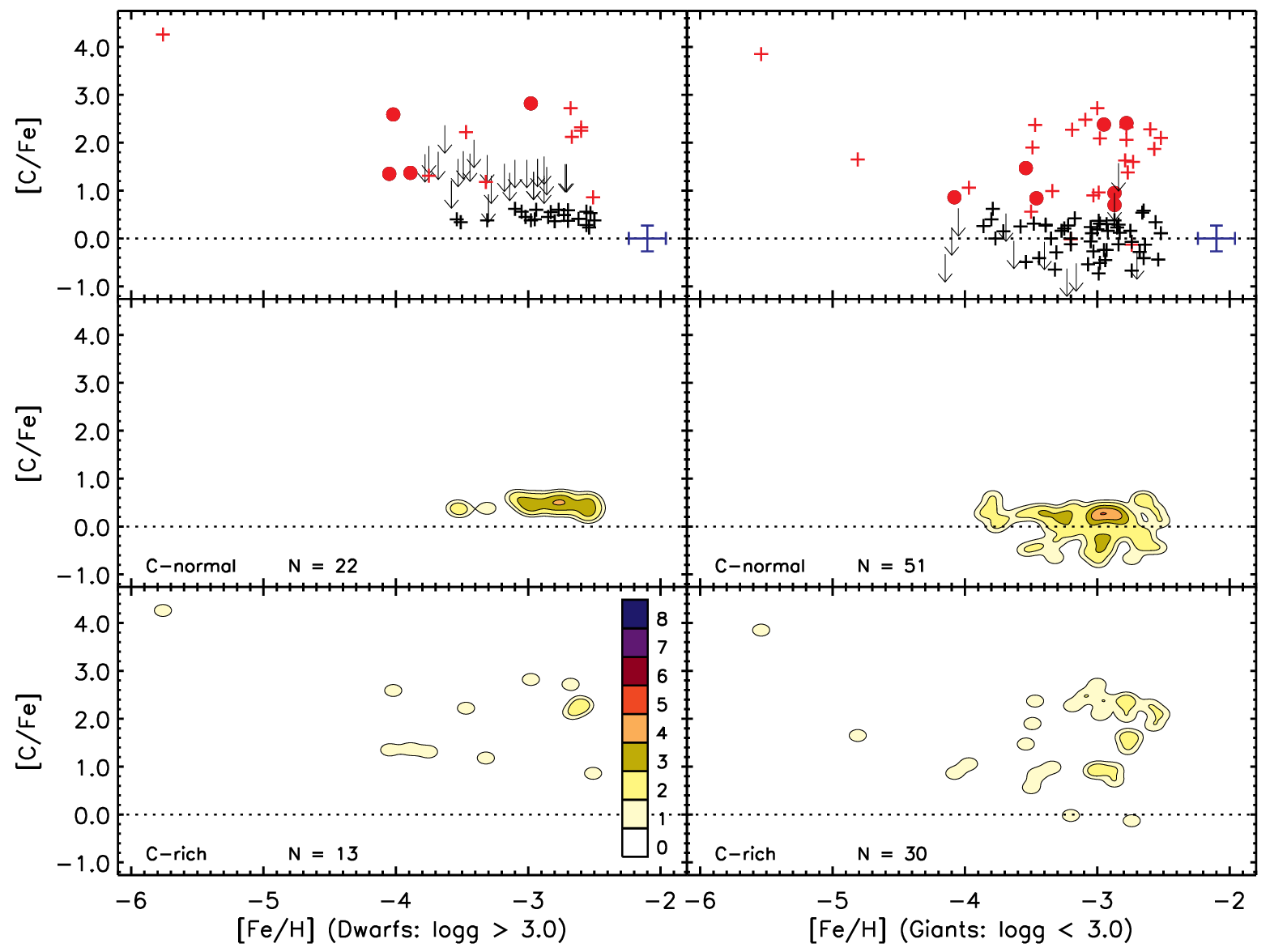} 
\caption{[C/Fe] vs.\ [Fe/H] for stars with [Fe/H] $\le$ $-$2.5. 
The left panels present only dwarf stars ($\log g$ $>$ 3.0) 
and the right panels only giants ($\log g$ $<$ 3.0), respectively. 
In the upper panels, our 
program stars are shown as circles, and the literature sample as plus signs. 
Red symbols denote C-rich objects (i.e., CEMP stars), 
while black symbols represent C-normal stars. 
In the top panels we plot a representative uncertainty 
(the average ``total error'' for the program stars). 
In the middle panels we present contour plots of the data for 
C-normal dwarfs and giants, while the lower panels show contour 
plots for C-rich dwarfs and giants. 
\label{fig:c}}
\end{figure*}

\begin{figure*}[t!] 
\epsscale{0.90}
\plotone{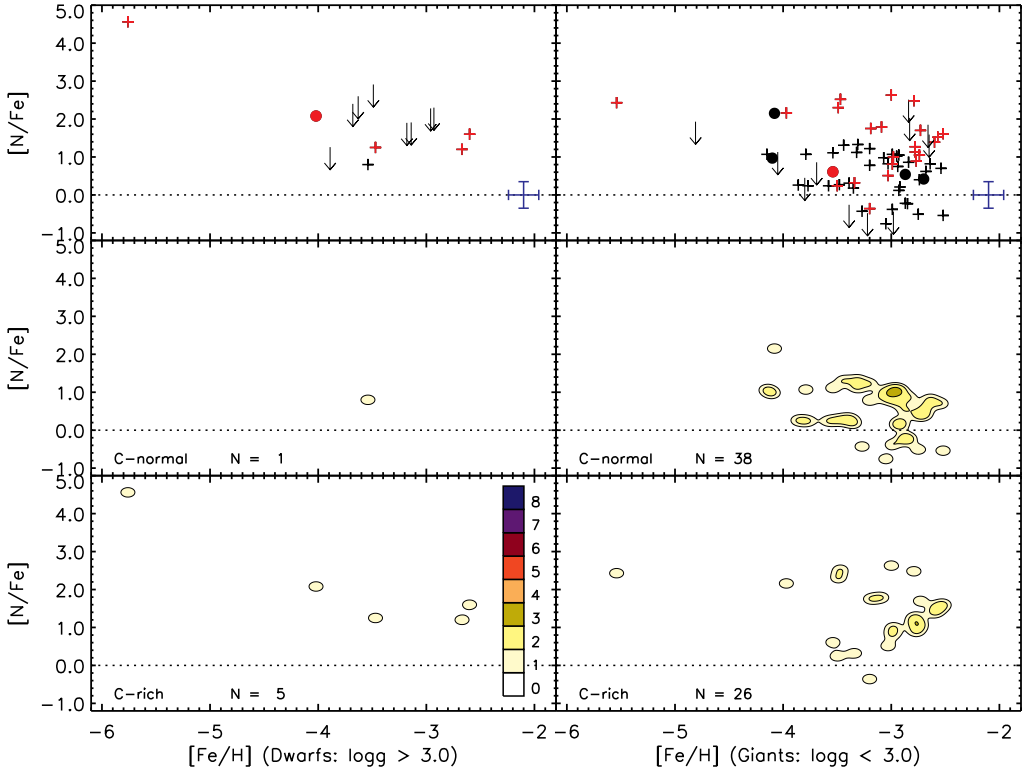} 
\caption{Same as Figure \ref{fig:c}, but for [N/Fe]. Here the red symbols 
mark objects with [N/Fe] $\ge$ +1.0. 
\label{fig:n}}
\end{figure*}

\begin{figure*}[t!] 
\epsscale{0.90}
\plotone{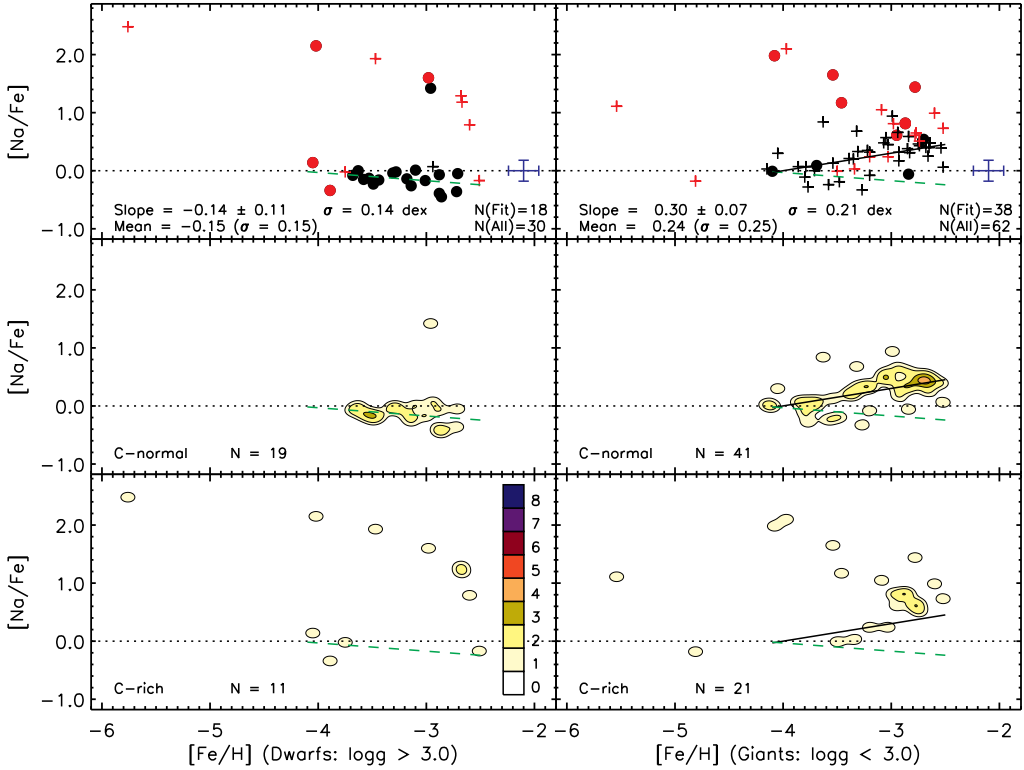} 
\caption{Same as Figure \ref{fig:c}, but for [Na/Fe]. 
In the left panels, we present the linear fit to the dwarf data 
(green dashed line), 
excluding CEMP 
objects 
and 2-$\sigma$ outliers. The slope 
(and associated error) of this fit are shown along with the dispersion 
about the best fit. 
The right panels contain the linear fit to the giant data 
(solid black line) again excluding CEMP objects and 2-$\sigma$ outliers. 
In the right panels, we overplot the linear fit to the dwarf sample. 
\label{fig:na}}
\end{figure*}

\begin{figure*}[t!]
\epsscale{0.90}
\plotone{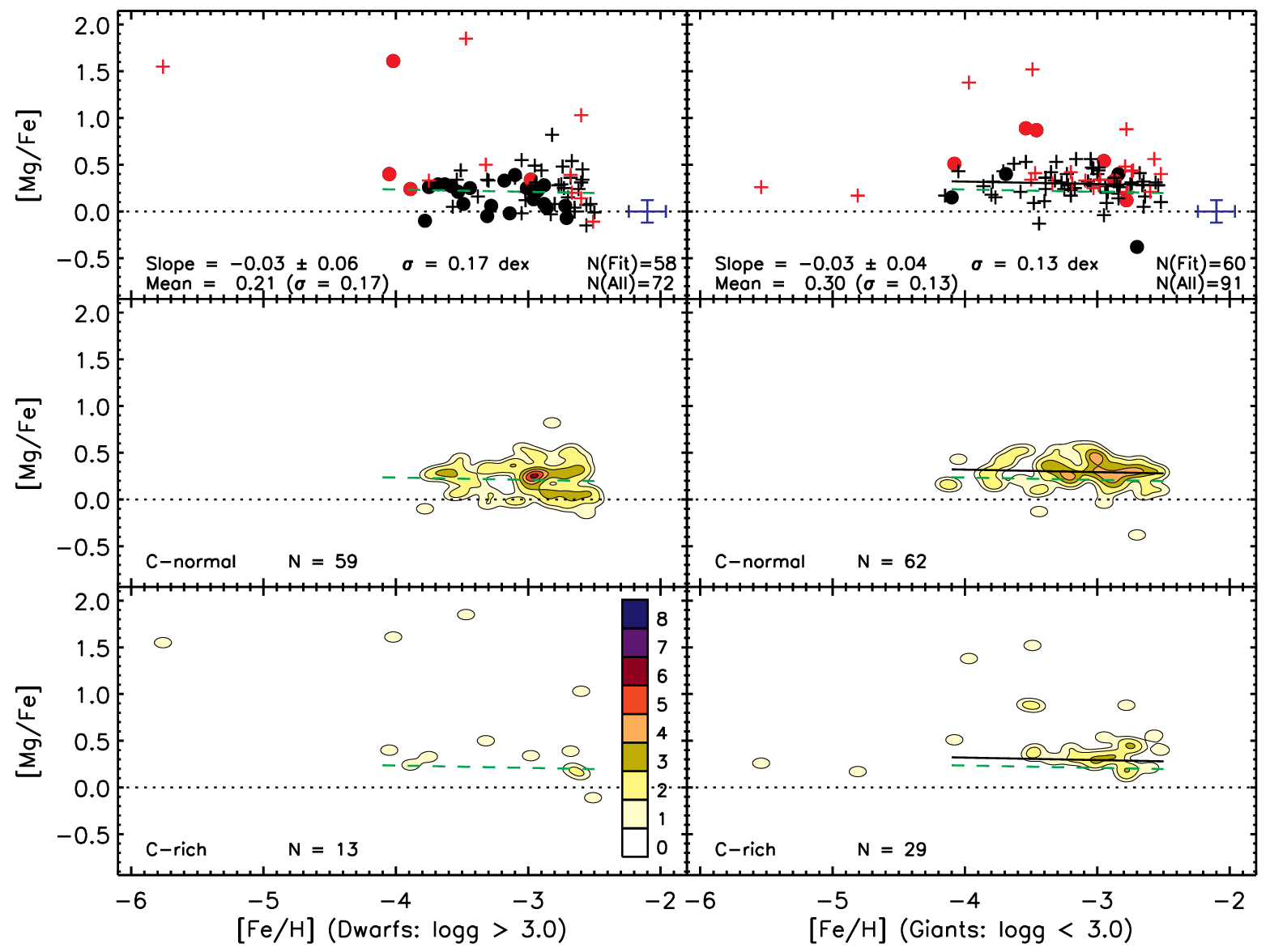} 
\caption{Same as Figure \ref{fig:na}, but for [Mg/Fe].
\label{fig:mg}}
\end{figure*}

\begin{figure*}[t!]
\epsscale{0.90}
\plotone{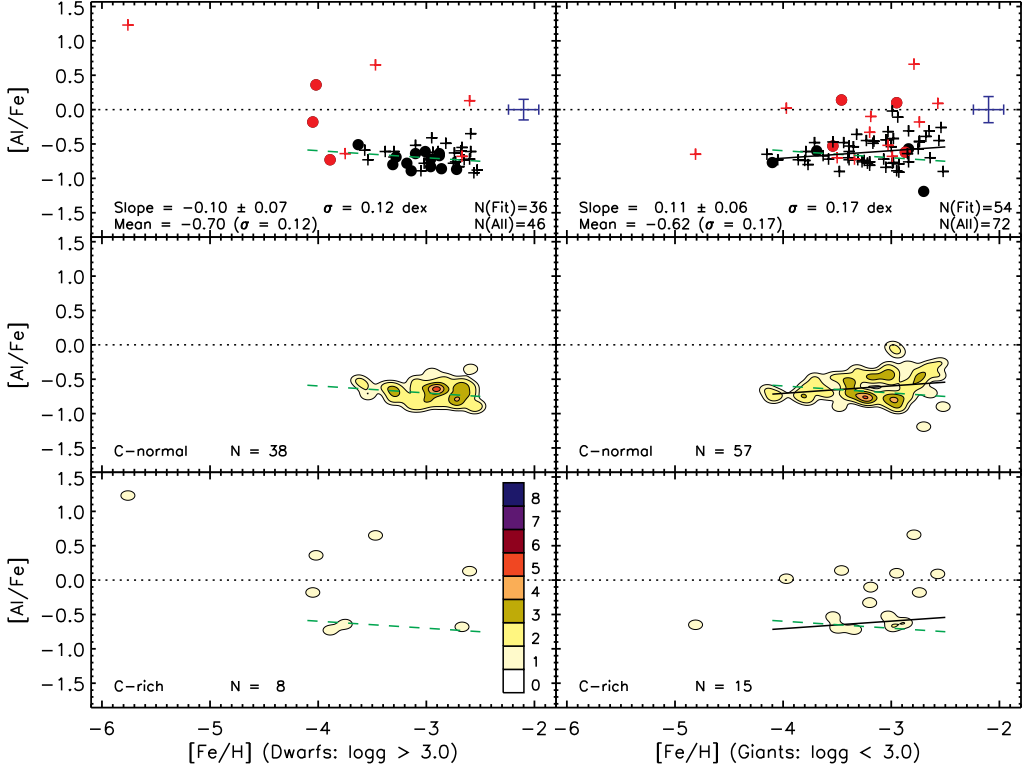} 
\caption{Same as Figure \ref{fig:na}, but for [Al/Fe].
\label{fig:al}}
\end{figure*}

\begin{figure*}[t!]
\epsscale{0.90}
\plotone{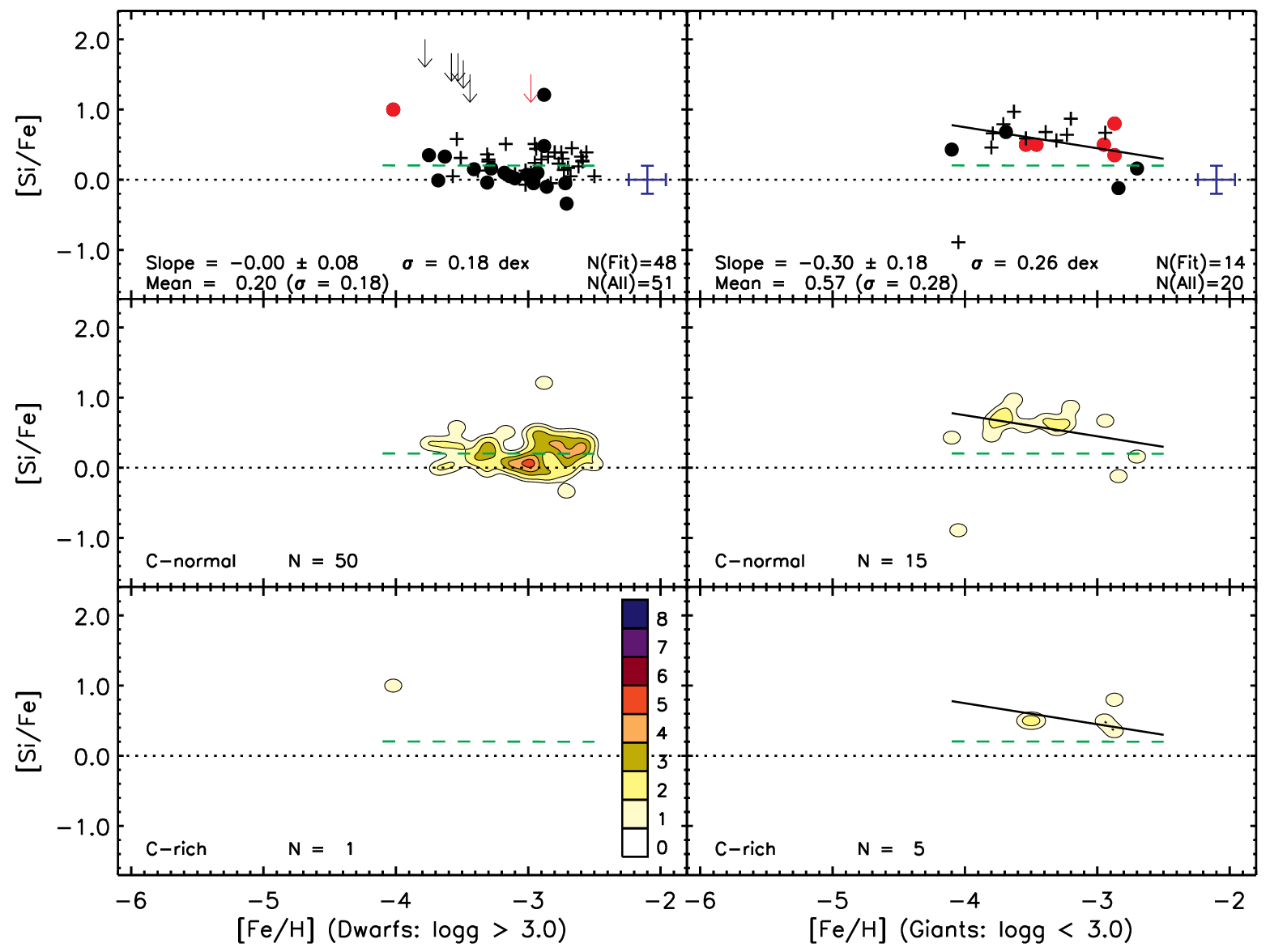} 
\caption{Same as Figure \ref{fig:na}, but for [Si/Fe].
\label{fig:si}}
\end{figure*}

\begin{figure*}[t!] 
\epsscale{0.90}
\plotone{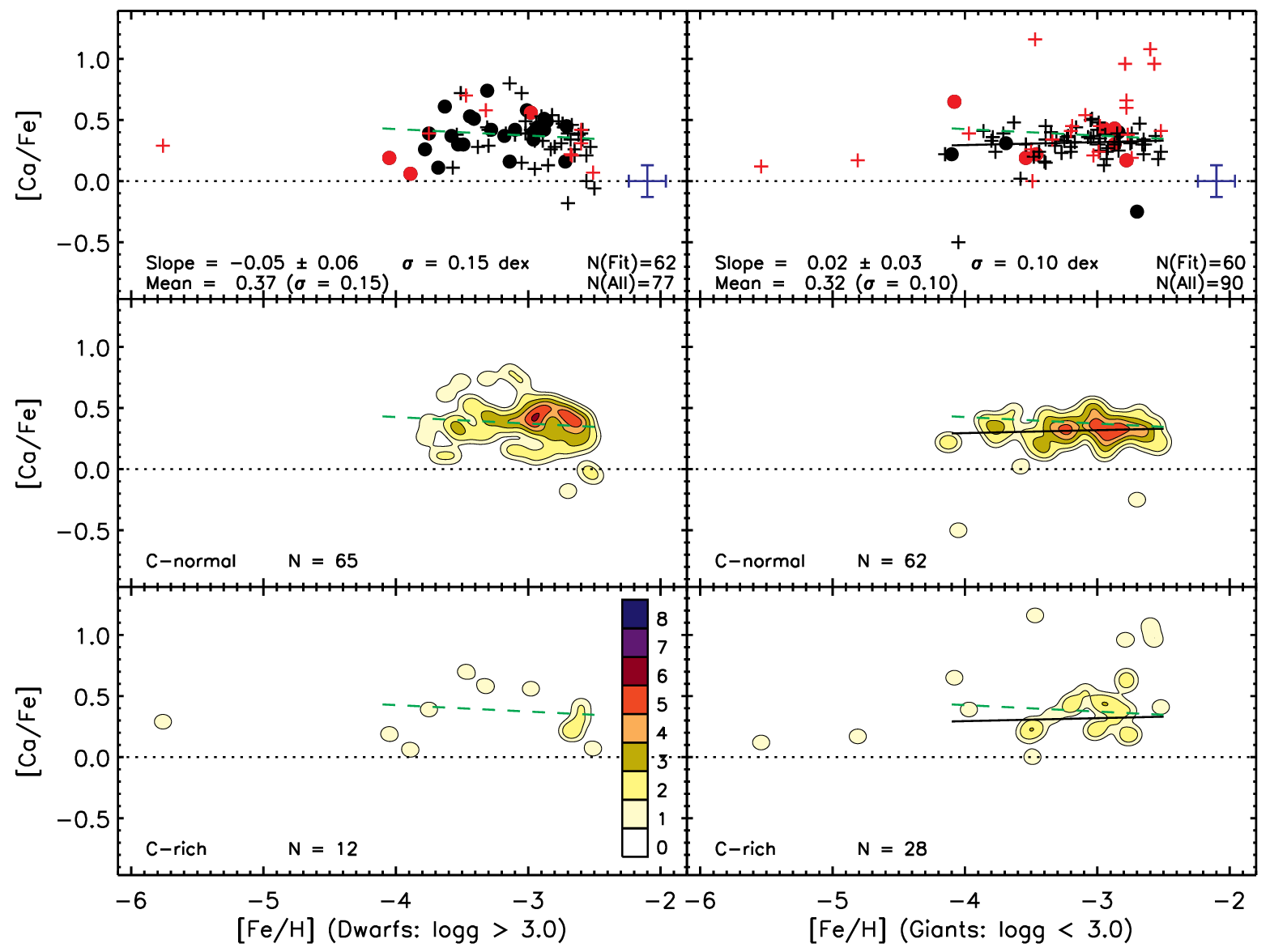} 
\caption{Same as Figure \ref{fig:na}, but for [Ca/Fe].
\label{fig:ca}}
\end{figure*}

\begin{figure*}[t!] 
\epsscale{0.90}
\plotone{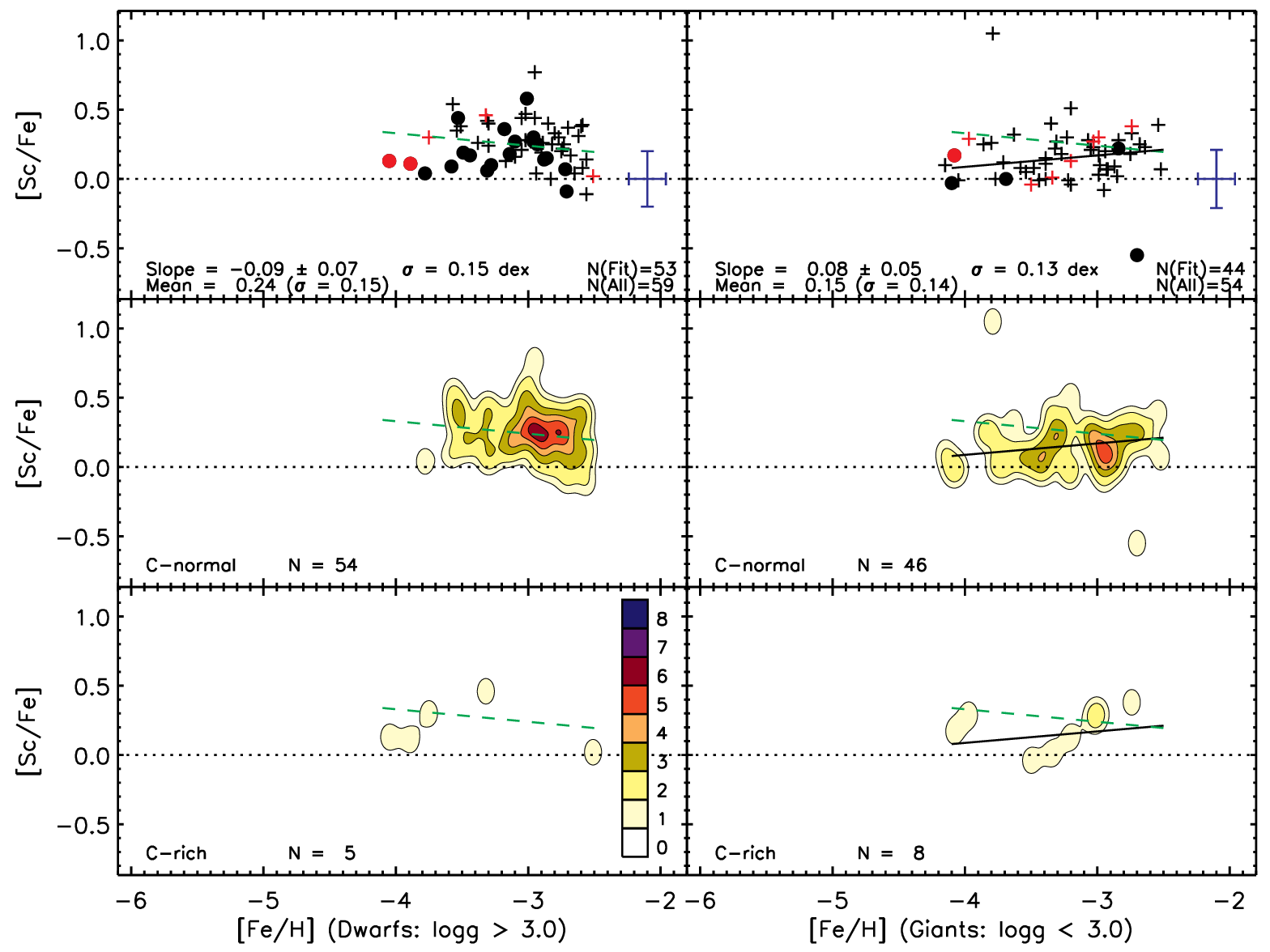} 
\caption{Same as Figure \ref{fig:na}, but for [Sc/Fe].
\label{fig:sc}}
\end{figure*}

\begin{figure*}[t!]
\epsscale{0.90}
\plotone{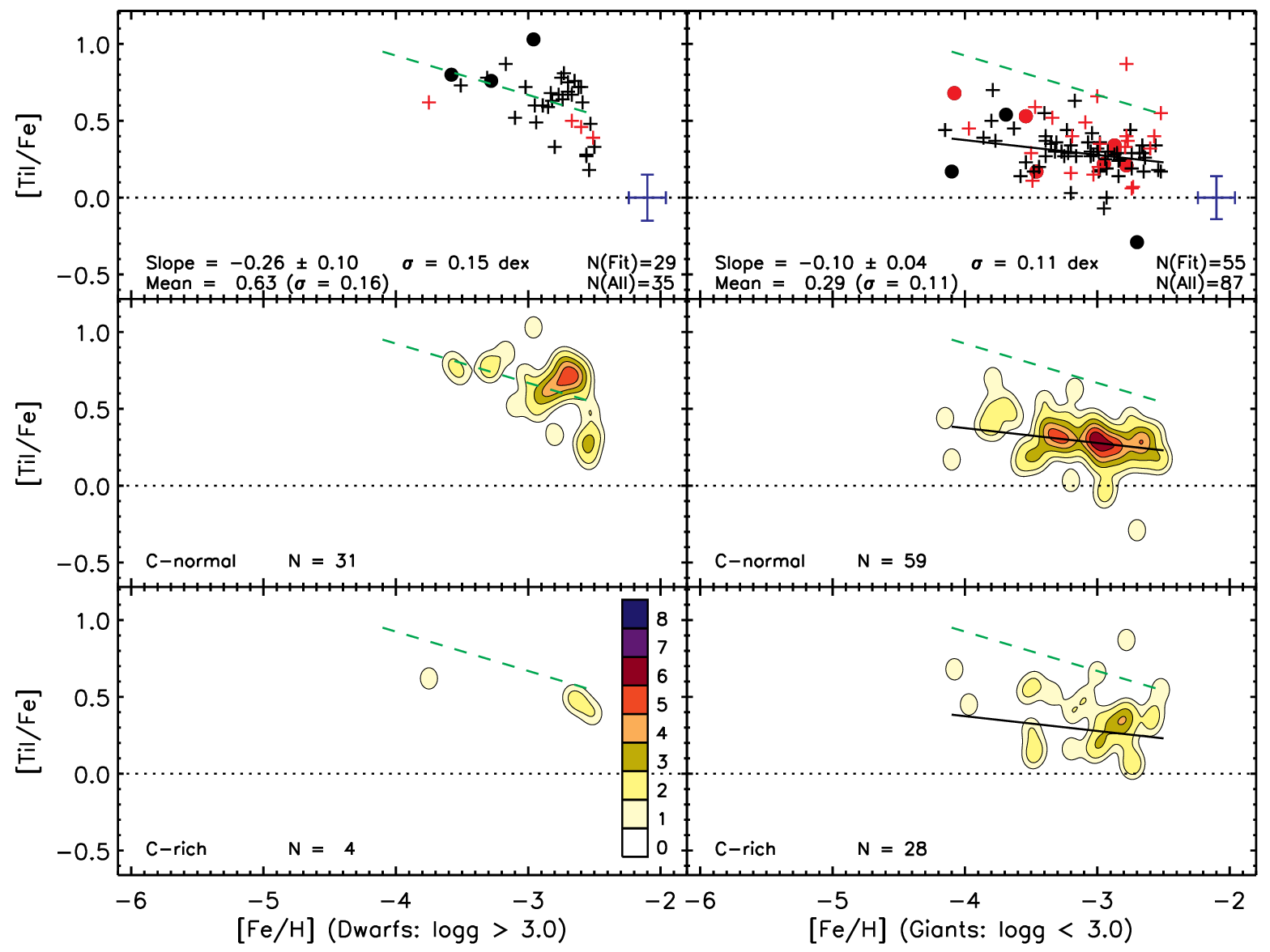} 
\caption{Same as Figure \ref{fig:na}, but for [\tii/Fe].
\label{fig:ti1}}
\end{figure*}

\begin{figure*}[t!]
\epsscale{0.90}
\plotone{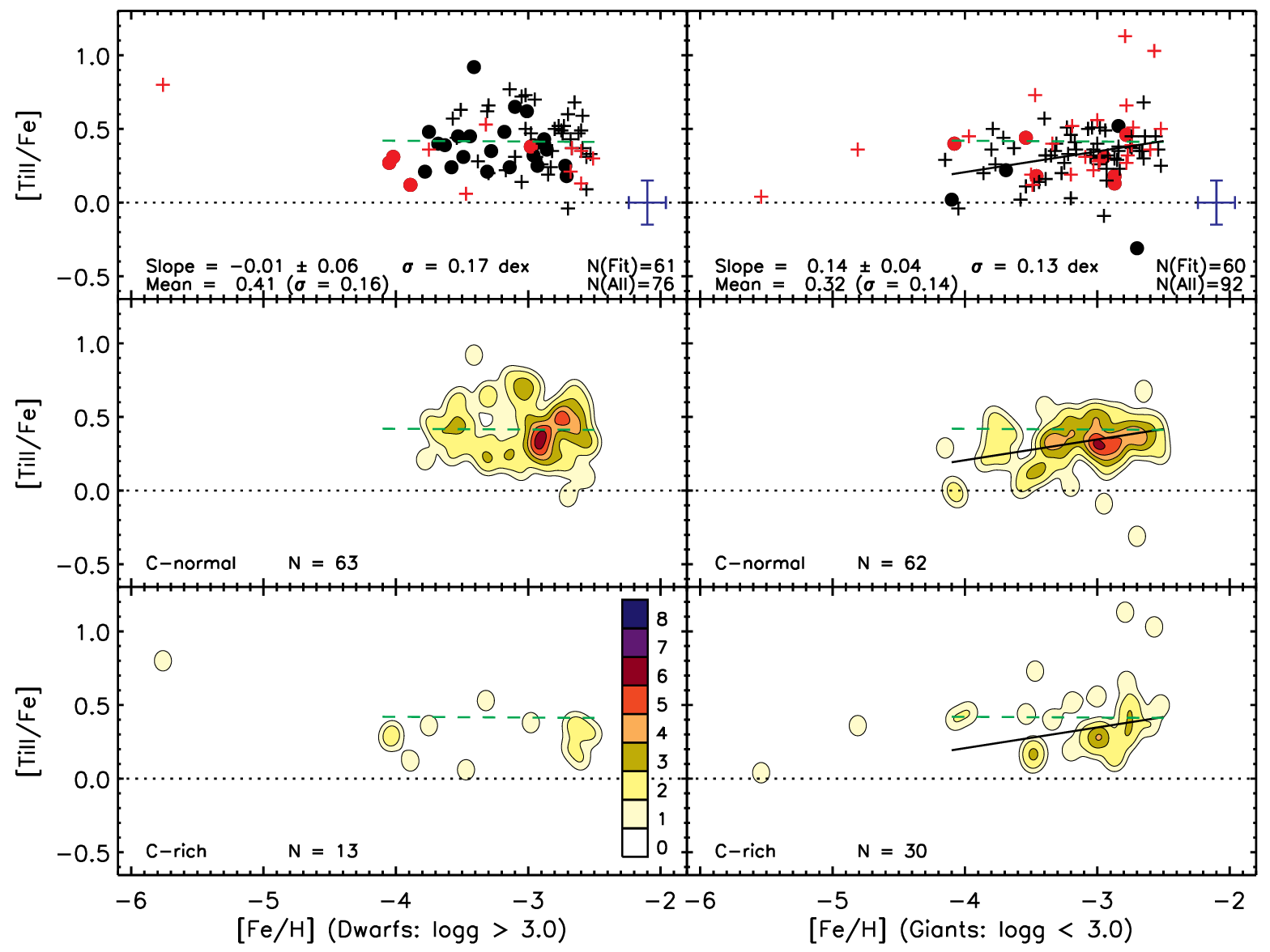} 
\caption{Same as Figure \ref{fig:na}, but for [\tiii/Fe].
\label{fig:ti2}}
\end{figure*}

\begin{figure*}[t!] 
\epsscale{0.90}
\plotone{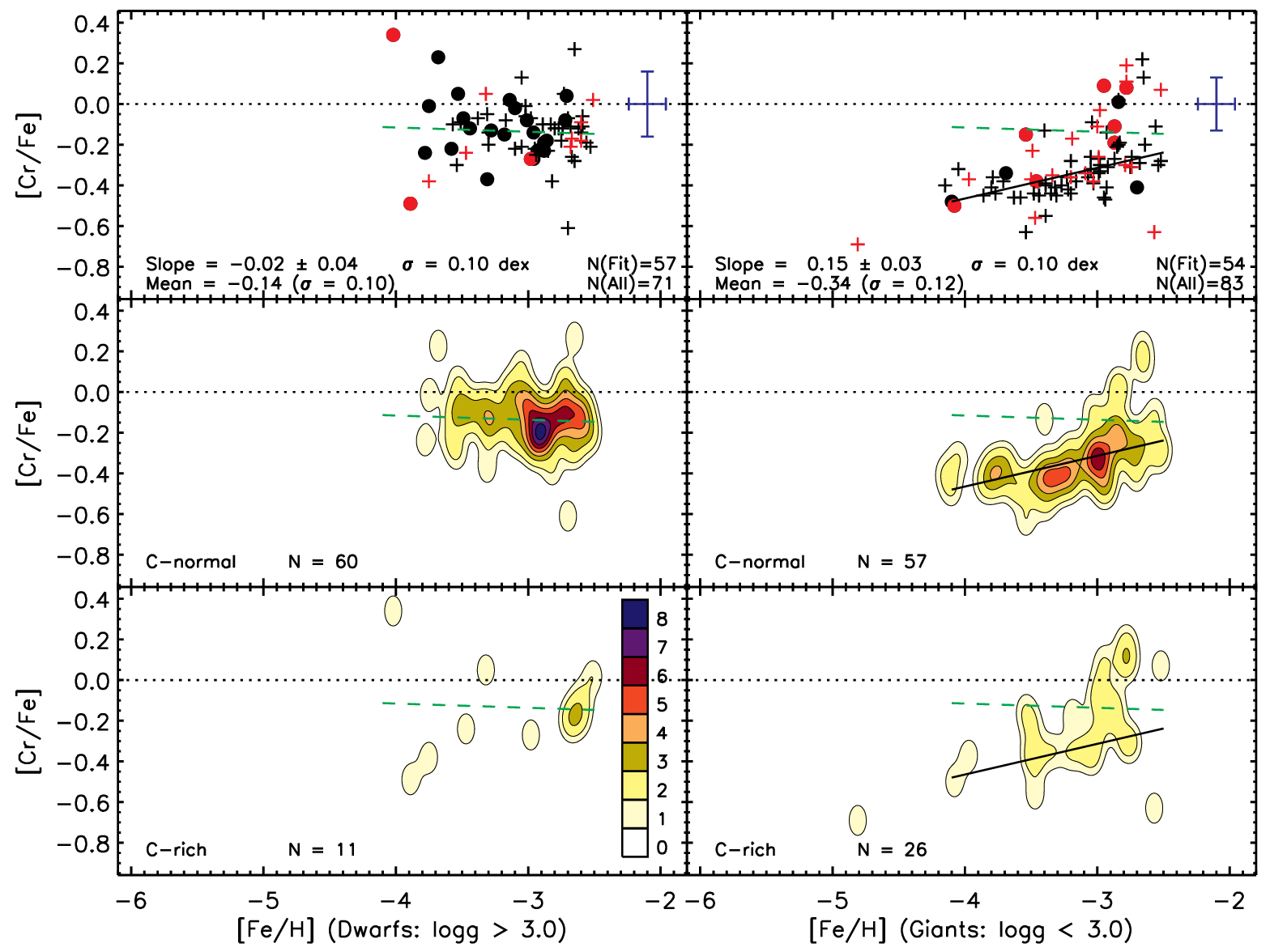} 
\caption{Same as Figure \ref{fig:na}, but for [Cr/Fe].
\label{fig:cr}}
\end{figure*}

\begin{figure*}[t!] 
\epsscale{0.90}
\plotone{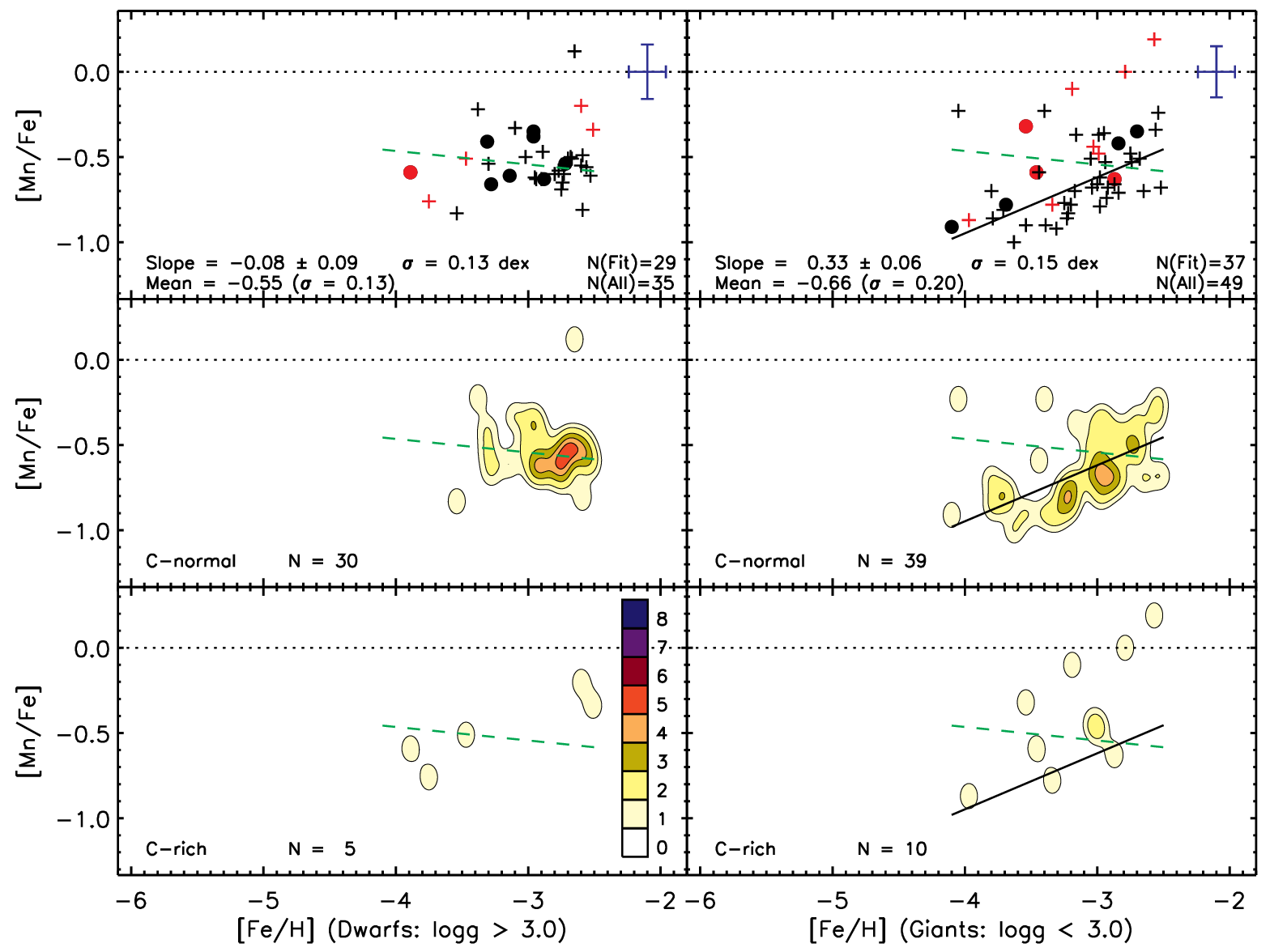} 
\caption{Same as Figure \ref{fig:na}, but for [Mn/Fe].
\label{fig:mn}}
\end{figure*}

\begin{figure*}[t!] 
\epsscale{0.90}
\plotone{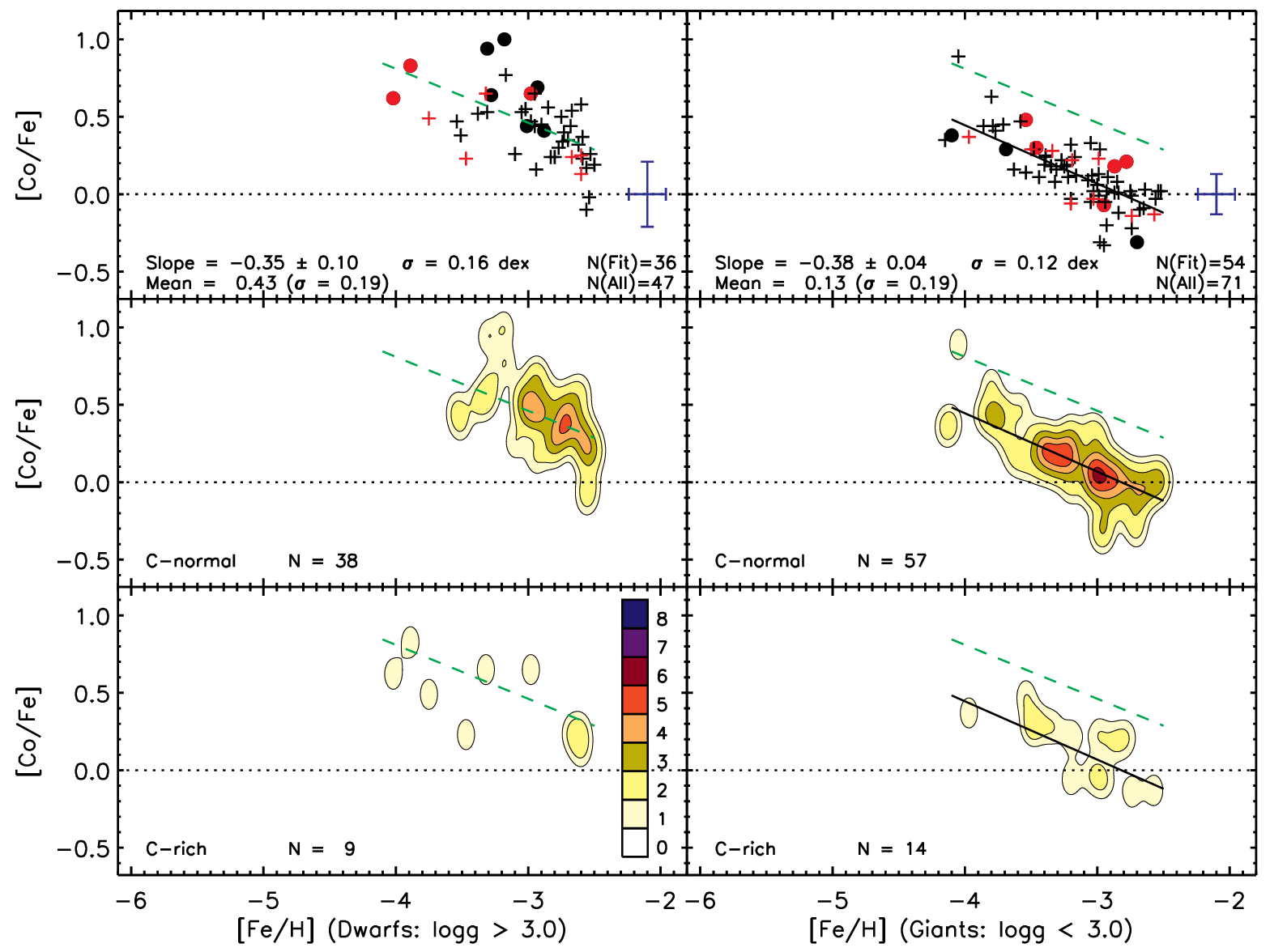} 
\caption{Same as Figure \ref{fig:na}, but for [Co/Fe].
\label{fig:co}}
\end{figure*}

\begin{figure*}[t!] 
\epsscale{0.90}
\plotone{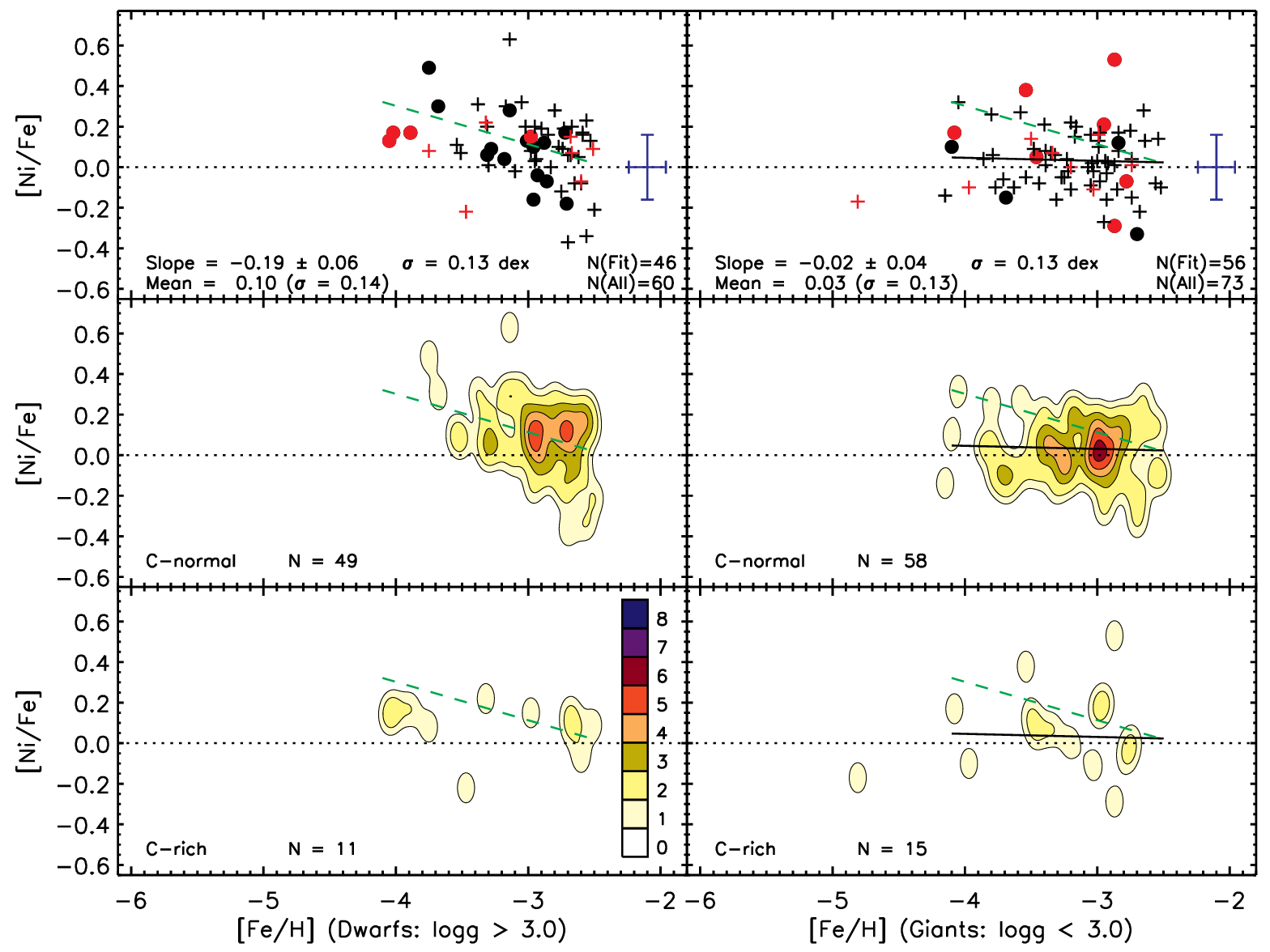} 
\caption{Same as Figure \ref{fig:na}, but for [Ni/Fe].
\label{fig:ni}}
\end{figure*}

\begin{figure*}[t!] 
\epsscale{0.90}
\plotone{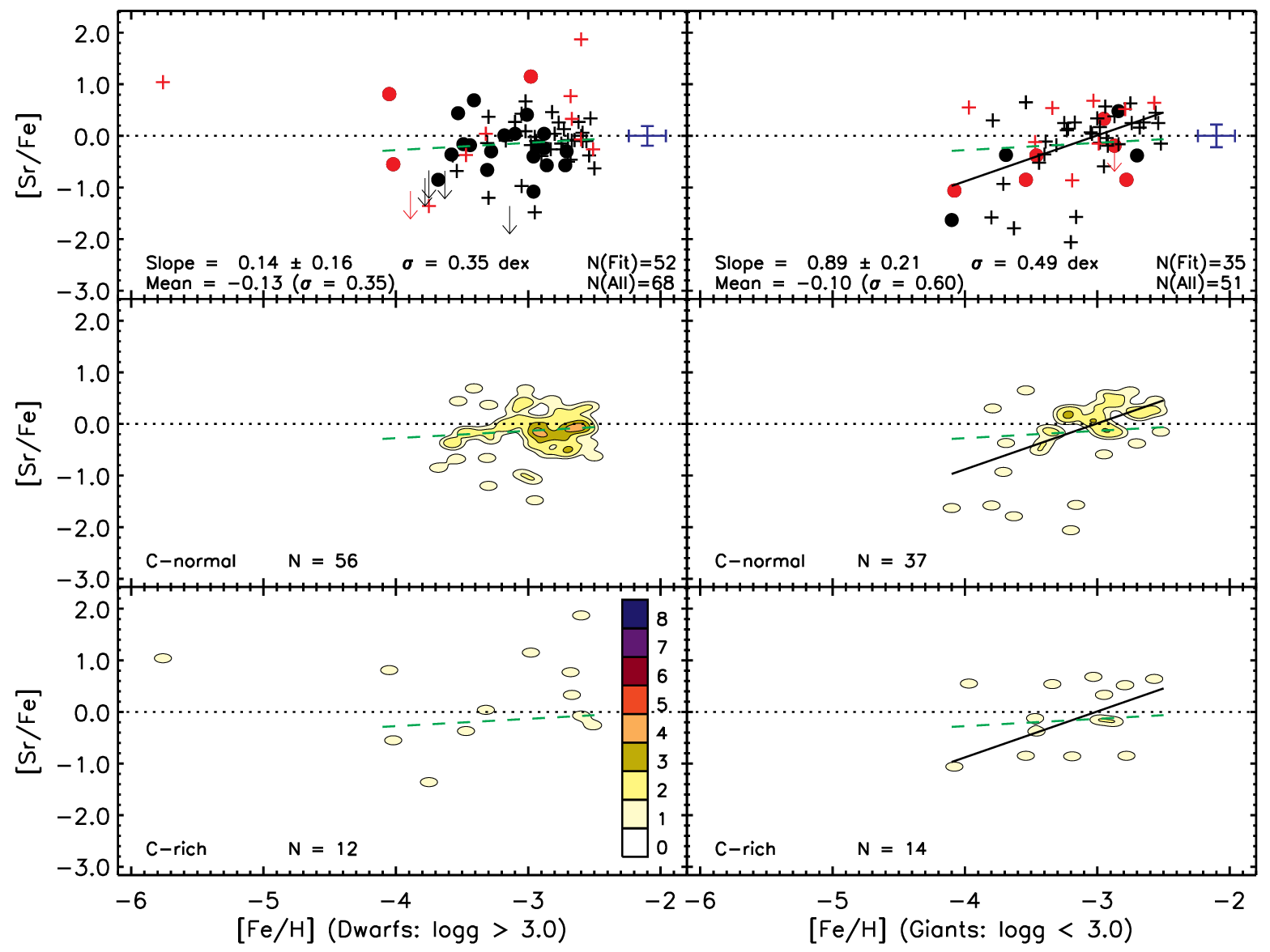} 
\caption{Same as Figure \ref{fig:na}, but for [Sr/Fe].
\label{fig:sr}}
\end{figure*}

\begin{figure*}[t!] 
\epsscale{0.90}
\plotone{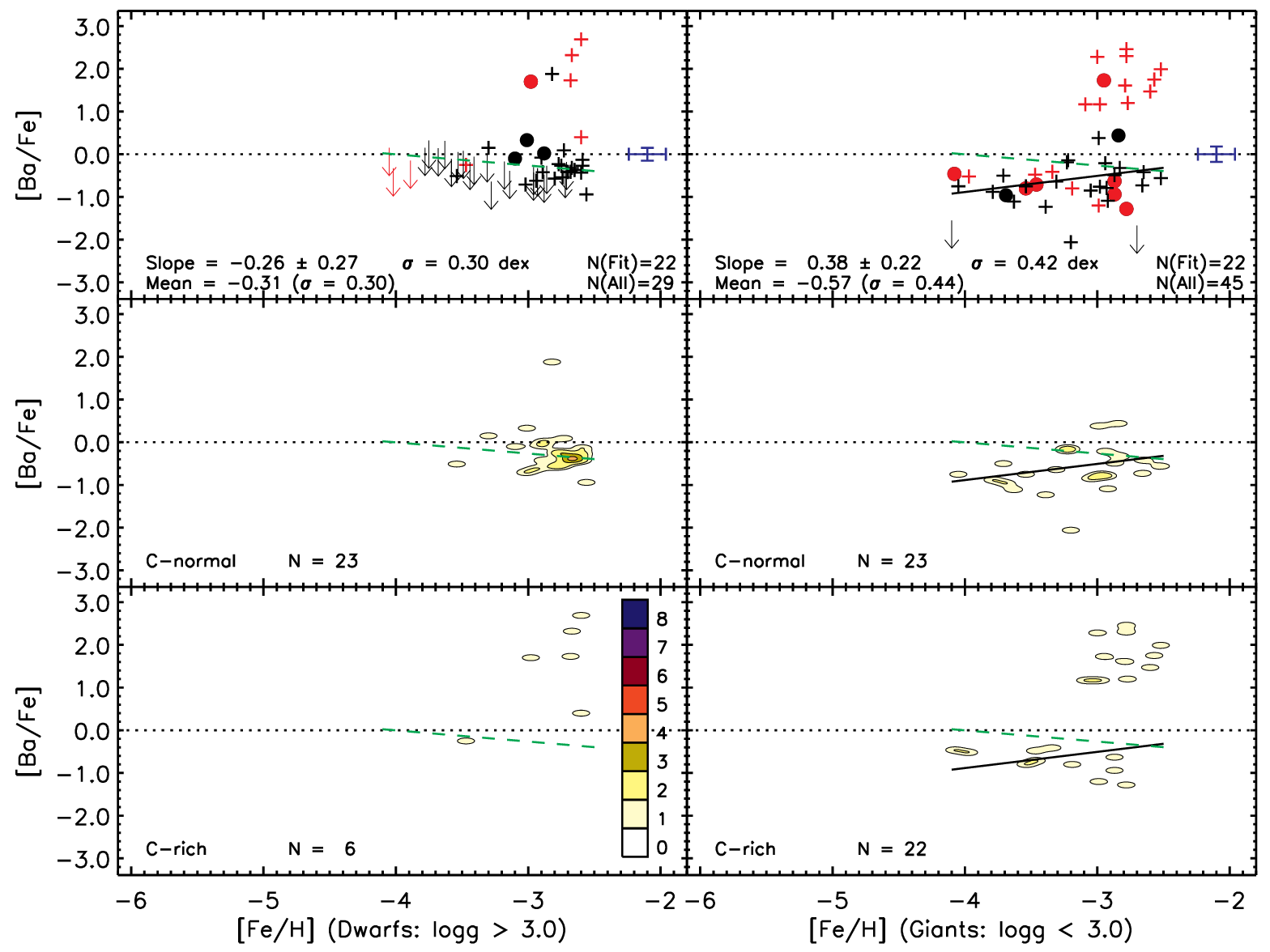} 
\caption{Same as Figure \ref{fig:na}, but for [Ba/Fe].
\label{fig:ba}}
\end{figure*}

For each element from C to Ba, we plot the abundance ratio [X/Fe] 
vs.\ [Fe/H] for the combined sample, i.e., program stars and literature 
stars (Figures \ref{fig:c} to \ref{fig:ba}). 
In all figures, 
the left panels show only the dwarf stars ($\log g$ $>$ 3.0),  
and the right panels show only the giant stars ($\log g$ $<$ 3.0).  
In Figures \ref{fig:na} to \ref{fig:ba}, 
we determine the linear fit to the data in the following manner. 
First, we exclude all CEMP stars from the fit. 
Secondly, we determine the linear fit and measure the dispersion about 
the mean trend. Thirdly, we eliminate 2-$\sigma$ outliers from the fit. 
Fourthly, we re-determine the linear fit and show in the plots 
(a) the slope of the fit and its associated uncertainty, 
(b) the dispersion about the fit, and 
(c) the mean abundance, [X/Fe], and the standard deviation. 
The reason for excluding C-rich objects and 2-$\sigma$ outliers 
was that we were seeking to (a) identify a ``normal'' population of 
metal-poor stars and (b) characterize the 
mean trend between [X/Fe] and [Fe/H] for this ``normal'' population. 
We emphasize that while CEMP objects and 2-$\sigma$ outliers 
were not included in determining the best fits to the data, these 
objects are included in all plots. 

Our motivation for attempting to define a normal population comes from
the {\sc First Stars} analyses \citep{cayrel04,spite05,francois07,bonifacio09}. As
mentioned, these studies (along with \citealt{arnone05}) 
found extremely small scatter, which is likely
due in part to the high-quality data and analysis, and to the fact that
their samples included only two CEMP giants (CS~22949$-$037 and
CS~22892$-$052) and one CEMP dwarf (CS~29527$-$015). Thus, they  
are ``biased against carbon-rich objects, and cannot
be used to constrain the full dispersion of carbon abundances at the
lowest metallicities'' \citep{cayrel04}. 
As identified in the literature, and confirmed in the
present series, CEMP stars often have anomalous abundance ratios
for elements other than C. However, it has also become evident that
some C-normal stars show peculiar abundances for other elements (e.g.,
\citealt{cohen08}).  Therefore, when searching for a ``normal''
population of metal-poor stars, we eliminate CEMP objects as well as 
2-$\sigma$ outliers from the fit, but retain them in the plots.

In these figures, we also present contour plots illustrating the density in the 
[X/Fe] vs.\ [Fe/H] plane. We consider (a) C-normal dwarfs, 
(b) CEMP dwarfs, (c) C-normal giants, 
and (d) CEMP giants. 
For each sample, we represent a given data point with a two-dimensional  
Gaussian, for which the FWHM in the [Fe/H] and [X/Fe] directions corresponds 
to our estimates of the typical measurement 
uncertainties for [Fe/H] and [X/Fe], respectively. The 
height of each Gaussian is set to 1.0. The Gaussians 
are then summed and a contour plot is generated. In the event that a 
given panel has $N$ data points with a single value of [Fe/H] and [X/Fe], 
the contour would have a maximum height of $N$. 
One advantage of using such plots is that the density of data 
points can be more readily seen, which may be useful given our sample size. 

Although we determine linear fits to the data throughout this paper,
we are not suggesting that this is the appropriate function to
use. Rather, we consider this a first pass to begin to understand how
the abundance ratios, [X/Fe], evolve with metallicity, [Fe/H]. Such an
approach has the advantage of enabling comparisons between different
objects (dwarfs vs.\ giants), different elements, and different
studies.  Without knowing what is the appropriate function to use,
another option might be to follow \citet{norris01}, and use LOESS  
regression lines \citep{cleveland79}.

The reason for separating the dwarfs from giants is that we want 
to compare, as best we can, stars with similar stellar parameters 
to minimize systematic errors 
(e.g., \citealt{cayrel04,melendez08,bonifacio09,alvesbrito10,nissen10})\footnote{We reiterate that our ``dwarf'' definition, $\log g$ $>$ 3.0 will 
include subgiants as well as stars at the base of the giant branch. In reality, 
the two groups which we refer to as ``giants'' and ``dwarfs'' 
represent a ``low gravity'' and a ``high gravity'' group. Recall 
that the temperatures for the literature sample are internally consistent 
within each group.}. 
Additionally, 
we distinguish dwarfs from giants because, for some elements (notably 
C and N), we anticipate abundance differences due to 
stellar evolution. 
Thus, by 
considering dwarfs and giants separately, we hope to minimize 
such effects and thereby more clearly identify a ``normal'' population 
of stars, and quantify the abundance trends [X/Fe] vs.\ [Fe/H] for 
the ``normal'' dwarf sample and the ``normal'' giant sample. 

\begin{figure}[t!]
\epsscale{1.2}
\plotone{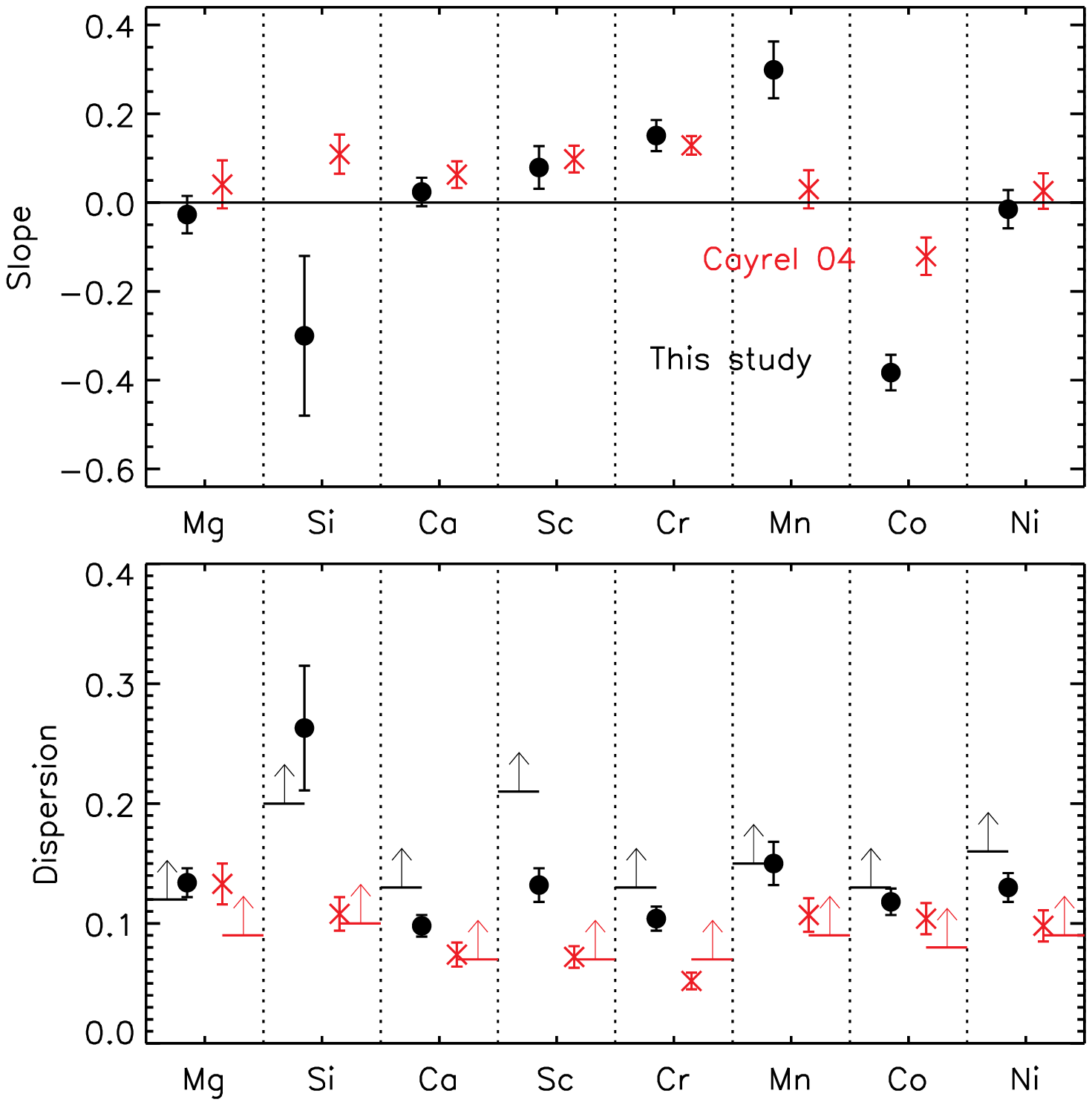} 
\caption{Comparison of the slopes of the linear fit to the data, [X/Fe] 
vs.\ [Fe/H], (upper) and the dispersion about the linear fit (lower) 
for our study (black circles) and the \citet{cayrel04} sample 
(red crosses). The linear fit excludes CEMP objects and 2-$\sigma$ outliers. 
The horizontal bars (and upwards facing arrows) 
in the lower panel indicate the representative 
measurement uncertainty. 
\label{fig:slope_comp}}
\end{figure}

A measure of (a) whether a ``normal'' population indeed exists, and (b)
whether our selection criteria are able to identify such a population, 
is to compare the dispersion about the linear fit to the
representative measurement uncertainty (the average ``total error''
for the program stars).  (Such a comparison is only meaningful 
if, as we assume, the dependence of [X/Fe] versus [Fe/H] is linear. 
We reiterate that while we use linear functions, we are not 
suggesting that they are correct. Instead, this assumption represents 
a first step to understanding trends between [X/Fe] and [Fe/H].) 
In Table \ref{tab:disp}, we compare the observed dispersion about the linear 
fit to the typical, i.e., average, measurement uncertainty (the lower panel 
of Figure \ref{fig:slope_comp} shows the comparison for a subset of elements 
in giant stars).  
In many cases (e.g., Al, Ca, Ti, Mn, Co, Ni),
the dispersion in [X/Fe] about the linear fit is in good agreement
with the representative measurement uncertainty. There are examples in
which the representative uncertainty exceeds the dispersion (e.g., Sc,
Cr), which may indicate that the uncertainties are overestimated.  For
other cases (e.g., Sr, Ba), the dispersion about the linear trend far
exceeds the measurement uncertainty, suggesting that the
uncertainties are underestimated, there is a large abundance
dispersion, and/or the ``normal'' population, if present,
was not successfully identified.

\begin{deluxetable}{lccccccc}
\tablecolumns{8} 
\tablewidth{0pc} 
\tabletypesize{\footnotesize}
\tablecaption{Comparison of Abundance Dispersions and 
Measurement Uncertainties \label{tab:disp}}
\tablehead{ 
\colhead{Species} & 
\colhead{$\sigma$\tablenotemark{a}} &
\colhead{Measurement} &
\colhead{N$_{\rm stars}$} & 
\colhead{} & 
\colhead{$\sigma$\tablenotemark{a}} &
\colhead{Measurement} &
\colhead{N$_{\rm stars}$} \\
\colhead{} & 
\colhead{} &
\colhead{Uncertainty\tablenotemark{b}} &
\colhead{} & 
\colhead{} & 
\colhead{} &
\colhead{Uncertainty\tablenotemark{b}} \\ 
\cline{2-4} 
\cline{6-8} 
\colhead{} & 
\multicolumn{3}{c}{Dwarfs} & 
\colhead{} & 
\multicolumn{3}{c}{Giants} \\ 
\colhead{(1)} &
\colhead{(2)} &
\colhead{(3)} &
\colhead{(4)} &
\colhead{} &
\colhead{(5)} &
\colhead{(6)} &
\colhead{(7)} 
}
\startdata 
{\rm [Na/Fe]}    & 0.14 & 0.18 & 18 &  & 0.21 & 0.18 & 38 \\
{\rm [Mg/Fe]}    & 0.17 & 0.12 & 58 &  & 0.13 & 0.12 & 60 \\
{\rm [Al/Fe]}    & 0.12 & 0.15 & 36 &  & 0.17 & 0.19 & 54 \\
{\rm [Si/Fe]}    & 0.18 & 0.20 & 48 &  & 0.26 & 0.20 & 14 \\
{\rm [Ca/Fe]}    & 0.15 & 0.13 & 62 &  & 0.10 & 0.13 & 60 \\
{\rm [Sc/Fe]}    & 0.13 & 0.20 & 53 &  & 0.13 & 0.21 & 44 \\
{\rm [\tii/Fe]}  & 0.15 & 0.15 & 29 &  & 0.10 & 0.14 & 55 \\
{\rm [\tiii/Fe]} & 0.17 & 0.15 & 61 &  & 0.13 & 0.15 & 60 \\
{\rm [Cr/Fe]}    & 0.10 & 0.16 & 57 &  & 0.10 & 0.13 & 54 \\
{\rm [Mn/Fe]}    & 0.13 & 0.16 & 29 &  & 0.15 & 0.15 & 37 \\
{\rm [Co/Fe]}    & 0.17 & 0.21 & 36 &  & 0.12 & 0.13 & 54 \\
{\rm [Ni/Fe]}    & 0.13 & 0.16 & 46 &  & 0.13 & 0.16 & 56 \\
{\rm [Sr/Fe]}    & 0.36 & 0.19 & 52 &  & 0.49 & 0.22 & 35 \\
{\rm [Ba/Fe]}    & 0.30 & 0.15 & 22 &  & 0.41 & 0.18 & 22 \\
\enddata 

\tablenotetext{a}{Dispersions about the linear fit to the data in the 
[X/Fe] vs.\ [Fe/H] plane, for stars with [Fe/H] $\le$ $-$2.5, after 
discarding 2-$\sigma$ outliers and CEMP objects, i.e., the values shown 
in Figures \ref{fig:na} to \ref{fig:ba}.} 
\tablenotetext{b}{Average measurement uncertainty for the program stars 
based on the data presented in Table \ref{tab:parvar}.} 

\end{deluxetable}

\subsection{Dwarf vs.\ Giant Differences}

\citet{bonifacio09} conducted a detailed analysis of metal-poor dwarf
stars, and then compared their results with abundances from the
\citet{cayrel04} metal-poor giants.  For Ca, Ni, Sr, and Ba, they
found very good agreement between the abundances from dwarfs and
giants.  However, for C, Sc, Ti, Cr, Mn, and Co, the abundances from
dwarfs were roughly 0.2 dex higher than those for giants; for Mg and
Si, the abundances from dwarfs were approximately 0.2 dex lower than
for giants. Carbon was the only element in which the abundance difference
between dwarfs and giants could be attributed to an astrophysical
cause, namely mixing and nucleosynthesis in giants \citep{iben64}. For
some elements, the abundance discrepancies could be the result of
neglecting non-LTE and/or 3D effects in the analysis.
\citet{bonifacio09} advocated using the abundances from giants for
      comparisons with chemical evolution models, because
      their 3D corrections were typically smaller than for dwarfs. 
On the other hand, \citet{asplund05} found that non-LTE effects tend to be 
larger for giants. Also, \citet{collet11} argued that the CO$^5$BOLD 
3D models used by \citet{bonifacio09} suffer from 
systematic errors in the high atmospheric layers, and thus 
underestimate the 3D effects. 

For the complete sample, program stars + literature stars, 
we now use Figures \ref{fig:na} to \ref{fig:ba} 
to compare abundances between giants and dwarfs. 
(For C and N, we defer 
to \citealt{spite05} and \citealt{bonifacio09}, who have 
more accurate measurements for homogeneous samples of dwarfs and giants.) 
We consider 
the slope in [X/Fe] vs.\ [Fe/H] and the mean abundance, for each 
element, having excluded CEMP stars and 2-$\sigma$ outliers. 
For Na, Al, Sc, \tiii, Cr, Mn, Ni, and Sr, 
we note that the slope 
differs between dwarfs and giants at the 
2-$\sigma$ level or higher. 
The remaining elements, Mg, Si, Ca, \tii, Co, and Ba, exhibit slopes 
in [X/Fe] vs.\ [Fe/H] that agree between dwarfs and giants.  
For Na, Si, \tii, Cr, Co, and Ba, the mean abundance between 
dwarfs and giants differs by more than 0.20 dex. For Na and Si, the 
mean abundance for giants exceeds the mean abundance for dwarfs,  
while for \tii, Cr, Co, and Ba, the mean abundance for dwarfs 
is higher than for giants. The sign of the differences for Si, Ti, 
and Co is the same as that found by \citet{bonifacio09}. 
Consideration of the standard error of the mean would indicate that, 
for all elements except Al, Ca, Ni, Sr, and Ba, the differences 
in the average abundances between the dwarf and giant samples 
are significant at the 3-$\sigma$ 
level or higher. 
There is no obvious astrophysical cause for these abundance differences 
between dwarfs and giants; thus, 
we would attribute these abundance differences to non-LTE 
and/or 3D effects. We remind the reader that we have employed 
the same set of lines for giants and dwarfs, although the 
giants and dwarfs may use different subsets of lines for a given element. 

\begin{figure}[t!]
\epsscale{1.1}
\plotone{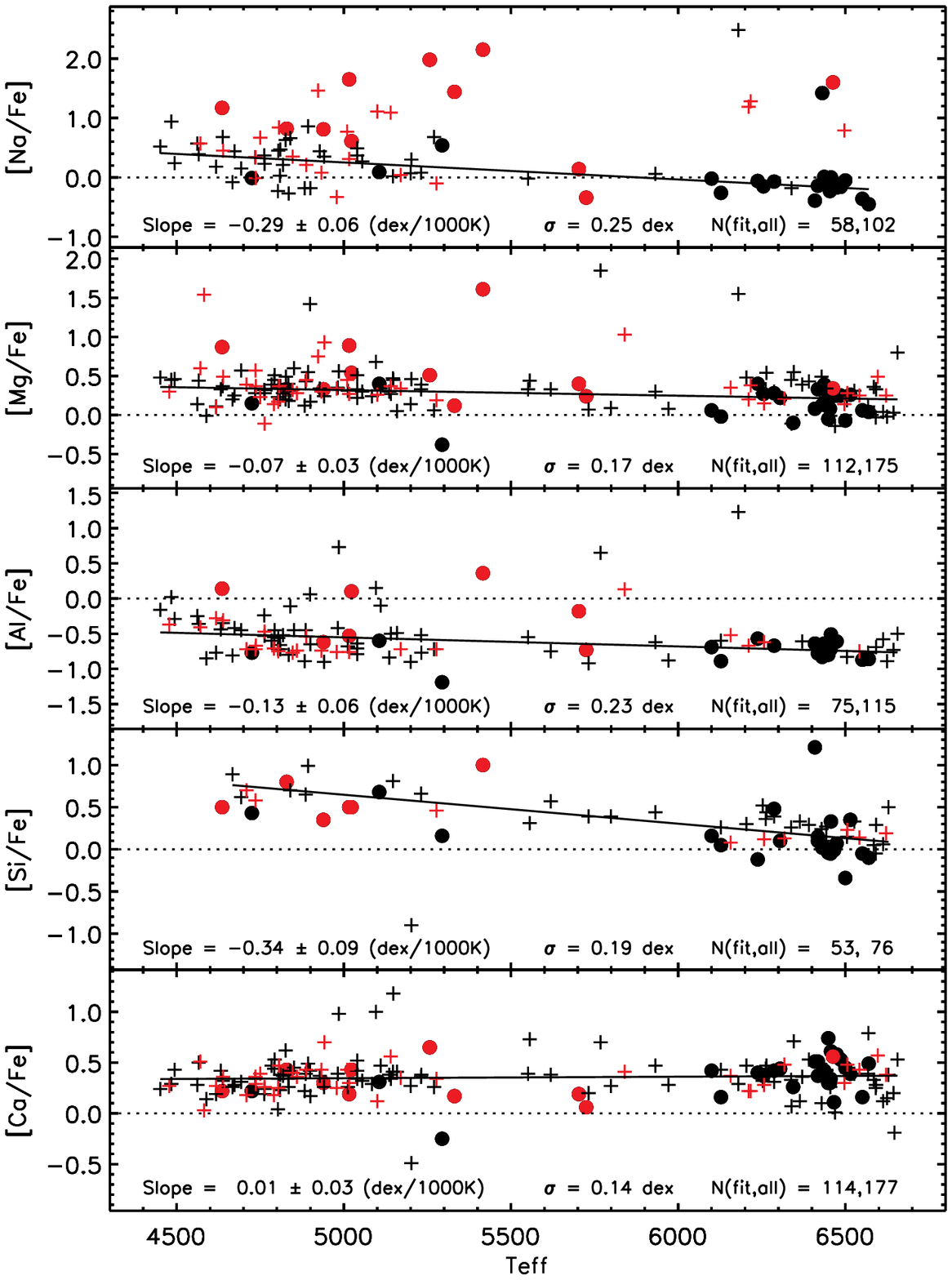} 
\caption{[X/Fe] vs.\ \teff\ for Na, Mg, Al, Si, and Ca. 
We only plot stars with [Fe/H] $\le$ $-$2.5. 
The symbols are the same as in Figure \ref{fig:c}. In each 
panel we show the linear fit to the data excluding CEMP objects 
and 2-$\sigma$ outliers. In each panel we present the slope, uncertainty, 
dispersion about the linear fit, the number of stars involved in the fit, 
and the total number of stars plotted. 
\label{fig:xfeteffa}}
\end{figure}

\begin{figure}[t!]
\epsscale{1.1}
\plotone{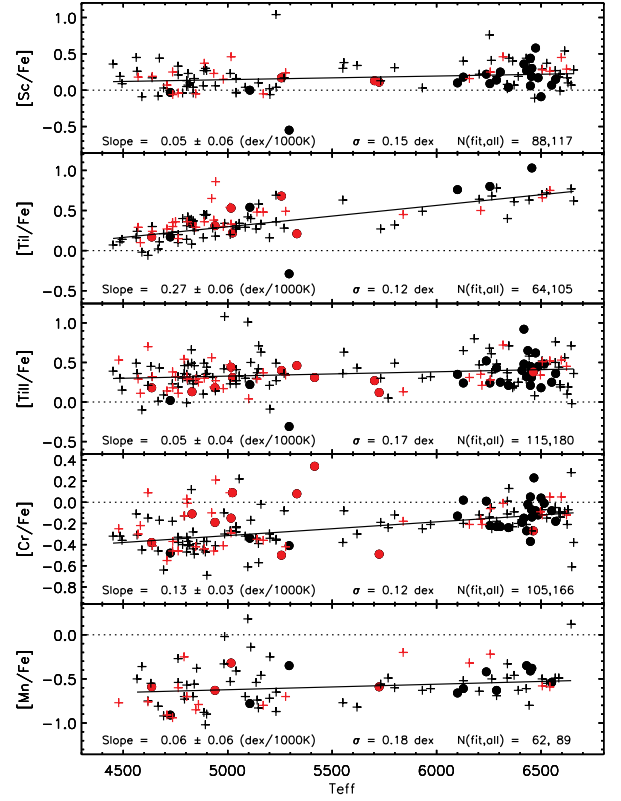} 
\caption{Same as Figure \ref{fig:xfeteffa}, but for Sc, \tii, \tiii, 
Cr, and Mn. 
\label{fig:xfeteffb}}
\end{figure}

\begin{figure}[t!]
\epsscale{1.1}
\plotone{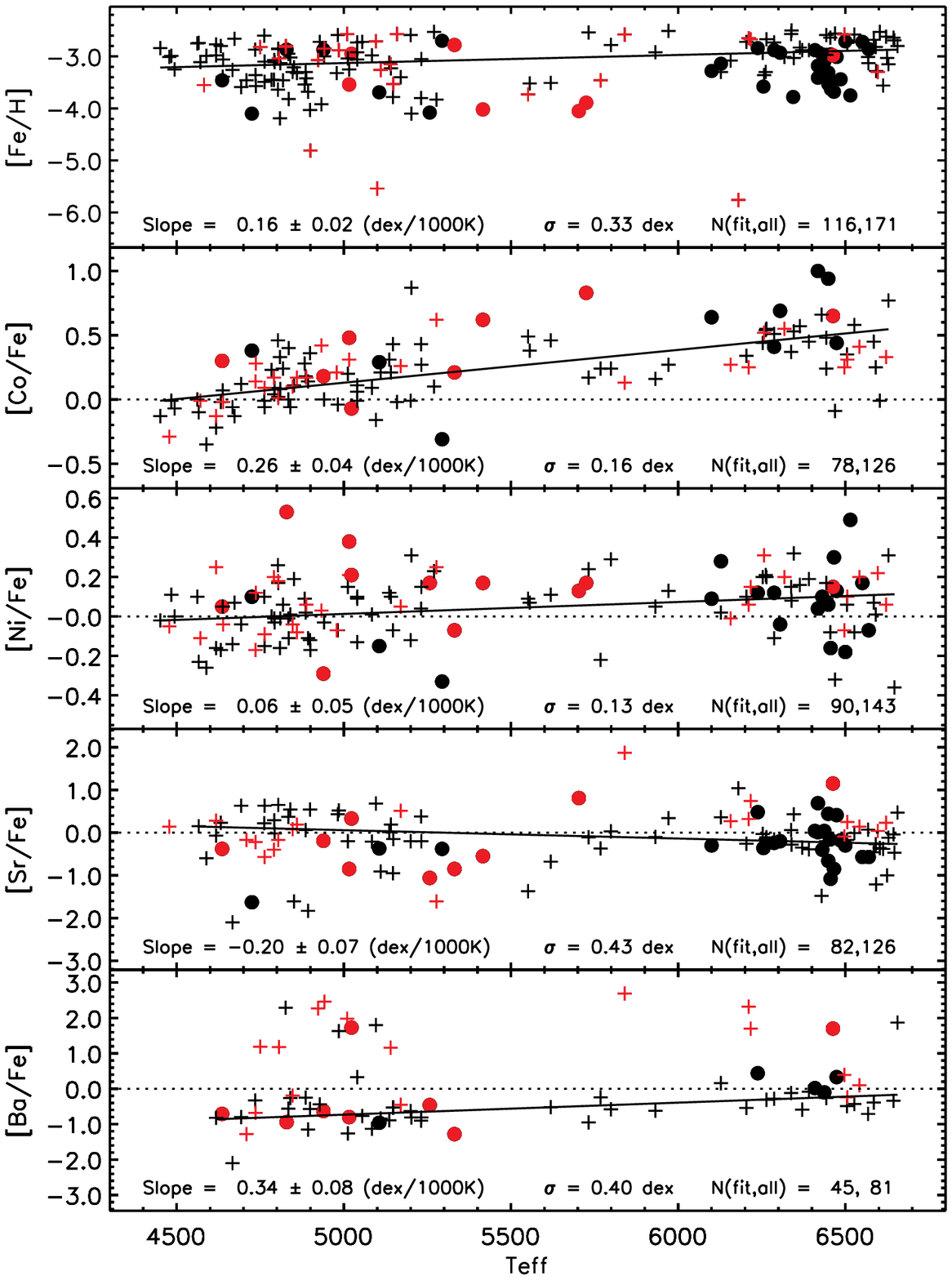} 
\caption{Same as Figure \ref{fig:xfeteffa}, but for Fe, Co, Ni, Sr, and Ba. 
\label{fig:xfeteffc}}
\end{figure}

Another way to view the abundance differences between dwarfs and 
giants is to plot the abundance ratios [X/Fe] versus \teff\ 
(see Figures \ref{fig:xfeteffa} to \ref{fig:xfeteffc}). 
Many elements exhibit clear trends, in particular, Na, Si, \tii, 
Cr, Co, Sr, and Ba are significant at the 3-$\sigma$ level or higher. 
For Na, Si, and Sr, the trend is negative (decreasing [X/Fe] with
increasing \teff), while for the other elements, the trend is positive
(increasing [X/Fe] with increasing \teff).  \citet{lai08} found
similar results for Si, Ti, and Cr, and speculated that 
they could 
be due to blends, deficiencies
in the model atmospheres and/or inadequacies in spectral-line analysis
techniques (e.g., non-LTE, 3D).  They also caution that due care must
be taken when comparing abundances of these elements with chemical
evolution models or supernova yields.  On looking at the data in
\citet{barklem05}, considering only stars with [Fe/H] $\le$ $-$2.5, we
find that many elements (e.g., Ca, Co, Cr, Sc, and Ti) exhibit a very
significant trend (3-$\sigma$ or higher) between [X/Fe] versus \teff.
Therefore, we echo the \citet{lai08} concerns, and note the importance
of restricting the range in \teff\ when possible (e.g., comparing
dwarfs with dwarfs or giants with giants) in order to minimize
systematic uncertainties, as done successfully by \citet{arnone05}.

Finally, we note that radiative levitation and gravitational settling 
(sometimes called atomic diffusion) is suspected to play a role in
altering the photospheric abundances between dwarf and giant stars at
low metallicity (e.g., \citealt{richard02,richard02b}).  At present,
the observational tests have been limited to the moderately metal-poor
globular clusters NGC 6397 \citep{korn07,lind08,lind09b} and NGC 6752
\citep{korn10}. These analyses support the view that 
radiative levitation and gravitational settling can
play an important role in producing abundance differences between
dwarfs and giants. On looking at Figure 11 of \citet{richard02}, the
model with [Fe/H] = $-$3.31 predicts that [Na/Fe] will be $\sim$0.3
dex higher at \teff\ = 4500K compared with \teff\ = 6200K; 
this is in fair agreement with our observations. However, for Si
and Cr, the same models predict abundance differences between warmer
dwarfs and cooler giants that are in the opposite sense to
our findings. Ultimately, understanding the abundance
differences between dwarfs and giants will require a combination of 
improved modelling in terms of 
non-LTE, 3D, and/or radiative levitation and gravitational settling.  

\subsection{Non-LTE Na Abundances}

\begin{figure*}[t!] 
\epsscale{1.0}
\plotone{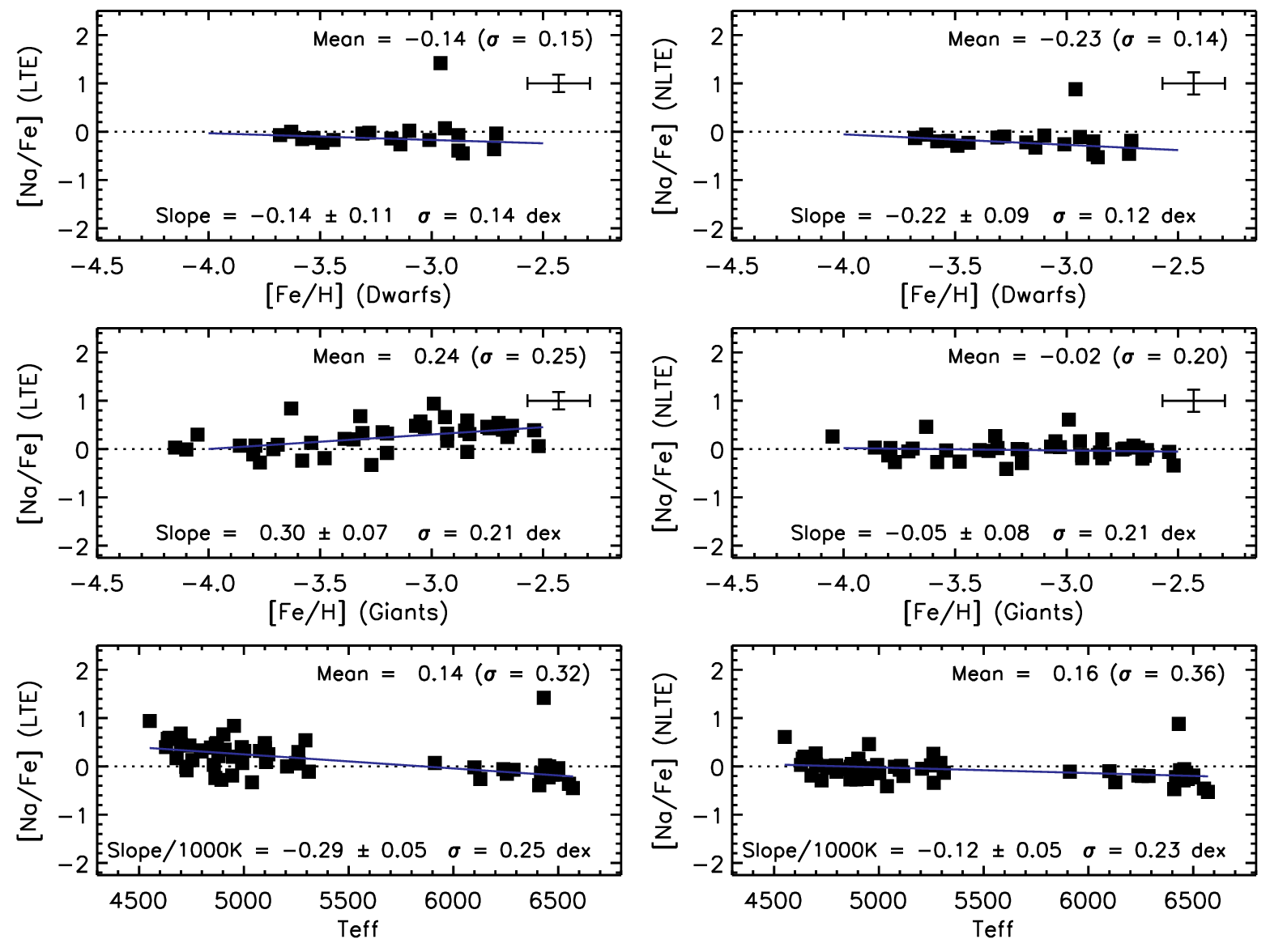} 
\caption{[Na/Fe] vs.\ [Fe/H] (upper and middle panels) and 
[Na/Fe] vs.\ \teff\ (lower panels), excluding CEMP objects. The left and right 
panels correspond to LTE and non-LTE Na abundances, 
respectively, where the non-LTE 
corrections are taken from \citet{lind11}. 
The upper panels present dwarf stars ($\log g$ $>$ 3.0) 
and the middle panels the giants ($\log g$ $<$ 3.0). 
In all panels we show the mean abundance and dispersion, 
the slope and uncertainty of the linear fit to the data 
(excluding 2-$\sigma$ outliers), and the dispersion about the linear fit. 
\label{fig:nlte.na}} 
\end{figure*}

For Na, \citet{lind11} computed non-LTE abundance corrections 
for a number of lines, including the resonance lines used in our analysis. 
Their non-LTE corrections covered a large range in stellar parameters 
(\teff, $\log g$, [Fe/H]), as well as a large range in [Na/Fe]. 
In Figure \ref{fig:nlte.na}, we apply the \citet{lind11} non-LTE
corrections to the C-normal sample presented in Figure \ref{fig:na},
for which Na abundances were all determined from the resonance lines.
To re-iterate, we exclude CEMP objects in this plot.  To obtain the
non-LTE corrections, we used linear interpolation for stars having
stellar parameters (Teff, $\log g$, [Fe/H]) within the grid. For stars
beyond the grid (those with $\log g$ < 1.0 or [Fe/H] $\le$ $-$5.0), we
applied the non-LTE correction at the nearest boundary of the grid; 
thus, the corrections for those stars 
are uncertain. (Had we excluded the stars that lie beyond the grid, 
our results would be unchanged.) 

In Figure \ref{fig:nlte.na}, the left columns show LTE abundances. 
The top panels show dwarf stars and the middle panels show giants. 
In these panels, we again determine the 
linear fit to the data (excluding 2-$\sigma$ outliers), 
and show the slopes and uncertainties, 
dispersions about the linear fit, and the mean abundance and dispersion. 
For the upper left and middle left panels, 
the numbers are the same as in Figure \ref{fig:na}, 
as they should be. Furthermore, we note that Na is an element in which 
dwarfs and giants exhibit significant 
differences in their mean abundance, [Na/Fe], and 
for the slope, [Na/Fe] vs.\ [Fe/H]. 

In the right panels of Figure \ref{fig:nlte.na}, we apply the
\citet{lind11} non-LTE corrections\footnote{We note that we applied
  the corrections for the 5895\AA\ line to our average abundance and
  that the corrections for the 5889\AA\ line are essentially
  identical.}, and determine the linear fit.  (For clarity, we stress
that only the Na abundances have been corrected, not the Fe
abundances.)  These corrections generally result in lower Na
abundances, and therefore lower [Na/Fe] ratios, than in the LTE case.
For the giant sample, the slope changes from 0.30 dex/dex to $-$0.05 dex/dex, while for
the dwarf sample, it changes from $-$0.14 dex/dex to
$-$0.22 dex/dex. Therefore, application of the non-LTE Na abundance
corrections results in an improved agreement for the slope of [Na/Fe]
vs.\ [Fe/H] between dwarfs and giants.  The slopes only differ
at the 1.4-$\sigma$ level, although the mean abundance difference is
0.21 dex.  Similar results are obtained when using the
\citet{andrievsky07} Na non-LTE corrections, although we note that
their corrections
assume an equivalent width corresponding to 
an LTE abundance of [Na/Fe] = 0. 
The \citet{lind11} non-LTE corrections cover a large range 
in [Na/Fe] at a given \teff/$\log g$/[Fe/H], and 
we expect (and find) the magnitude of the non-LTE correction to be 
a function of 
LTE Na abundance. 
For this sample, the non-LTE corrections 
are larger for giants than for dwarfs. 

In the bottom panels of Figure \ref{fig:nlte.na}, we plot the 
[Na/Fe] abundances (LTE and non-LTE) vs.\ \teff. 
Application of the \citet{lind11} 
Na non-LTE corrections results in a significantly 
shallower slope between [Na/Fe] vs.\ \teff. 
This plot serves to highlight the importance of taking into 
account non-LTE effects, when 
possible, 
and as a useful exercise 
in assessing the importance of non-LTE 
corrections to the (i) mean abundance, (ii) slope of [X/Fe] vs.\ [Fe/H], 
and (iii) trends between [X/Fe] vs.\ \teff. 
We await with great interest 
more detailed 
grids of non-LTE abundance corrections for additional elements as well as 
grids of 3D abundance corrections, although we recognize the 
magnitude of such efforts currently underway 
(e.g., \citealt{andrievsky08,andrievsky11,collet11,bergemann08,bergemann10,bergemann11,bergemann12,dobrovolskas12,lind09,lind11,lind12,spite12}). 

\subsection{Mg as the Reference Element}

As discussed by \citet{cayrel04}, one possibility to aid the interpretation 
of the abundances in metal-poor stars is to use Mg as the reference 
element rather than Fe. An advantage of such an approach is that the 
nucleosynthetic production of Mg during hydrostatic burning 
is well understood, 
whereas the synthesis of Fe is more complicated and not unique. 
On the other hand, a disadvantage of using Mg over Fe is that there are 
fewer lines from which the abundance is measured, thus the 
measurements are less accurate. Nevertheless, \citet{cayrel04} 
took this alternate approach and noted that, in the 
regime [Mg/H] $\le$ $-$3.0, there was a suggestion that all abundance 
ratios, [X/Mg], 
were flat. The plateau value of [X/Mg] at lowest [Mg/H] may 
therefore reflect yields from the first generation of supernovae. 

\begin{figure}[t!] 
\epsscale{1.1}
\plotone{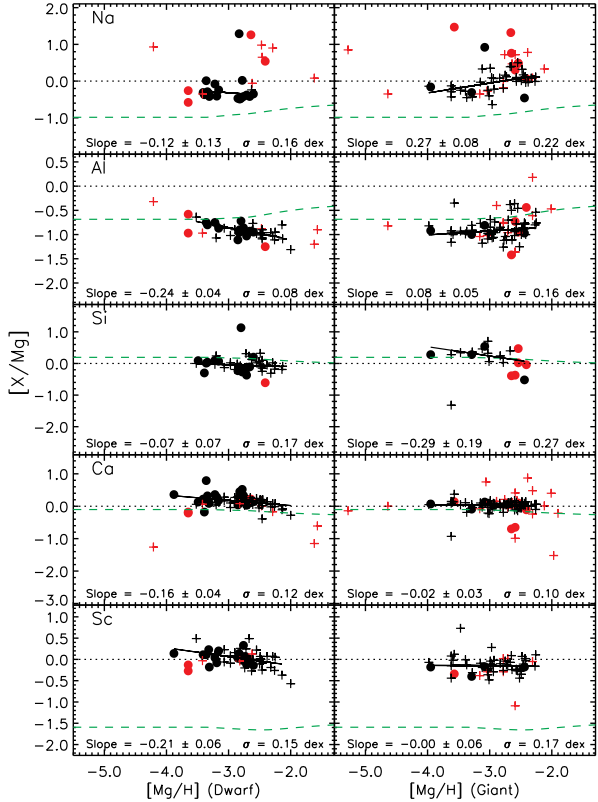} 
\caption{[X/Mg] vs.\ [Mg/H] for the elements Na to Sc.  The left
 panels present dwarf stars ($\log g$ > 3.0) and the
   right panels the giants ($\log g$ < 3.0).  Symbols are the same as
   in Figure \ref{fig:c}. In each panel we show the linear fit to the
   data below [Mg/H] = $-$2.0, excluding CEMP objects and 2-$\sigma$
   outliers.  The slope (and associated error) of this fit are
   presented as well as the dispersion about the best fit. The
 dashed green line represents the predictions from
 \citet{kobayashi06}. 
\label{fig:mgfe1}}
\end{figure}

\begin{figure}[t!] 
\epsscale{1.1}
\plotone{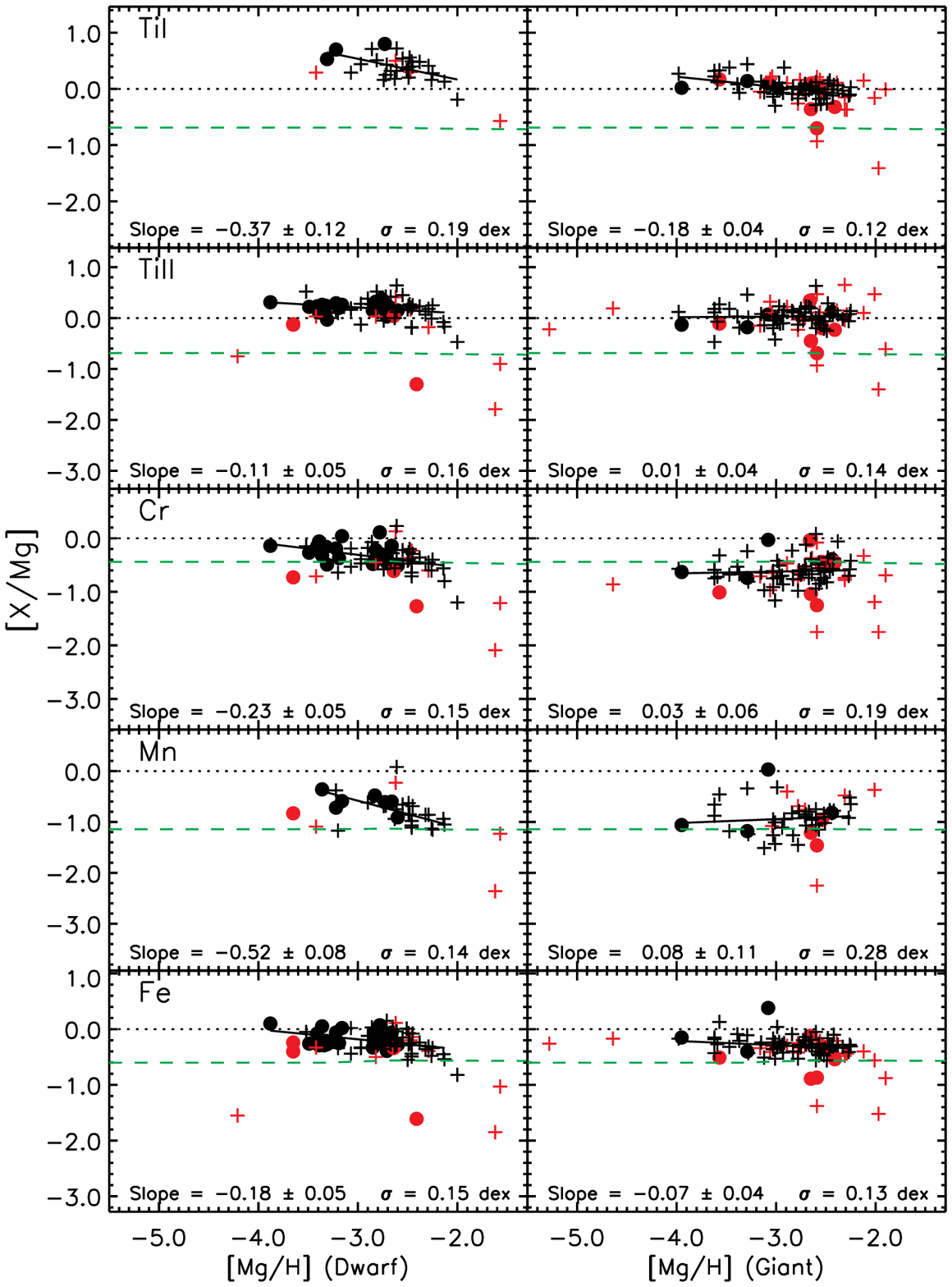} 
\caption{Same as Figure \ref{fig:mgfe1}, but for the elements \tii\ to Fe. 
\label{fig:mgfe2}}
\end{figure}

\begin{figure}[t!] 
\epsscale{1.1}
\plotone{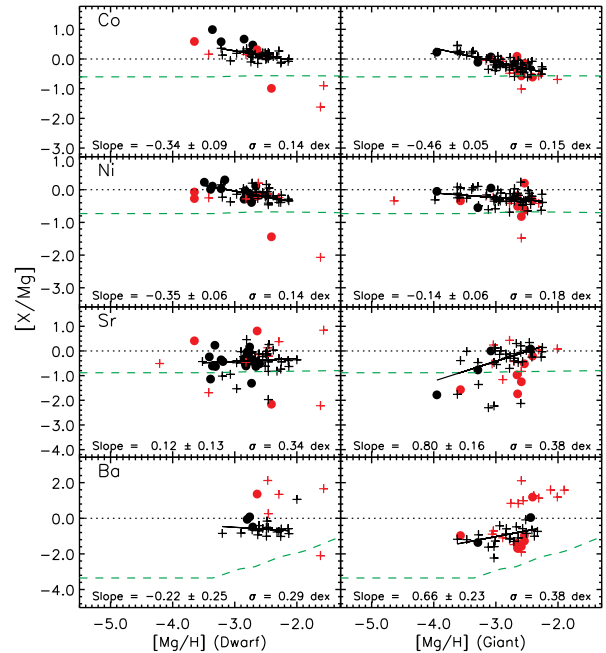} 
\caption{Same as Figure \ref{fig:mgfe1}, but for the elements Co to Ba. 
\label{fig:mgfe3}}
\end{figure}

In Figures \ref{fig:mgfe1} to \ref{fig:mgfe3}, we plot [X/Mg] 
vs.\ [Mg/H] for dwarfs and giants; for comparison with 
the work of \citet{cayrel04}, in the range [Mg/Fe] $\le$ $-$3.0, 
we fit the data excluding CEMP stars and 2-$\sigma$ outliers. 
In these figures we again 
include the predictions from \citet{kobayashi06}. 
We reach similar conclusions 
to those of \citet{cayrel04}. Namely, the dispersion about the 
linear fit is generally greater when plotting [X/Mg] vs.\ [Mg/H] than for 
[X/Fe] vs.\ [Fe/H] (although there are a few cases in which the 
opposite is true, e.g., Al, Si, and \tiii). 
This is presumably due to the Mg measurements 
being more uncertain than Fe due to the smaller number of lines.
(Note that we are only fitting stars with [Mg/H] $\le$ $-$3.0 
rather than the full abundance range.)  
For most elements, the linear fit to the data with [Mg/H] $\le$ $-$3.0 
shows zero slope (at the 2-$\sigma$ level), 
with notable exceptions including 
Al (dwarfs), \tii\ (giants), and Co (dwarfs and giants). 
Considering stars in the range [Mg/H] $\le$ $-$2.0, 
rather than [Mg/H] $\le$ $-$3.0, we note that the linear 
fit to the data is not consistent with zero slope for a larger 
number of elements. In general, 
the behavior of [X/Mg] vs.\ [Mg/H] exhibits a similar behavior at all
metallicities.  It is difficult to assess whether the
\citet{kobayashi06} predictions are a better match to the [X/Mg] or
the [X/Fe] plots. Ultimately, using Mg as the reference element does
not seem to offer any major advantages over Fe, at least in this
analysis.

Furthermore, we note that \citet{chieffi00} have suggested that 
Si, S, Ar, and Ca (or combinations of these elements) 
may be better reference 
elements and tracers of Galactic chemical evolution than 
Fe (or O). \citet{chieffi00} argue that the yields of 
these elements are not strongly dependent 
on the location of the mass cuts employed in the supernova explosion 
calculation, nor with the pre-explosive chemical composition. 
Of 
these elements, Ca has the most number of lines measured 
in our program stars, although the average 
number of lines is smaller than for Mg, and the average total error 
for Ca exceeds that of Mg. 

\subsection{Additional Comparisons with \citet{cayrel04}} 

In the upper panel of Figure \ref{fig:slope_comp}, we compare the
slope from the linear fit of [X/Fe] vs.\ [Fe/H] between our sample
(program + literature stars) and the \citet{cayrel04} study.  That is,
we are comparing the coefficient $a$ from the relation, [X/Fe] = $a$
$\times$ [Fe/H] + $b$, describing the best fit to our data and the
best fit to their data.  Since their sample consists exclusively of
giants, we use the slope as determined from our giant sample in the
comparison.  To ensure that this was a proper comparison of gradients,
we used our software to determine the slope (and uncertainty) to
the \citet{cayrel04} data, rather than relying upon the
values in their Table 9.  As with our sample, we exclude 
the two CEMP stars 
and 2-$\sigma$
outliers when determining the linear fit to the \citet{cayrel04}
data.  Figure \ref{fig:slope_comp} shows that, for Mg, Ca, Sc, Cr, and
Ni, the gradients measured in this study and \citet{cayrel04} are in
very good agreement. For Si, Mn, and Co, the slopes differ by more
than 2-$\sigma$. For Si, the sign of the gradient differs between the
two studies.

In the lower panel of this figure, 
we also compare the scatter about the linear fit between our giant 
sample and that of \citet{cayrel04}. 
For their data, we again used our software 
to determine the scatter about the linear fit after  
eliminating 2-$\sigma$ outliers and the two CEMP stars. 
(We used the relation given in \citet{taylor97} for the fractional uncertainty 
in the dispersion, 1/$\sqrt{2(N-1)}$.) 
Some elements (Mg, Ca, Mn, Co, and Ni) exhibit very good agreement. The
elements that differ by more than 2-$\sigma$ are Si, Sc, and Cr. It is
not surprising that, for all elements, the dispersions about the linear
fit to the \citet{cayrel04} data are always equal to or smaller than
for our sample of program and literature stars. This could be due, in
part, to the fact that the \citet{cayrel04} sample is very
homogeneous, and their spectra are of very high quality. While our
sample includes their stars, we have a more heterogeneous sample, 
albeit one that was 
analyzed in a homogeneous manner. 

In the lower panel of Figure \ref{fig:slope_comp}, 
we also plot 
a representative measurement uncertainty for each element, 
the average ``total error'' 
for the program stars. 
For the \citet{cayrel04} data, we also plot 
their estimate of the expected scatter for each element, based on their 
measurement uncertainties. 
These are shown as upward facing arrows, and 
we would expect the observed dispersions to 
lie on or above these values (for species having an 
astrophysically significant dispersion).
For the \citet{cayrel04} data,
the observed dispersions are in good agreement with the
expected scatter. For our data, some elements (Mg, Mn, Co) exhibit an
observed dispersion in good agreement with the expected
scatter. However, for Si, the observed dispersion exceeds the expected
value; for Sc, the observed dispersion is considerably smaller
than the expected value, which may indicate that our measurement
uncertainties are overestimated for this element.
Nevertheless, it is reassuring that in several cases (Mg, Ca, Co, and
Ni) our dispersions about the linear fit are comparable with 
the values of \citet{cayrel04}, suggesting that if indeed a ``normal''
population exists, our selection criteria were effective in
identifying this population, even though our sample of program and
literature stars is quite heterogeneous.

\subsection{Comparison with Predictions, and the 
Curious Case of Scandium and Titanium} 

In Figure \ref{fig:ab2} and Figures \ref{fig:mgfe1} to
\ref{fig:mgfe3}, we overplot the \citet{kobayashi06} predictions
    of Galactic chemical enrichment. Their chemical
    evolution model includes the following assumptions: (1) one-zone
    centered on the solar neighborhood, (2) no instantaneous recycling
    approximation, (3) contributions from hypernovae with large
    explosion energy ($E_{51}$ $>$ 10), Type II supernovae, and Type
    Ia supernovae, (4) no contributions from low- and
    intermediate-mass stars, and (5) infall of primordial gas.  Figure
    \ref{fig:ab2} shows the evolution of [X/Fe] against [Fe/H].  In
    general, the predictions provide a fair fit to the data in terms
    of the slope (or lack thereof). For many elements, the mean 
predictions differ from the mean observations by $\sim$0.5 dex or more. 

\begin{figure*}[t!] 
\epsscale{1.0}
\plotone{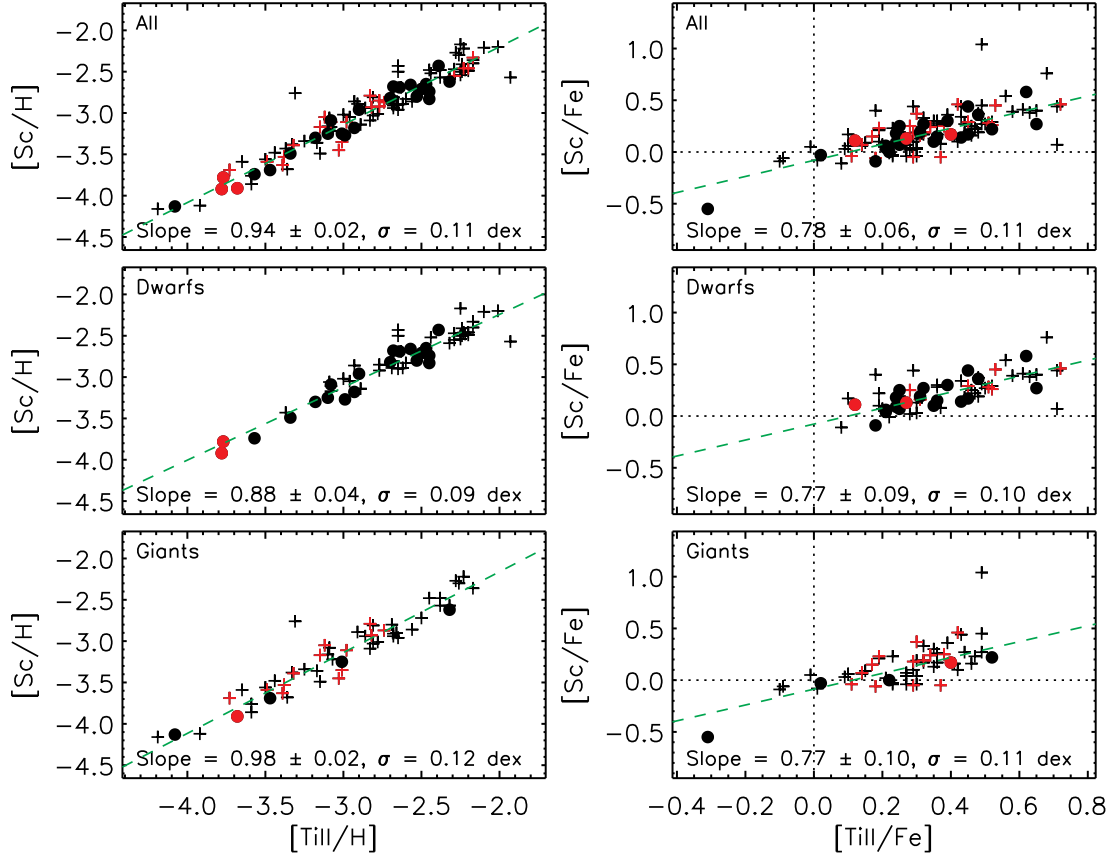} 
\caption{[Sc/H] vs.\ [\tiii/H] (left panels) and 
[Sc/Fe] vs.\ [\tiii/Fe] (right panels) for all stars (upper panels), 
dwarf stars (middle panels), and giant stars (bottom panels). 
The symbols are the same as in 
Figure \ref{fig:c}, and in each panel we present the linear fit 
to the data, excluding CEMP objects and 2-$\sigma$ outliers. The 
slope, uncertainty, and dispersion about the slope are shown. 
\label{fig:ti2sc}}
\end{figure*}

As reported in other investigations (e.g., \citealt{kobayashi06}), 
the predictions for Ti and Sc are underabundant relative to
the LTE measurements.  
In our study, Sc measurements are exclusively from Sc\,{\sc ii} lines. 
We find that the abundance ratios [\tiii/Fe]
and [Sc/Fe] (and [\tiii/H] vs.\ [Sc/H]) 
exhibit a highly significant, $\sim$10-$\sigma$,
correlation (see Figure \ref{fig:ti2sc}). In this figure, we show
all stars (upper panel), dwarfs (middle panel), and giants (lower
panel). For each sample, the scatter about the mean trend is only 
$\sim$0.10 dex, which is substantially lower than the average total
error for either [\tiii/Fe] (0.14 dex) or [Sc/Fe] (0.20 dex). We also
find a correlation between [\tii/Fe] and [Sc/Fe], but with a
lower significance ($\sim$ 2-$\sigma$) and shallower slope (0.19 dex/dex 
to 0.38 dex/dex).  Although Sc and Ti are produced via different processes,
the correlation we find suggests that the two elements might be
produced in similar conditions.  \citet{umeda05} suggest that the
yield of Sc in metal-poor supernovae can be greatly increased in 
low-density (i.e., high-entropy) regions. \citet{kobayashi11b} suggest
that the $\nu$-process in core-collapse supernovae may
produce Sc, although to our knowledge there have been few, if any,
studies of the yields of Sc from the $\nu$-process. The strong
correlation between Sc and Ti found here might suggest that
the $\nu$-process does not provide a complete explanation of 
the production of Sc at lowest metallicities.

Figures \ref{fig:mgfe1} to \ref{fig:mgfe3} show the evolution of
[X/Mg] against [Mg/H]. Again, the predictions provide a fair fit to
the slope and the mean abundance. In this case, the elements which are
poorly fit include Na, Sc, 
and Ti. 

Presumably, the \citet{kobayashi06} model, 
or other chemical evolution models, could be fine-tuned to provide a better 
fit to this set of observations. 
We note in particular that the inclusion of yields from intermediate-mass stars 
cannot account for the underproduction of Na at low metallicity \citep{kobayashi11c}. 
As discussed earlier, we caution that 
non-LTE and 3D effects should be taken into consideration. 
Chemo-dynamical 
models of Galactic formation and evolution 
(e.g., \citealt{kobayashi11}), and/or inhomogeneous chemical enrichment 
models (e.g., \citealt{karlsson05}) will enable more 
comprehensive comparisons with the available observations, including the 
predicted dispersion in abundance ratios 
as a function of stellar population characteristics such as 
metallicity, kinematics, age, and spatial distribution. 

\subsection{Chemically Unusual Stars}

\begin{figure}[t!] 
\epsscale{1.2}
\plotone{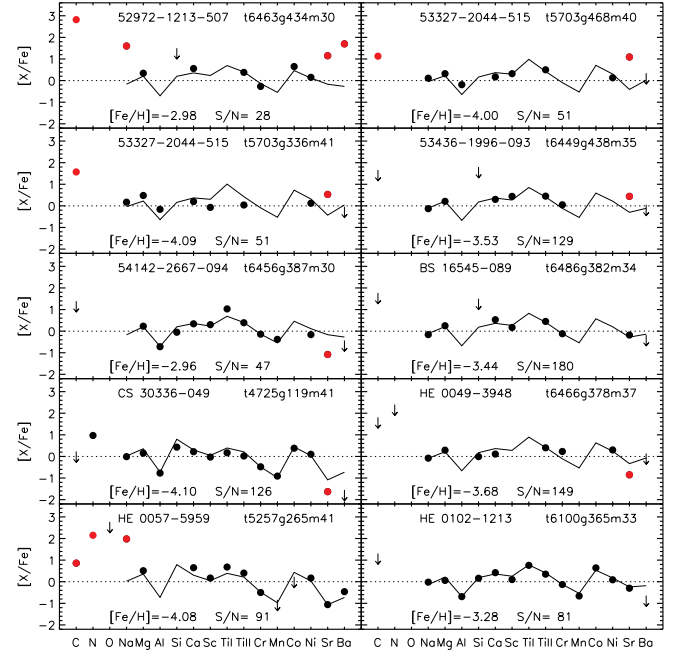} 
\caption{Abundance patterns [X/Fe] for each element in our program stars. 
(Arrows indicate abundance limits.) 
In each panel, the solid line represents the ``normal'' [X/Fe] abundance 
ratio for a giant (or dwarf) at the metallicity of each program star 
(see text for details). The S/N and model parameters are also shown. 
Red points are for [C,N,O/Fe] $\ge$ +1.0, or when 
[X/Fe] differs from the solid line by more than 0.5 dex. 
\label{fig:ind1}}
\end{figure}

\begin{figure}[t!] 
\epsscale{1.2}
\plotone{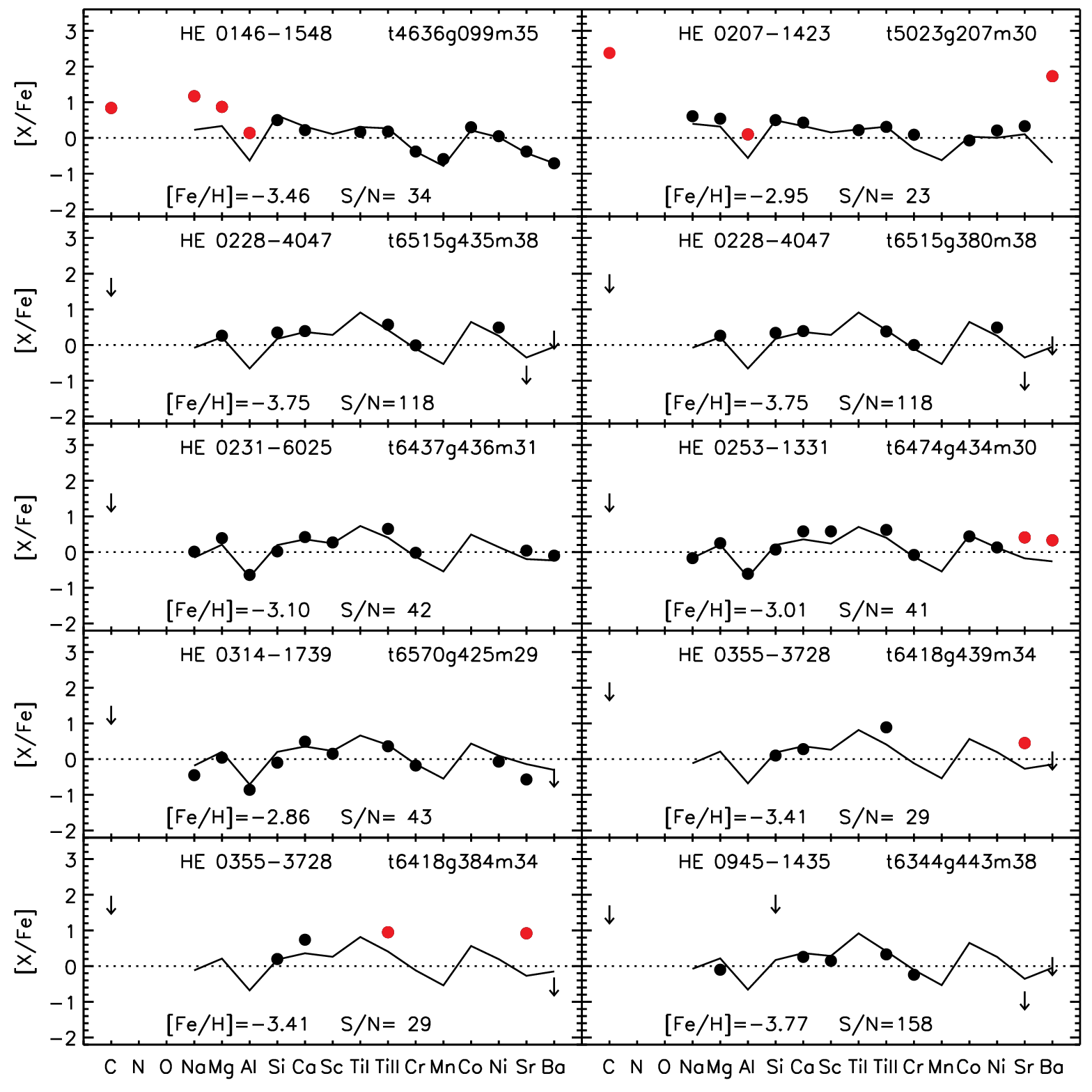} 
\caption{Same as Figure \ref{fig:ind1}, but for the next 10 stars. 
\label{fig:ind2}}
\end{figure}

\begin{figure}[t!] 
\epsscale{1.2}
\plotone{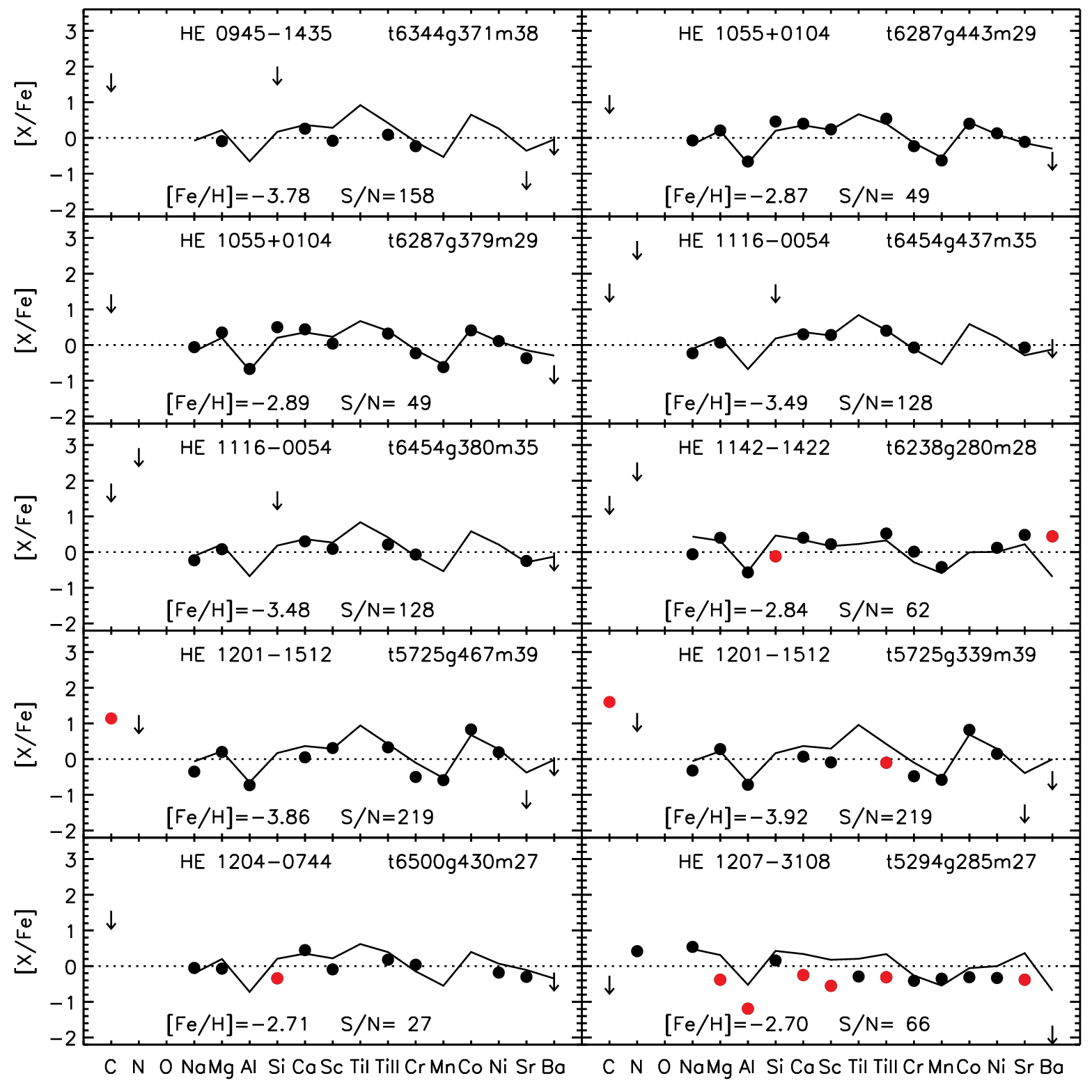} 
\caption{Same as Figure \ref{fig:ind1}, but for the next 10 stars. 
\label{fig:ind3}}
\end{figure}

\begin{figure}[t!] 
\epsscale{1.2}
\plotone{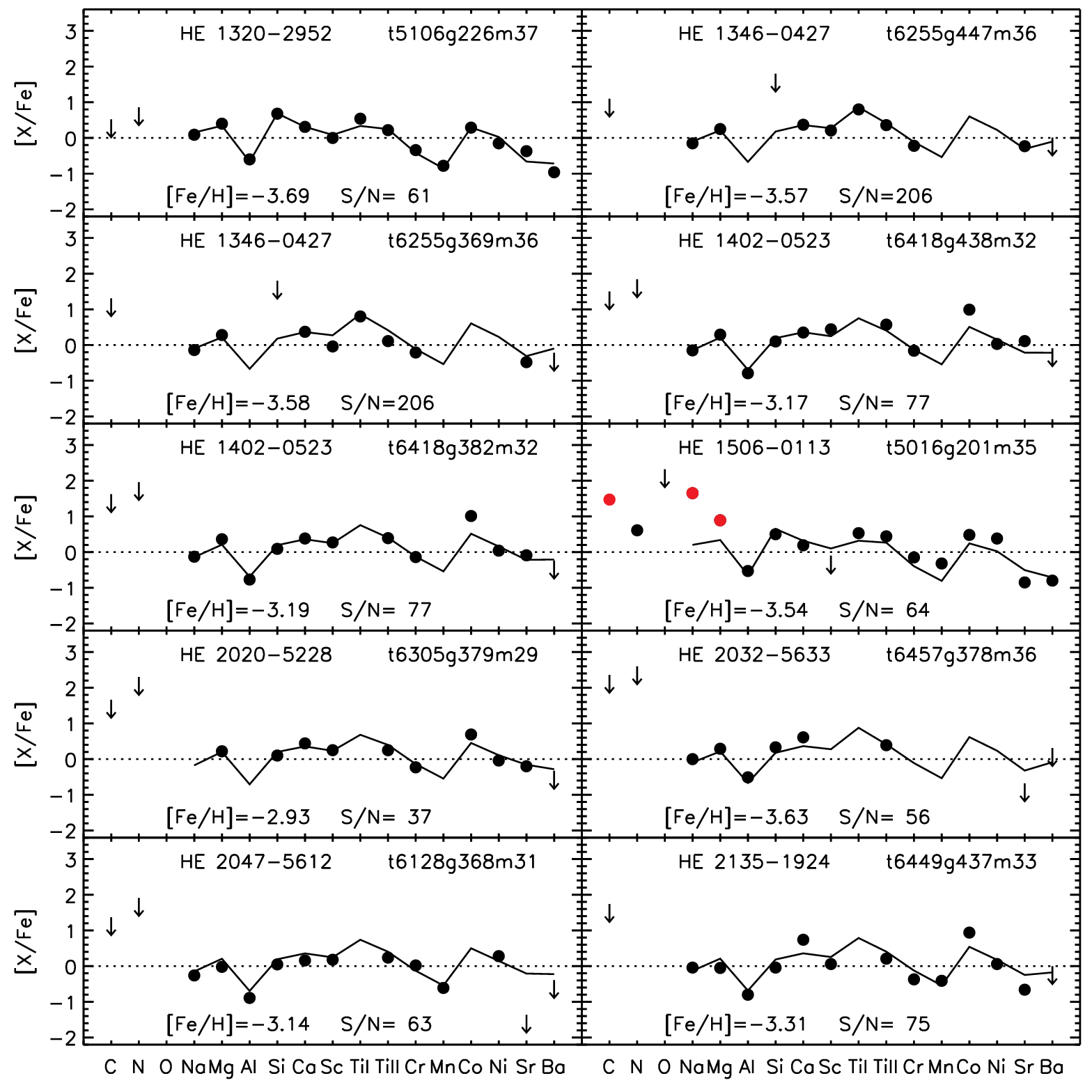} 
\caption{Same as Figure \ref{fig:ind1}, but for the next 10 stars. 
\label{fig:ind4}}
\end{figure}

\begin{figure}[t!] 
\epsscale{1.2}
\plotone{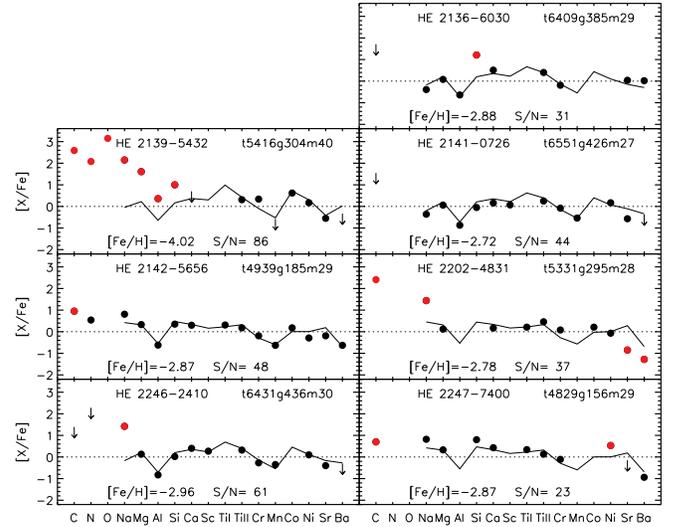} 
\caption{Same as Figure \ref{fig:ind1}, but for the final seven stars. 
\label{fig:ind5}}
\end{figure}

In Figures \ref{fig:ind1} to \ref{fig:ind5}, we plot 
for each program star the abundance pattern [X/Fe] vs.\ element.  (In
these figures, we include, as double entries, results of
both the dwarf and subgiant analyses of the nine stars for which there
was disagreement between spectrophotometric and Balmer-line analyses
of the evolutionary status.)
The solid line in each panel represents the abundance ratio [X/Fe]
that a ``normal'' dwarf ($\log g$ > 3.0) or giant ($\log g$ < 3.0)
star would have at the metallicity of the program star. The ``normal''
star abundance was taken from the linear trends described above, and
plotted in Figures \ref{fig:na} to \ref{fig:ba}.  For elements that
deviate from the solid line by more than 0.50 dex, we regard these
abundances to be peculiar, and mark them in red.  (Note that some
peculiar abundances lie above the solid line, while
others lie below.)  For CEMP objects, we mark their C abundance in red. For N and
O, we regard ratios [X/Fe] $\ge$ +1.0 to be unusual, 
and also mark them in red. 
Inspection of Figures \ref{fig:ind1} to \ref{fig:ind5} then readily
highlights whether a given star has elements that may be regarded as
peculiar.  Any star with several such elements can be considered as a
chemically peculiar star. 

\begin{figure}[t!] 
\epsscale{1.2}
\plotone{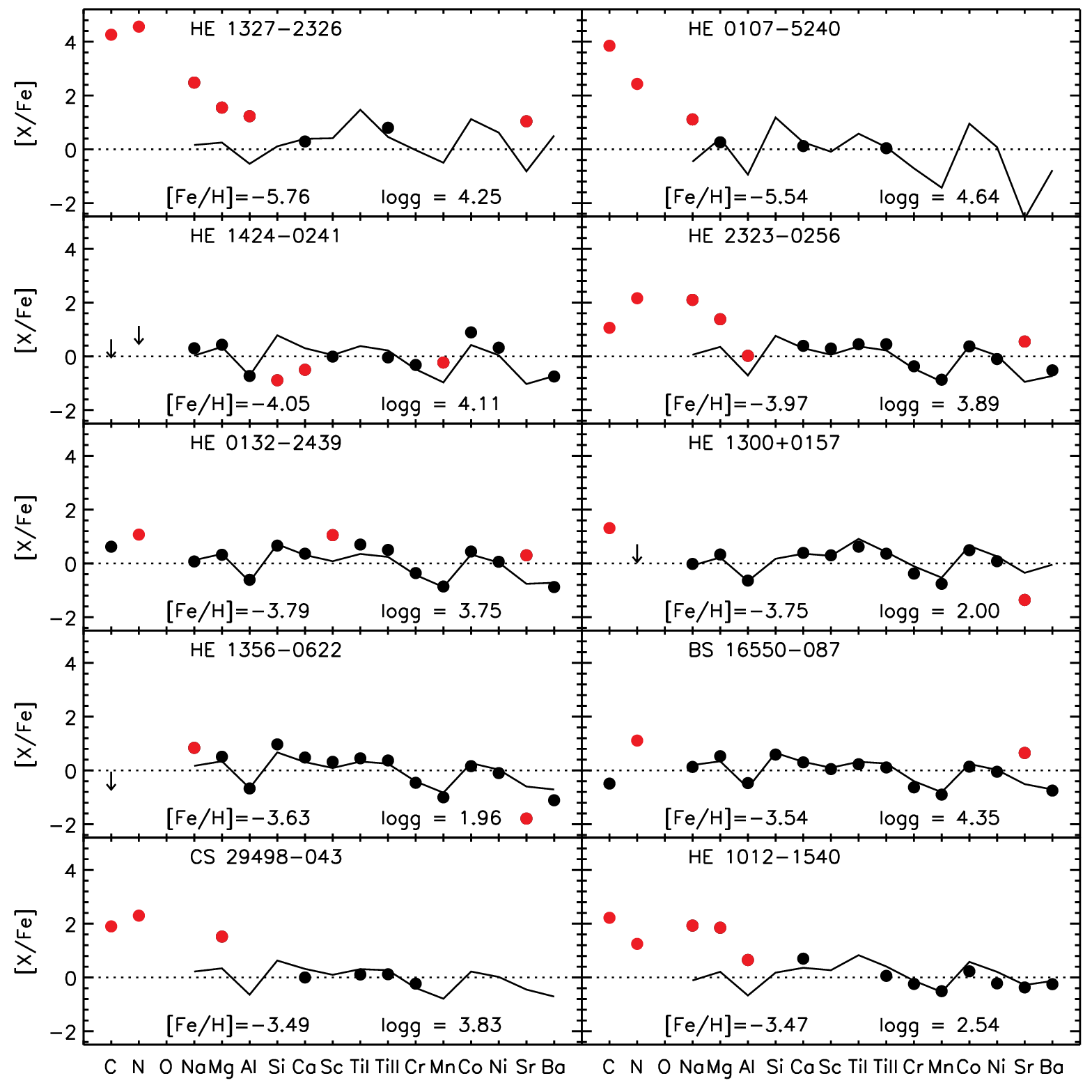} 
\caption{Same as Figure \ref{fig:ind1}, but for the literature sample. 
The stars are ordered by increasing metallicity, and only stars with 
[Fe/H] $\le$ $-$2.55 and with two 
or more unusual elements are plotted. For the two most Fe-poor stars, 
the ``normal'' [X/Fe] abundance ratios are for [Fe/H] = $-$4.2 
(see text for details). \label{fig:lit1}}
\end{figure}

\begin{figure}[t!] 
\epsscale{1.2}
\plotone{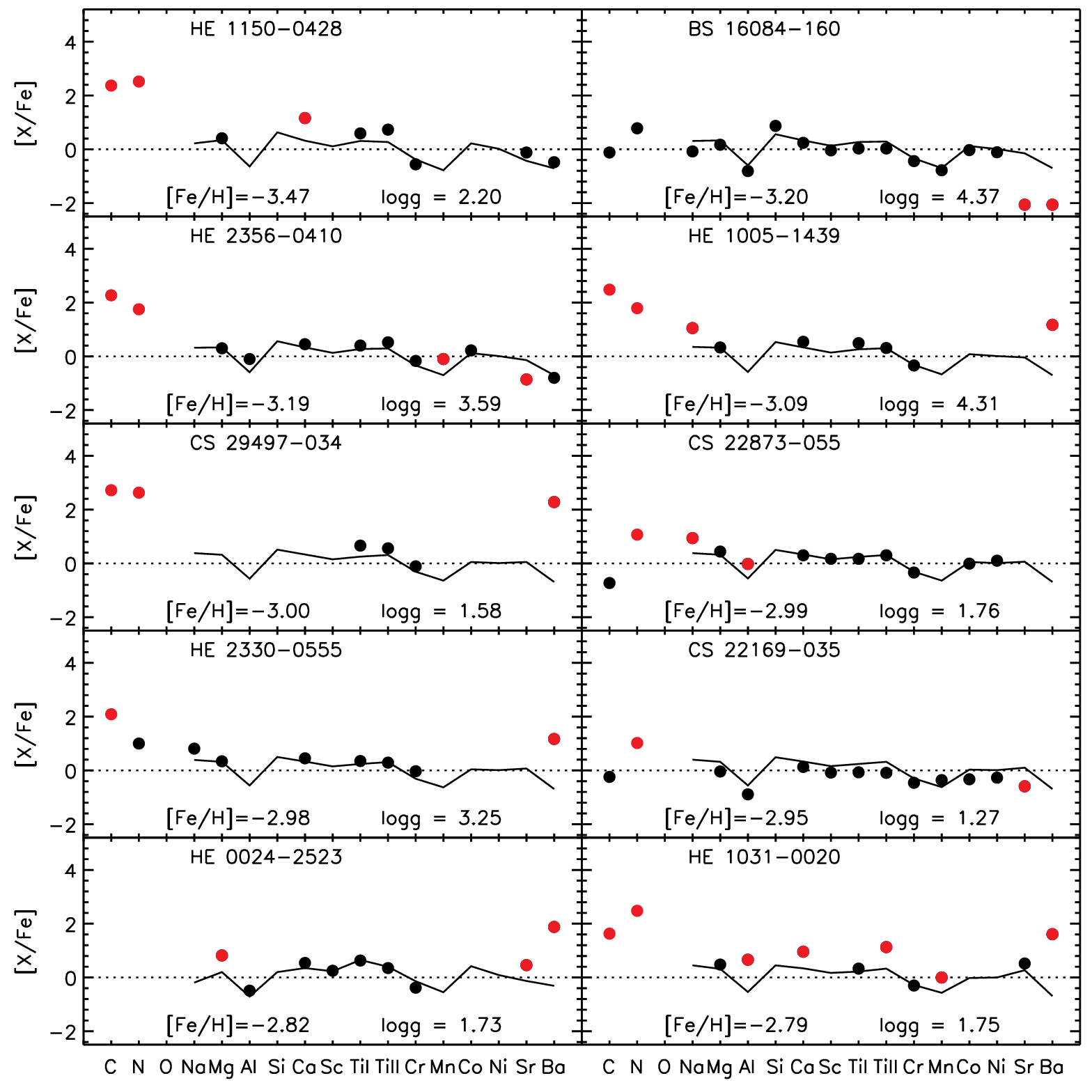} 
\caption{Same as Figure \ref{fig:ind1}, but for the next 10 stars. 
\label{fig:lit2}}
\end{figure}

\begin{figure}[t!] 
\epsscale{1.2}
\plotone{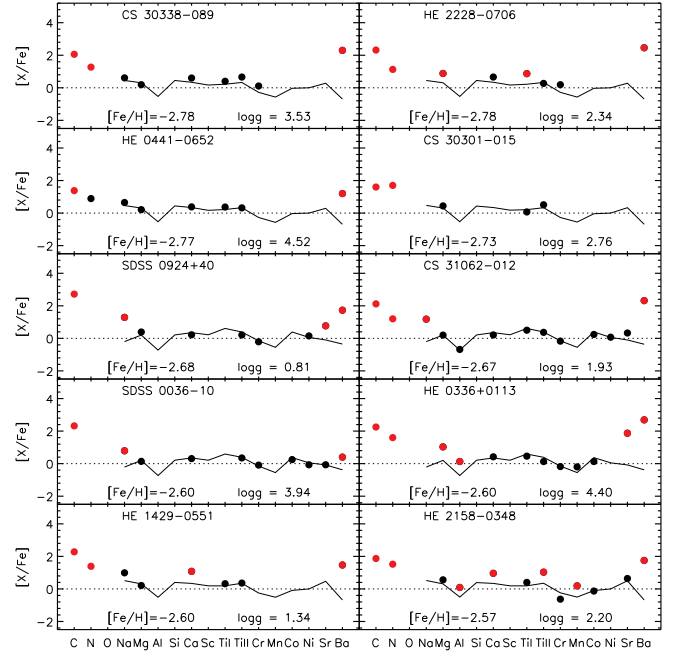} 
\caption{Same as Figure \ref{fig:ind1}, but for the final 10 stars. 
\label{fig:lit3}}
\end{figure}

Similarly, we examined the 152 star literature sample in order 
to identify the chemically peculiar objects. 
In Figures \ref{fig:lit1} to \ref{fig:lit3}, we plot the abundance pattern 
[X/Fe] vs.\ element for the subset of program stars with 
[Fe/H] $\le$ $-$2.55 in which there are 
at least two elements that are unusual (i.e., elements that are at least 
0.5 dex above, or below, the [X/Fe] ratio of a ``normal'' dwarf or 
giant at the same [Fe/H]). For the two most Fe-poor stars in these figures, 
the solid lines showing the [X/Fe] ratios of a normal star 
are for [Fe/H] = $-$4.2. 

In Section 7.2, we noted that in the [X/Fe] vs.\ [Fe/H] plane, 
abundance outliers were often, but not always, CEMP objects. 
We now comment briefly on the C-normal population (i.e., 
only those stars with [C/Fe] measurements, or limits, that enable us 
to confirm that they are not CEMP stars). 
We find that of the 109 C-normal stars, some 32 are chemically 
peculiar (i.e., these 32 stars have at least one element, 
from Na to Ba, for which 
the [X/Fe] ratio is at least 0.5 dex above, or below, that of a
normal star at the same metallicity). These 32 objects have a 
mean metallicity of [Fe/H] = $-$3.02. 
If we exclude the neutron-capture 
elements Sr and Ba (these elements exhibit a very large dispersion such 
that many stars will have [X/Fe] ratios at least 0.5 dex from the 
``normal'' star abundance), 
there are 23 C-normal stars that are chemically peculiar, 
and these stars have a mean metallicity of [Fe/H] = $-$2.91. 

We note that for the majority of these objects, only one element in a 
given star may be regarded as unusual. If we consider only those C-normal stars that 
are chemically peculiar for two or more elements, there are seven 
such objects with a mean metallicity of [Fe/H] = $-$3.31. 
When excluding Sr and Ba, there are only four C-normal stars, 
with an average metallicity of [Fe/H] = $-$3.14, that 
are chemically peculiar for two or more elements -- 
HE~1207$-$3108 (This Study), HE~0024$-$2523 \citep{carretta02,cohen02}, 
HE~1424$-$0241 \citep{cohen08}, and CS~22873$-$055 \citep{cayrel04}. 
The main point of this analysis is to note that indeed there are 
C-normal objects that are chemically peculiar, although the 
fraction is small, four of 109 objects (4\% $\pm$ 2\%).
We now discuss some interesting examples of chemically 
unusual stars.

\subsubsection{Stars with enhanced C, N, O, Na, Mg, and/or Al}

HE~0057$-$5959, HE~1506$-$0113, and HE~2139$-$5432 are extremely metal-poor 
stars, [Fe/H] $\le$ $-$3.5, with large enhancements of C, N, O, Na, Mg, 
and/or Al. All are CEMP-no objects, i.e., they are a subclass of CEMP 
stars that have ``no strong overabundances of 
neutron-capture elements'', [Ba/Fe] < 0 \citep{beers05}. 
HE~1506$-$0113 and HE~2139$-$5432 bear a striking resemblance to 
HE~1327$-$2326, the most Fe-poor star known \citep{frebel05,aoki06}, 
as well as to 
the CEMP-no stars CS~22949$-$037 and CS~29498$-$043 (e.g., \citealt{aoki04}).
HE~0057$-$5959 appears to be an extremely rare 
nitrogen-enhanced metal-poor star (NEMP; \citealt{johnson07}). 
\citet{johnson07} identified only four stars in the recent 
literature that could potentially be classified as NEMP objects. 
HE~0146$-$1548 also exhibits enhancements of C, Na, Mg, and Al. 
In Paper IV of this series (Norris et al.\ 2012b, in
preparation), we shall explore the nature of the CEMP-no objects in more detail.

\subsubsection{A star with enhanced Si}

\begin{figure}[t!] 
\epsscale{1.2}
\plotone{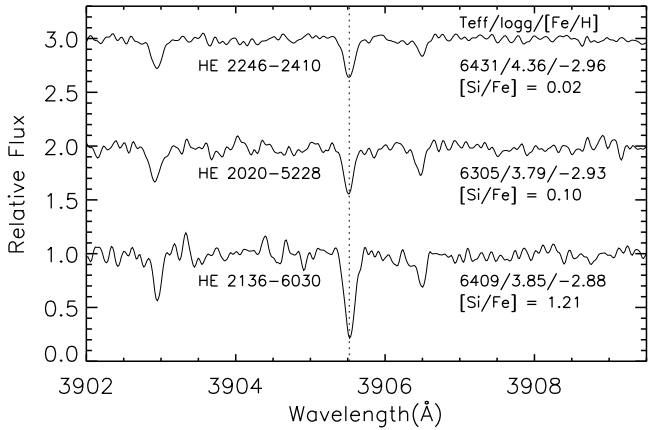} 
\caption{Spectra in the region of the 3905\AA\ Si line for 
two Si-normal stars (HE~2246$-$2410 and HE~2020$-$5228), and 
for the Si-enhanced star (HE~2136$-$6060). The \teff, \logg, 
[Fe/H], and [Si/Fe] are displayed. 
\label{fig:sispec}}
\end{figure}

HE~2136$-$6030 is a C-normal star with 
an unusually high Si abundance, [Si/Fe] = +1.20. 
However, for all other elements measured, the abundance ratios are 
in good agreement with a ``normal'' star at the same metallicity. 
This object has \teff\ = 6409K and [Fe/H] = $-$2.88. 
Figure \ref{fig:si} shows that for dwarf stars, the [Si/Fe] ratio 
is almost constant as [Fe/H] evolves from $-$4.0 to $-$2.5. 
While Figure \ref{fig:xfeteffa} shows a strong trend between 
[Si/Fe] and \teff, stars with \teff\ similar to that of HE~2136$-$6030 all 
have [Si/Fe] $\lesssim$ +0.5 dex. 
In Figure \ref{fig:sispec}, we plot the spectra of HE~2136$-$6030,  
along with two Si-normal stars with similar stellar parameters. 
The Si line is substantially stronger in HE~2136$-$6030, relative 
to the two comparison stars. Examination of our spectra 
shows that CH blending of the 3905.52\AA\ Si line is unlikely, and that 
the subordinate 4102.94\AA\ Si line is likely present. 
Spectrum synthesis 
of the 4102.94\AA\ Si line returns a [Si/Fe] ratio in good agreement 
with the abundance from the 3905.52\AA\ line. We 
are thus confident that the Si abundance is particularly high in this star. 

At the other extreme, \citet{cohen07} found a star with an unusually low Si 
abundance, [Si/Fe] = $-$1.01. This star, HE~1424$-$0241, also has 
low abundances of [Ca/Fe] and [Ti/Fe] but a normal [Mg/Fe] ratio. 
These two stars, HE~2136$-$6030 and HE~1424$-$0241, reveal that 
the [Si/Fe] ratio in metal-poor stars can vary by a factor of 100. 

\subsubsection{''Fe-enhanced'' metal-poor stars}

\begin{figure}[t!] 
\epsscale{1.2}
\plotone{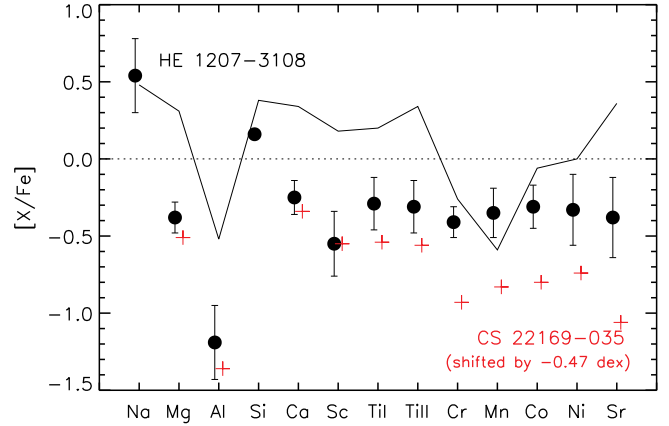} 
\caption{
[X/Fe] for the two ``Fe-enhanced'' metal-poor stars 
HE~1207$-$3108 (black circles) and 
CS~22169$-$035 (red plus signs). 
By normalizing the abundances to Sc, we shift the abundance 
ratios for CS~22169$-$035 by $-$0.43 dex. The solid line represents 
the ``normal'' [X/Fe] ratio for a giant at the metallicity of 
HE~1207$-$3108. 
\label{fig:fepoor}}
\end{figure}

HE~1207$-$3108 (\teff/\logg/[Fe/H] = 5294/2.85/$-$2.70) 
is notable for having unusually low abundance ratios
[X/Fe] for Mg, Al, Ca, Sc, Ti, and Sr. (With the exception of Ti, and
possibly Sr, none of these elements exhibit significant trends with
\teff, thus it is highly unlikely that the peculiar abundance
pattern can be attributed to any \teff-dependent trends.)  
The effect is shown in 
Figure \ref{fig:fepoor}, where the filled circles represent [X/Fe] for HE~1207$-$3108
as a function of atomic species.  Indeed, one sees here that 
for all elements other than Na, the abundance ratios [X/Fe] lie below that
of a ``normal'' star at the same metallicity. 
(We note here that the two Na lines yield abundances that differ by 0.48 dex. 
Without knowing which of the lines to reject, we retain both 
lines and the large error reflects the discordant measurements.) 
\citet{cayrel04}
identified another star, CS~22169$-$035, as being deficient
in Mg, Si, Ca, Ti, Co, Ni, and Zn with respect to Fe.  They suggested
that ``the abundance anomalies are most simply characterized as an
enhancement of Fe''; the same description may be applied to
HE~1207$-$3108. Inspection of Figure \ref{fig:ind3} suggests that if
the Fe abundance were lowered, by say 0.6 dex, HE~1207$-$3108 would have
normal abundance ratios [X/Fe] for Mg and all heavier
elements, but [Na/Fe] might be regarded as being unusually high.  

In Figure \ref{fig:fepoor}, 
we also present the data for CS~22169$-$035. Here, we 
we arbitrarily select Sc as the
element to which we normalize the abundances; this requires a shift
of $-$0.47 dex for CS~22169$-$035. The solid line shows the abundance
ratio that a ``normal'' giant would have at the metallicity of
HE~1207$-$3108.  (An equivalent line for CS~22169$-$035 would be
essentially identical, given that both stars are giants with comparable
metallicity, [Fe/H] = $-$2.70 and $-$2.95 for HE~1207$-$3108 and
CS~22169$-$035, respectively.)  HE~1207$-$3108 and CS~22169$-$035 appear to
have similar and unusual abundance patterns, i.e., they are chemically
distinct from the bulk of the halo stars at the same metallicity.
We note that their chemical abundance patterns are not identical, namely, there are 
differences for Cr, Co, Ni, and Sr. 
That said, their abundance ratios for Mg, Al, Ca, Ti, and Mn are
indistinguishable and unusually low.  We suggest that
  HE~1207$-$3108 and CS~22169$-$035 may belong to a relatively new and
  growing class of ``Fe-enhanced metal-poor'' stars. 

A small number of field halo stars, both dwarfs and giants, are known to
have low [$\alpha$/Fe] ratios relative to field stars at the same
metallicity (e.g.,
\citealt{carney97,nissen97b,fulbright02,stephens02,ivans03,nissen10}).  These
stars cover the metallicity range $-$2 $\lesssim$ [Fe/H] $\lesssim$ $-$0.8.  One
suggestion is that these stars likely formed from regions in which the
interstellar gas was unusually enriched in the products of Type Ia
supernovae, relative to Type II supernovae, and as such they may be
regarded as having an unusually high Fe-content relative to their
$\alpha$-content. One explanation is that these stars may have been
accreted from dwarf galaxies, many of which are known to contain stars
that are chemically distinct from the majority of field halo stars
(e.g., \citealt{tolstoy09}).  While it may be tempting to assign the
low [$\alpha$/Fe] stars and HE~1207$-$3108 and CS~22169$-$035 into a
single group, there are differences in [Fe/H] and other
  [X/Fe] ratios.  Nevertheless, this intriguing possibility exists.

\citet{venn12} have identified a giant star in the
Carina dwarf galaxy with a similar chemical pattern to that of
HE~1207$-$3108 and CS~22169$-$035.  Their star, Car-612, with [Fe/H] =
$-$1.3, is considerably more metal-rich than HE~1207$-$3108 and
CS~22169$-$035, with [Fe/H] = $-$2.70 and $-$2.95, respectively.
  \citet{venn12} suggest that this star formed from gas with
  an unusually high ratio of Type Ia supernovae to Type II supernovae
  products. They estimate the magnitude of the ``excess of Fe'' to be
  0.7 dex, a value comparable to that of HE~1207$-$3108 and
  CS~22169$-$035.  Had we overplotted their data for this star in Figure
  \ref{fig:fepoor}, again normalizing to the Sc abundances, there
  would have been striking similarities in the some abundance ratios
  (Mg, Ti, and Ni), as well as notable differences for other elements
  (Ca, Cr, and Mn). As before, we do not overplot these data, since we
  do not wish to conduct inhomogeneous comparisons.  Nevertheless, it
  is tempting to associate the three stars (HE~1207$-$3108,
  CS~22169$-$035, and Car-612) as belonging to the same class of
  object. We tentatively speculate that any differences in [X/Fe]
  ratios between Car-612, HE~1207$-$3108, and CS~22169$-$035 may reflect
  differences in the currently unknown formation processes for such
  stars.

\subsubsection{A CEMP-s star with enhanced [Ba/Sr]}

\begin{figure}[t!] 
\epsscale{1.2}
\plotone{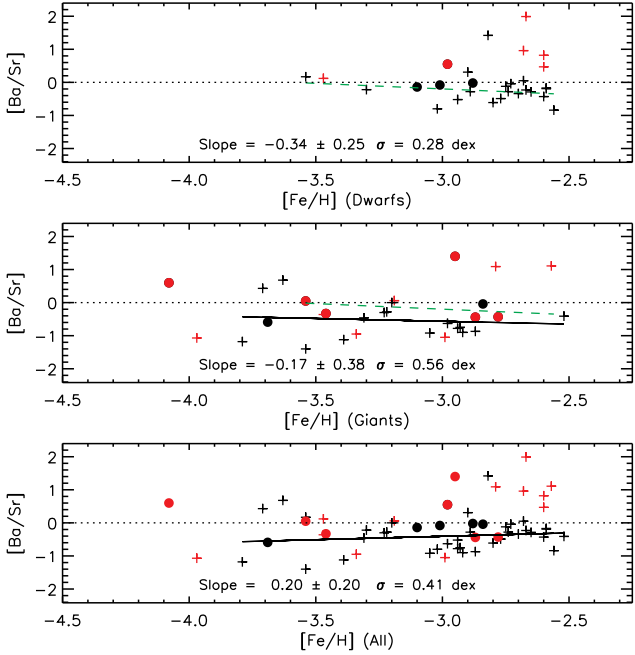} 
\caption{[Ba/Sr] vs.\ [Fe/H] for dwarfs (upper), giants (middle), 
and all stars (lower). 
The symbols are the same as in Figure \ref{fig:c}. 
In each panel we present 
the linear fit to the data 
(excluding CEMP stars and 2-$\sigma$ outliers). The linear 
fit to the dwarf data is superimposed upon the giant data. 
The slopes and uncertainties 
are shown, along with the dispersion about the slope. 
\label{fig:srba}}
\end{figure}

 HE~0207$-$1423 ([Fe/H] = $-$2.95) is a CEMP-s star with the
 unusually large Ba/Sr ratio of [Ba/Sr] = +1.40.  (CEMP-s stars are a
 subclass of CEMP objects that exhibit overabundances of the
 $s$-process elements, defined by [Ba/Fe] > +1.0 and [Ba/Eu] > +0.5; 
 \citealt{beers05}.)  That is to say, the nucleosynthetic site(s) that
 produced the large Ba abundance, [Ba/Fe] = +1.73, in this object,
 yielded only [Sr/Fe] = +0.33, considerably smaller than one might
 have expected.  Inspection of Figure \ref{fig:srba} presents the
 dependence of [Ba/Sr] on [Fe/H] (which we shall discuss in the
 following section), and shows there are other (literature) stars that
 exhibit unusually large [Ba/Sr] values (or low [Sr/Ba]). These stars are
 all CEMP stars with [Fe/H] $\ge$ $-$3.0.

\subsection{Neutron-capture Elements and CEMP Stars}

Although our spectra are of moderate-to-high quality, we were only
able to measure the abundance of two neutron-capture elements, Sr and
Ba. The absence of other neutron-capture elements was not for lack of
effort: in the course of our analysis, we tried to measure the
equivalent widths of numerous lines of La\,{\sc ii}, Ce\,{\sc ii},
Nd\,{\sc ii}, and Eu\,{\sc ii}. That said, our 
measurements of Sr and Ba enable us to comment on three issues.

First, previous studies have found a very large scatter in 
Sr and Ba abundances at low metallicity. 
In Figures \ref{fig:sr} and \ref{fig:ba}, there is a large 
dispersion about the linear fit to the [Sr/Fe] vs.\ [Fe/H] and 
[Ba/Fe] vs.\ [Fe/H] trends, for both dwarfs 
and giants, even after excluding CEMP stars and 2-$\sigma$ outliers. 
In all four cases, the dispersion (ranging from 0.30 dex to 0.60 dex) 
exceeds the typical measurement 
uncertainties (ranging from 0.15 dex to 0.22 dex). 
In Figure \ref{fig:srba}, we consider 
[Ba/Sr] vs.\ [Fe/H]. 
For dwarfs, the dispersion about the linear fit 
to the data (with the usual exclusions) is 0.28 dex. This is comparable to 
the dispersions for [Sr/Fe] (0.36 dex, Figure \ref{fig:sr}) and 
[Ba/Fe] (0.30 dex, Figure \ref{fig:ba}). 
For giants, the dispersion 
about the linear fit to [Ba/Sr] vs.\ [Fe/H] 
is 0.56 dex. Again, the value 
is comparable to the dispersions for 
[Sr/Fe] (0.60 dex, Figure \ref{fig:sr}) and 
[Ba/Fe] (0.43 dex, Figure \ref{fig:ba}). Therefore, 
our data set suggests that the dispersion about the linear 
fit to [Ba/Sr] vs.\ [Fe/H] is not 
smaller than the individual dispersions about the linear fit 
to [Sr/Fe] or [Ba/Fe] vs.\ [Fe/H], 
as might have been expected if a correlation existed between 
the abundances of the two elements. 

\begin{figure}[t!] 
\epsscale{1.2}
\plotone{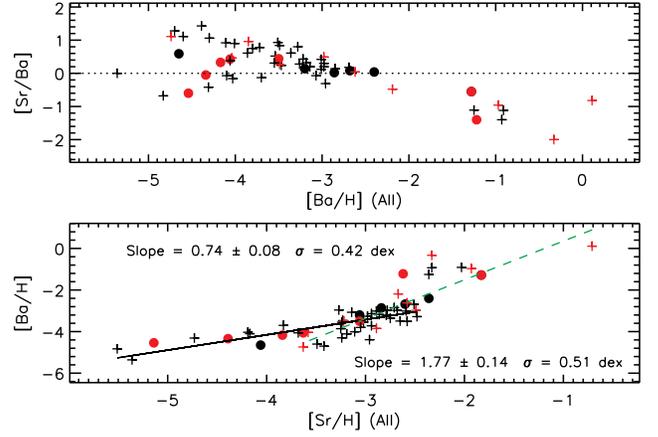} 
\caption{[Sr/Ba] vs.\ [Ba/H] (upper) and 
[Ba/H] vs.\ [Sr/H] (lower). 
The symbols are the same as in Figure \ref{fig:c}. 
In the lower panel, we present 
the linear fit to the data 
(excluding 2-$\sigma$ outliers) with 
[Sr/H] $\le$ $-$2.4 (black solid line) and [Sr/H] $\ge$ $-$3.65 
(green dashed line). For the two lines, the slopes and uncertainties 
are shown, along with the dispersion about the slope. 
\label{fig:neutrona}}
\end{figure}

Secondly, there is evidence that an additional process (or
processes) dominates the production of the lighter neutron-capture
elements.  \citet{francois07} plotted [(Sr,Y,Zr)/Ba] vs.\ [Ba/H] (see
their Figure 15), and noted that (i) for [Ba/H] $\ge$ $-$2.5, all
ratios are close to the solar value, (ii) for $-$4.5 $\le$ [Ba/H]
$\le$ $-$2.5 [(Sr,Y,Zr)/Ba] increases as [Ba/H] decreases, and (iii)
for [Ba/H] $\le$ $-$4.5, all ratios drop to solar. They concluded that
such observations are inconsistent with a single $r$-process.  In
Figure \ref{fig:neutrona} (upper panel), we consider the ratio [Sr/Ba]
vs.\ [Ba/H]. Our data are qualitatively consistent with those of
\citet{francois07} and their results, had we overplotted
 them, would have provided a very good match to ours. 
However, such an 
inhomogeneous comparison would be contrary to the spirit of 
the homogeneous analysis presented herein. 
Considering only our data below [Ba/H] = $-$2.5, 
the dispersion in [Sr/Ba] appears to increase with decreasing [Ba/H]. 
Indeed, there is a hint of two populations below [Ba/H] = $-$4.0, one with 
high [Sr/Ba] and the other with solar, or sub-solar, [Sr/Ba], although 
the statistics are still relatively poor. 

The lower panel of Figure \ref{fig:neutrona} shows [Ba/H] vs.\ [Sr/H].
Within the limited data, there is a hint of two (overlapping)
populations, with the boundary at [Sr/H] $\simeq$ $-$3.2. 
Excluding 2-$\sigma$ outliers (and in this case
retaining CEMP stars), we find that the two populations have
significantly different slopes, but very similar dispersions about the
linear fit.  These data may indicate that at lowest Sr and Ba
abundances, the nucleosynthetic process(es) that create Sr and Ba
produces a different ratio of Sr/Ba than the nucleosynthetic
process(es) that operate when the Sr and Ba abundances are higher. The
scaled-solar $r$-process distribution would be represented by a line
of gradient 1.00 in the [Sr/H] vs.\ [Ba/H] plane.  For the two
populations that we identify, neither slope
matches the main $r$-process line, which may suggest that there are two
(or more) nucleosynthetic sites producing Sr and Ba at low metallicity
that are distinct from the main $r$-process.  Ignoring the sole point
with [Sr/H] $\ge$ $-$1.5, one might still argue that the data with
highest [Sr/H] (admittedly mainly CEMP objects) do not appear to lie
on a linear extrapolation of the line fitting the data with lower
[Sr/H]. Again, we note in passing that the \citet{francois07} data
confirm the trends seen in our data. Had we included their measurements 
in the linear fits to the data described above, we would have obtained
essentially identical slopes, uncertainties, and dispersions about the
linear fits. However, once again we do not overplot these data to
avoid inhomogeneous comparisons.  \citet{travaglio04} noted the need
for a primary process to produce Sr, Y, and Zr at low metallicity that
was different from the $s$- and $r$-process. They referred to this as
a lighter element primary process (LEPP). At face value, 
  our results are consistent with the suggestion there may be two
components to the \citet{travaglio04} LEPP. 
\citet{roederer10} suggest that high-entropy neutrino winds from 
core-collapse supernovae can explain the diversity of neutron-capture element 
abundances found at low metallicity. It would be interesting to examine whether 
our Sr and Ba measurements can be explained by this scenario. 

Thirdly, the relation between the abundances of C and the neutron
capture elements Ba and Sr may shed light on the process(es) that
created the CEMP-s and CEMP-no classes of stars.  As already noted,
\citet{beers05} defined CEMP-s stars as those with [C/Fe] $\ge$ +1.0,
      [Ba/Fe] $>$ +1.0, and [Ba/Eu] $>$ +0.5, and CEMP-no stars as
      those with [C/Fe] $\ge$ +1.0 and [Ba/Fe] $<$ 0.  The CEMP-s
      stars are the majority population, $\sim$80\% of all CEMP
      objects \citep{aoki10}, and radial-velocity studies have
      revealed that the observed binary frequency in this subclass is
      consistent with a binary fraction of 100\% \citep{lucatello05b}.
      Mass transfer between an asymptotic giant branch (AGB) primary
      onto the currently CEMP-s star is believed to be responsible for
      the C and Ba enhancements, a process that also produces the more
      metal-rich classes of CH and Ba stars.  Additionally, the metallicity
      distributions differ, with CEMP-s being generally more metal
      rich, [Fe/H] $>$ $-$3.0, and CEMP-no being generally more metal
      poor, [Fe/H] $<$ $-$3.0 (e.g., \citealt{cohen06,aoki07,aoki10}).
      \citet{cohen06} and \citet{masseron10} have
        suggested that the CEMP-no stars are likely the extremely
        metal-poor counterparts of the CEMP-s stars. 

\begin{figure}[t!] 
\epsscale{1.2}
\plotone{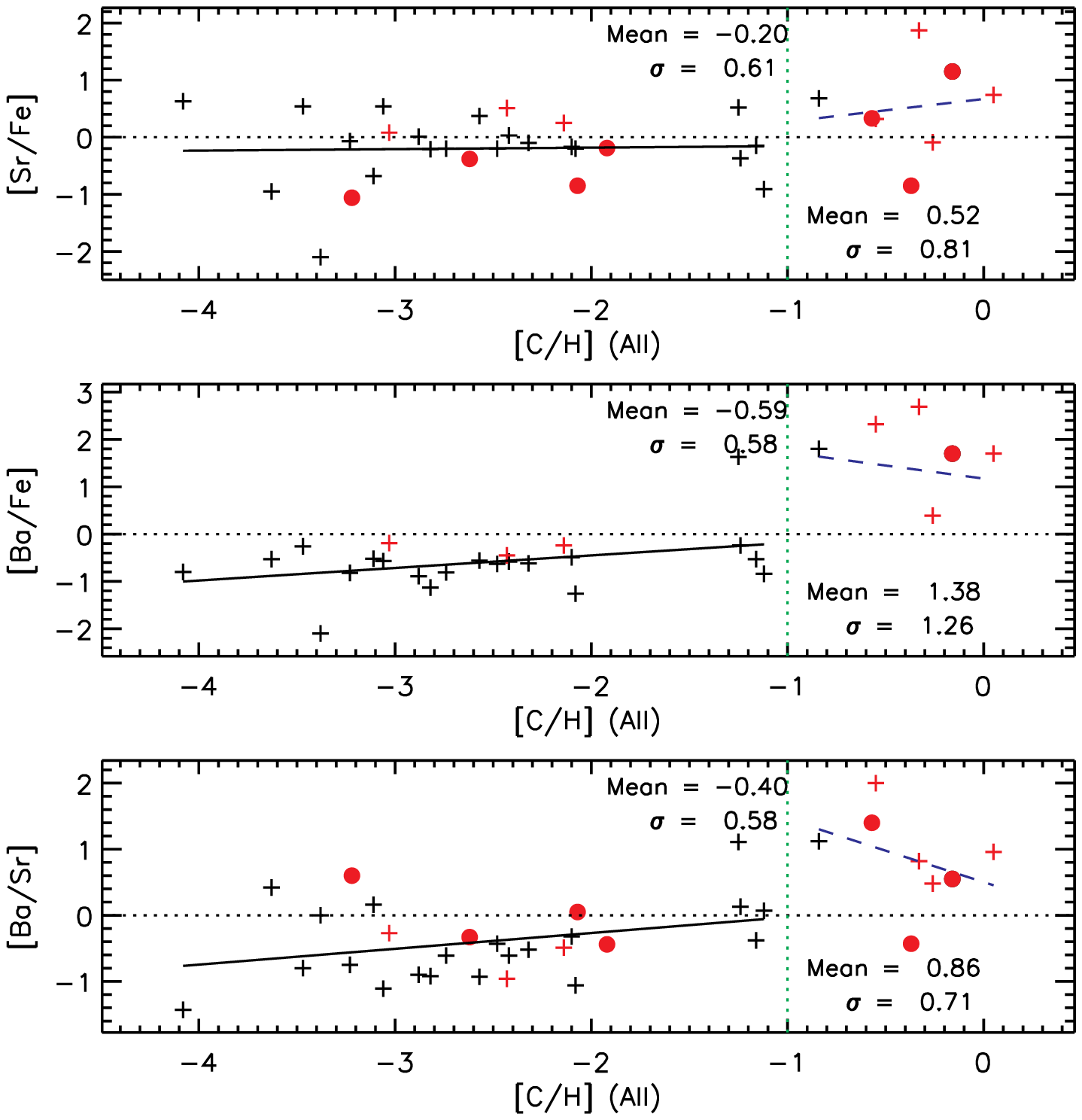} 
\caption{[Sr/Fe] (upper), [Ba/Fe] (middle), and [Ba/Sr] (lower) vs.\ [C/H]. 
The symbols are the same as in Figure \ref{fig:c}. 
\label{fig:neutronb}}
\end{figure}

       Figure \ref{fig:neutronb}
      shows [Sr/Fe], [Ba/Fe], and [Ba/Sr] vs.\ [C/H]. In this figure,
      the CEMP stars with the highest [Sr/Fe] and [Ba/Fe] ratios tend
      to have the highest [C/H] values.
While more data are urgently needed, we speculate that 
the stars which produce CEMP objects with [C/H] $\ge$ $-$1.0 
synthesize large amounts of Sr and Ba. 
The diverse set of abundance ratios seen in the observations 
calls for production sites capable of diverse yields. 
Yields of C and $s$-process elements from AGB stars are 
strongly dependent on mass and metallicity, such that they 
may be good candidates for explaining part, or perhaps most, 
of the large range 
of C and $s$-process element abundances 
(e.g., \citealt{karakas07,karakas10,cristallo11,lugaro12}). 
Regardless, the limited data indicate that stars with 
[C/H] $>$ $-$1.0 have Sr and Ba enhancements and [Ba/Sr] $>$ 0.  
See Papers III and IV in this series for additional discussions of CEMP stars. 

\section{CONCLUDING REMARKS}
\label{sec:conc} 

We present a homogeneous chemical abundance analysis of 16 elements in
190 metal-poor stars, which comprise 38 program stars and 152 
literature stars. The sample includes 86 extremely metal-poor stars,
[Fe/H] $\le$ $-$3.0, and ten new stars with [Fe/H] $\le$ $-$3.5. To
our knowledge this represents one of the largest homogeneous chemical
abundance analyses of extremely metal-poor stars based on a model
atmosphere analysis of equivalent widths measured in high-resolution,
high signal-to-noise ratio spectra.

We find strong evidence for large chemical diversity at low
metallicity.  For a given abundance ratio, [X/Fe], the outliers are
often, but not always, CEMP objects.  Considering dwarfs and giants
separately, we define the linear fit to [X/Fe] vs.\ [Fe/H] excluding
CEMP stars and 2-$\sigma$ outliers. We regard these trends as defining
the ``normal'' population of metal-poor stars. For many elements, the
dispersions about the linear fits are in good agreement with the
scatter expected from measurement uncertainties. Therefore, we believe
that (a) a ``normal'' population exists, and (b) that our crude
selection criteria were able to identify this population.

For several elements, there are clear discrepancies between dwarfs and
giants. The evidence for abundance differences include significantly
different slopes in the [X/Fe] vs.\ [Fe/H] linear fit (Na, Al, Sc, \tiii,
Cr, Mn, and Ni) as well as a hint of differences in the mean
abundances (Na, Si, \tii, Cr, Co, and Ba).  Similar results were found
by \citet{lai08} and \citet{bonifacio09}.  
Another way to identify abundance differences
between dwarfs and giants is to consider [X/Fe] versus \teff. We find
statistically significant trends for many elements studied in this
work. These effects (differences in the slopes of linear fits to
[X/Fe] vs.\ [Fe/H] between dwarfs and giants, differences in mean
abundances between dwarfs and giants, and trends between [X/Fe]
vs.\ \teff) likely signify the presence of non-LTE and/or 3D
effects \citep{asplund05}. 
Therefore, we stress the importance of comparing dwarfs with
dwarfs and giants with giants as well as caution when comparing
abundance trends with nucleosynthesis and/or chemical evolution
predictions. Additionally, abundance differences between dwarfs and
giants may be due, in part, to radiative levitation and gravitational settling.

Using linear fits between [X/Fe] vs.\ [Fe/H] for dwarfs and giants, 
we identified many examples of individual stars with abundance peculiarities, 
including CEMP-no objects, one star with [Si/Fe] = +1.2, 
  one with large [Ba/Sr], and a star with unusually low [X/Fe] for
all elements heavier than Na. While many CEMP stars exhibit peculiar
abundances for elements other than C, we note that there are
chemically peculiar stars which are not CEMP objects.
We find a hint for two nucleosynthetic processes for the production 
of Sr and Ba at lowest metallicity, neither of which match the 
main $s$-process or $r$-process. 

Although the present sample is substantial, there is
clear need for considerably more data at the lowest metallicities.
Further mining of existing data sets (HK, HES, SDSS, etc.) as well as
new and upcoming surveys and facilities (e.g., SkyMapper; \citealt{keller07},
LAMOST; \citealt{zhao06}) should increase the numbers of the most metal-poor stars.
Additionally, future analyses should incorporate non-LTE and 3D
effects and processes.

\acknowledgments

We thank B.\ Edvardsson for computing additional 
MARCS model atmospheres and 
A.\ Chieffi, P.\ Fran{\c c}ois, C.\ Kobayashi, A.\ Korn, and K.\ Venn for providing
electronic data and/or helpful discussions.  
We thank the referee for helpful comments. 
 D.\ Y., J.\ E.\ N., M.\ S.\ B., and M.\ A.\ gratefully acknowledge
 support from the Australian Research Council (grants DP03042613,
 DP0663562, DP0984924, and FL110100012) for studies of the Galaxy's most
 metal-poor stars and ultra-faint satellite systems.  
J.\ E.\ N.\ and
 D.\ Y.\ acknowledge financial support from the Access to Major
 Research Facilities Program, under the International Science
 Linkages Program of the Australian Federal Government. 
Australian
 access to the Magellan Telescopes was supported through the Major
 National Research Facilities program.  
 Observations with the Keck Telescope were made under Gemini exchange
 time programs GN-2007B-C-20 and GN-2008A-C-6. 
N.\ C.\ acknowledges financial support for this work through the
 Global Networks program of Universit\"at Heidelberg and
 Sonderforschungsbereich SFB 881 "The Milky Way System" (subproject
 A4) of the German Research Foundation (DFG). 
T.\ C.\ B.\ acknowledges partial funding of this work from grants
 PHY 02-16783 and PHY 08-22648: Physics Frontier Center/Joint
 Institute for Nuclear Astrophysics (JINA), awarded by the
 U.S. National Science Foundation. 
P.\ S.\ B.\ acknowledges support from the Royal Swedish Academy 
of Sciences and the Swedish Research Council. P.\ S.\ B.\ is a 
Royal Swedish Academy of Sciences Research Fellow supported by 
a grant from the Knut and Alice Wallenberg Foundation. 
   The authors wish to recognize
and acknowledge the very significant cultural role and reverence that
the summit of Mauna Kea has always had within the indigenous Hawaiian
community. We are most fortunate to have the opportunity to conduct
observations from this mountain. 
Finally, we are pleased to acknowledge support from the
European Southern Observatory's Director's Discretionary Time Program.

\noindent{\it Facilities:} {ATT(DBS); Keck:I(HIRES);
 Magellan:Clay(MIKE); VLT:Kueyen(UVES)}

\appendix

\section{Appendix} 
\subsection{Stellar Parameters and Chemical Abundances for the Complete Sample} 

In Table \ref{tab:app}, we provide the coordinates, stellar parameters, and
abundance ratios for all of the program stars and literature stars 
presented in this work. Below is a description of the columns in the 
table. 

(1) Star; 
(2) RA2000; 
(3) DE2000; 
(4) Effective Temperature (\teff); 
(5) Surface Gravity ($\log g$); 
(6) Stellar Metallicity [Fe/H]$_{\rm derived}$; 
(7) CEMP (0 = no, 1 = yes); 
(8) [C/Fe]; 
(9) [N/Fe]; 
(10) [Na/Fe]; 
(11) [Mg/Fe]; 
(12) [Al/Fe]; 
(13) [Si/Fe]; 
(14) [Ca/Fe]; 
(15) [Sc/Fe]; 
(16) [\tii/Fe]; 
(17) [\tiii/Fe]; 
(18) [Cr/Fe]; 
(19) [Mn/Fe]; 
(20) [\feii/H]; 
(21) [Co/Fe]; 
(22) [Ni/Fe]; 
(23) [Sr/Fe]; 
(24) [Ba/Fe]; 
(25) Source. 

The C and N abundances for literature stars were taken directly from the 
literature sources. 
For the abundances of Sr and Ba, in the cases in which the literature 
sources did not provide an equivalent width (i.e., they determined 
abundances using spectrum synthesis), we include (when available) the abundance 
measurements or limits. These literature values are flagged appropriately 
in the table. 

\begin{deluxetable*}{lrrrrrrrrrrrrrrr}
\voffset=1.0in
\hoffset=-0.25in
\tablecolumns{16} 
\tablewidth{0pc} 
\tabletypesize{\tiny}
\tablecaption{Model Atmosphere Parameters and Chemical Abundances, [X/Fe], for the Complete Sample 
(Including Literature Values for Sr and Ba) \label{tab:app}}
\tablehead{ 
\colhead{Star} & 
\colhead{RA2000\tablenotemark{a}} & 
\colhead{DEC2000\tablenotemark{a}} &
\colhead{\teff} & 
\colhead{$\log g$} & 
\colhead{[Fe/H]} &
\colhead{C-rich\tablenotemark{b}} &
\colhead{C\tablenotemark{c}} &
\colhead{N\tablenotemark{c}} &
\colhead{Na} &
\colhead{Mg} &
\colhead{Al} &
\colhead{Si} &
\colhead{Ca} &
\colhead{Sc} &
\colhead{\tii} \\ 
\colhead{} & 
\colhead{} & 
\colhead{} & 
\colhead{(K)} & 
\colhead{(cgs)} & 
\colhead{} &
\colhead{} &
\colhead{} &
\colhead{} &
\colhead{} &
\colhead{} &
\colhead{} &
\colhead{} &
\colhead{} &
\colhead{} &
\colhead{} \\ 
\colhead{(1)} &
\colhead{(2)} &
\colhead{(3)} &
\colhead{(4)} &
\colhead{(5)} &
\colhead{(6)} &
\colhead{(7)} &
\colhead{(8)} &
\colhead{(9)} & 
\colhead{(10)} &
\colhead{(11)} &
\colhead{(12)} &
\colhead{(13)} &
\colhead{(14)} &
\colhead{(15)} &
\colhead{(16)} \\ 
}
\startdata 
CS~22957-022                         &   00 01 45.5 &$-$05 49 46.6 & 5146 & 2.40 & $-$2.92 & 0 &     0.16 &     0.21 &   \ldots &     0.21 &   \ldots &   \ldots &     0.27 &     0.07 &     0.30 \\ 
CS~29503-010                         &   00 04 55.4 &$-$24 24 19.3 & 6570 & 4.25 & $-$1.00 & 1 &     1.07 &  $<$1.28 &  $-$0.04 &  $-$0.06 &   \ldots &   \ldots &     0.11 &   \ldots &     0.28 \\
CS~31085-024                         &   00 08 27.9 &  +10 54 19.8 & 5778 & 4.64 & $-$2.80 & 0 &     0.36 & $<-$0.24 &   \ldots &     0.08 &   \ldots &     0.39 &     0.28 &     0.33 &     0.33 \\ 
BS~17570-063                         &   00 20 36.2 &  +23 47 37.7 & 6233 & 4.46 & $-$2.95 & 0 &     0.40 &   \ldots &   \ldots &     0.28 &   \ldots &     0.51 &     0.33 &     0.77 &     0.60  \\
HE~0024-2523                         &   00 27 27.7 &$-$25 06 28.2 & 6635 & 4.11 & $-$2.82 & 0 &   \ldots &   \ldots &   \ldots &     0.82 &  $-$0.49 &   \ldots &     0.54 &     0.25 &     0.63 
\enddata

\tablerefs{
1 = \citet{aoki02}; 
2 = \citet{aoki06};
3 = \citet{aoki07}; 
4 = \citet{aoki08}; 
5 = \citet{bonifacio07,bonifacio09}; 
6 = \citet{carretta02,cohen02}; 
7 = \citet{cayrel04,spite05,francois07}; 
8 = \citet{christlieb04}; 
9 = \citet{cohen04}; 
10 = \citet{cohen06}; 
11 = \citet{cohen08}; 
12 = \citet{frebel07}; 
13 = \citet{honda04}; 
14 = \citet{lai08}; 
15 = \citet{norris01}; 
16 = \citet{norris07}; 
}

\tablenotetext{a}{Coordinates are from the 2MASS database \citep{2mass}.}
\tablenotetext{b}{1 = CEMP object, adopting the \citet{aoki07} definition and 0 = C-normal (see Section 7.1 for details).}
\tablenotetext{c}{For literature stars, [C/Fe] and [N/Fe] values are taken from the literature reference.}
\tablenotetext{d}{For this set of results, a dwarf gravity is assumed (see Section 2.1 for details).}
\tablenotetext{e}{For this set of results, a subgiant gravity is assumed (see Section 2.1 for details).}
\tablenotetext{f}{These abundances, or limits, were taken from the literature source.} 

\tablerefs{
Note. Table 9 is published in its entirety in the electronic edition of The Astrophysical Journal. A portion is shown here for guidance regarding its form and content.
}

\end{deluxetable*}

\end{document}